\pdfoutput=1
\documentclass[11pt]{article}
\usepackage[pdftex]{graphicx,color} 
\usepackage{jheppub}
\usepackage{amsmath}
\usepackage{amssymb}
\usepackage{ytableau}
\usepackage{comment}

\usepackage{multirow}
\usepackage{mathtools}
\usepackage{bm}

\newcommand{\bali}{\begin{align}}
\newcommand{\eali}{\end{align}}
\newcommand{\bea}{\begin{equation}\begin{aligned}}
\newcommand{\eea}[1]{\label{#1}\end{aligned}\end{equation}}
\newcommand{\beg}{\begin{equation}\begin{gathered}}
\newcommand{\eeg}[1]{\label{#1}\end{gathered}\end{equation}}

\renewcommand\Re{\operatorname{Re}}
\renewcommand\Im{\operatorname{Im}}

\graphicspath{ {./figures/} }

\newcommand{\beq}{\begin{equation}}
\newcommand{\eeq}{\end{equation}}

\newcommand{\D}{\Delta}
\newcommand{\e}{\text{e}}

\newcommand{\bra}{\big \langle}
\newcommand{\ket}{\big \rangle}

\newcommand{\calO}{\mathcal{O}}
\newcommand{\calN}{\mathcal{N}}

\newcommand{\calD}{\mathcal{D}}

\newcommand{\barz}{{\bar z}}
\newcommand{\barx}{{\bar x}}
\newcommand{\hatx}{{\hat x}}
\newcommand{\hatp}{{\hat p}}

\newcommand{\muh}{{\hat\mu}}
\newcommand{\nuh}{{\hat\nu}}
\newcommand{\rhoh}{{\hat\rho}}
\newcommand{\sigmah}{{\hat\sigma}}
\newlength\Colsep	
\title{Bounds for OPE coefficients on the Regge trajectory}
\author{Miguel S. Costa$ ^{\dagger}$,}
\author{Tobias Hansen$ ^{\dagger}$,}
\author{Jo\~ao Penedones$ ^{\Diamond}$}
\affiliation{$ ^{\dagger}$Centro de F\'\i sica do Porto,
Departamento de F\'\i sica e Astronomia\\
Faculdade de Ci\^encias da Universidade do Porto\\
Rua do Campo Alegre 687,
4169--007 Porto, Portugal}
\affiliation{$ ^{\Diamond}$Institute of Physics, \'Ecole Polytechnique F\'ed\'erale de Lausanne (EPFL), \\
Rte de la Sorge, BSP 728, CH-1015 Lausanne, Switzerland}
\emailAdd{miguelc@fc.up.pt, thansen@posteo.de, joao.penedones@epfl.ch}


\abstract{
We consider the Regge limit of the CFT correlation functions $\langle {\cal J} {\cal J} {\cal O}{\cal O}\rangle$ and $\langle TT {\cal O}{\cal O}\rangle$, where ${\cal J}$ is a vector current, $T$ is the stress tensor and ${\cal O}$ is some scalar operator. These correlation functions are related by a type of  Fourier transform to the AdS phase shift of the dual 2-to-2 scattering process. AdS unitarity was conjectured some time ago to be positivity of the  imaginary part of this bulk phase shift. This condition was recently proved using purely CFT arguments. For large $N$ CFTs we further expand on these ideas, by considering the phase shift  in the Regge limit, which is dominated by the leading Regge pole with spin $j(\nu)$, where $\nu$ is a spectral parameter. We compute  the phase shift as a function of the bulk impact parameter, and then use AdS unitarity to impose bounds on the analytically continued OPE coefficients $C_{{\cal J}{\cal J}j(\nu)}$ and $C_{TTj(\nu)}$ that describe the coupling to the leading Regge trajectory of the current ${\cal J}$ and stress tensor $T$. AdS unitarity implies that the OPE coefficients associated to non-minimal couplings of the bulk theory vanish at the intercept value $\nu=0$, for any CFT. Focusing on the case of large gap theories, this result can be used to show that the physical OPE coefficients $C_{{\cal J}{\cal J}T}$ and $C_{TTT}$, associated to non-minimal bulk couplings, scale with the gap $\Delta_g$ as $\Delta_g^{-2}$ or $\Delta_g^{-4}$. Also, looking directly at the unitarity condition imposed at the OPE coefficients $C_{{\cal J}{\cal J}T}$ and $C_{TTT}$ results precisely in the known conformal collider bounds, giving a new CFT derivation of these bounds. We finish with remarks on finite $N$ theories and show directly in the CFT that the spin function $j(\nu)$ is convex, extending this property to the continuation to complex spin.
}

\begin{document}
\maketitle

%%%%%%%%%%%%%%%%%%%%%%%%%%%%%%%%%%%%%%%%%%%%%%%%%%%%%%%%%%%%%%%%%%%%%%%%%%%%%%%%%%%%%%%%%%%
\section{Introduction}
%%%%%%%%%%%%%%%%%%%%%%%%%%%%%%%%%%%%%%%%%%%%%%%%%%%%%%%%%%%%%%%%%%%%%%%%%%%%%%%%%%%%%%%%%%%

Correlation functions of local operators in Conformal Field Theory (CFT) are determined by a set of numbers - scaling dimensions and Operator Product Expansion (OPE) coefficients - known as the CFT data. These numbers are not arbitrary because they must be compatible with OPE associativity, unitarity and the existence of a local stress energy tensor. It would be very useful to find an organizing principle for the CFT data. In this paper, we explore the idea of Regge trajectories as organizing principle.

\begin{figure}[t!]
\begin{centering}
\includegraphics[scale=0.4]{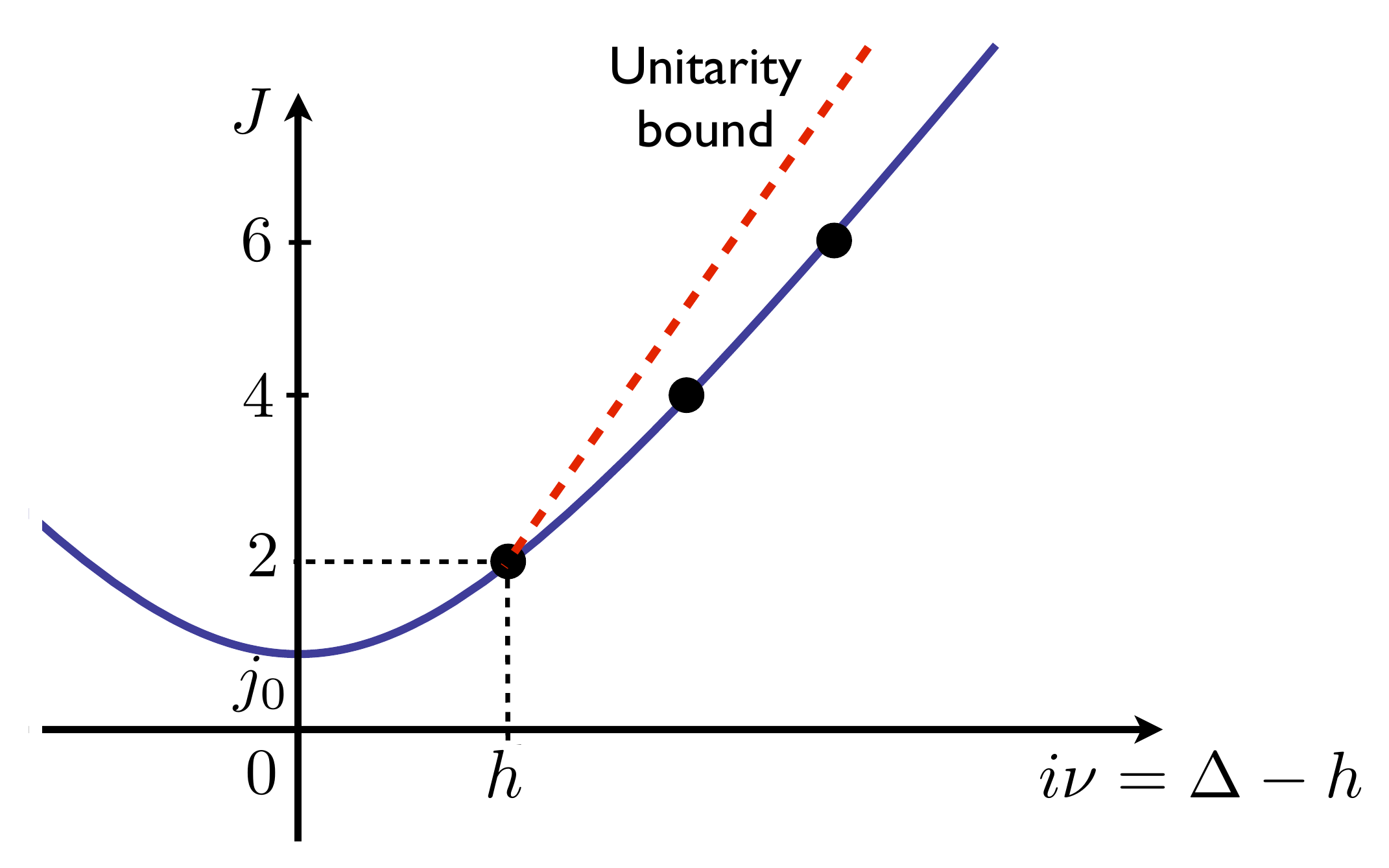}
\par\end{centering}
\caption{\label{fig:leadingReggeTraj}
Shape of the leading Regge trajectory $J=j(\nu)$ with vacuum quantum numbers in a CFT. The dimension
of operators $\Delta$ is related to the spectral parameter $\nu$ by $\Delta=h+i\nu $ where 
$h=d/2$. The function $j(\nu)$ is even and convex. The minimum (for imaginary $\nu$) is the intercept $j(0)\equiv j_0$.}
\end{figure}

Different kinematical limits focus on different subsets of the CFT data. One such example is the light-cone limit \cite{Fitzpatrick:2012yx,Komargodski:2012ek,Alday:2016njk,Alday:2016jfr}
which has recently been used to prove the conformal collider bounds \cite{Hofman:2008ar} and the CEMZ bounds \cite{Camanho:2014apa} on OPE coefficients of conserved currents and stress-tensors from the CFT side
\cite{Hartman:2015lfa,arXiv:1511.08025,arXiv:1603.03771,Hartman:2016lgu,Afkhami-Jeddi:2016ntf}.
Here we study the Regge limit of CFT four-point correlators \cite{Cornalba:2007fs,Costa:2012cb}, 
which are dominated by the leading Regge trajectory, i.e.  the set of operators of lowest dimension $\Delta(J)$ for each even spin $J$. 
We focus on  correlators for which the exchanged Regge trajectories have the 
vacuum quantum numbers. In this 
case the leading trajectory encodes a lot of interesting physics, since its
first operator is the stress tensor. In particular, in  \cite{Nachtmann:1973mr,Komargodski:2012ek}
it was shown that this trajectory is convex, as depicted in figure \ref{fig:leadingReggeTraj}.
The argument involves a deep inelastic scattering thought experiment in a gapped phase obtained by deforming the CFT with a relevant operator.  
However, it  has been shown recently \cite{Caron-Huot:2017vep}, that this trajectory admits a continuation into complex spin $J$.
Using this new result we shall be able to prove 
the convexity directly in the CFT, showing also that this property extends to the continuation to non-integer spin.
As we shall review, it is this continuation that controls the Regge limit of the four-point function. In particular, the high energy growth of the correlator is determined by the value of the intercept $j_0$ shown in figure \ref{fig:leadingReggeTraj} and defined by
$\Delta(j_0)=d/2$.

The leading Regge trajectory also plays a central role in holographic CFTs.
In this context, we assume a large $N$ expansion and consider the leading trajectory of single-trace operators. In the gravity limit, the absence of light higher spin fields in the bulk implies a large gap in the operator dimensions, i.e.\ that $\Delta_{g}\equiv \Delta(J=4) \gg 1$. Therefore, large $N$ and large $\Delta_{g}$ are necessary conditions for the emergence of a local bulk dual. It is also natural to conjecture that these conditions are sufficient for bulk locality \cite{Heemskerk:2009pn}. 
There has been a significant amount of work testing this conjecture. 
More concretely, we would like to prove that CFTs with large $N$ and large $\Delta_{g}$ have 
 other expected universal properties of gravitational theories in AdS.
 One such property is that tree-level high energy scattering is dominated by graviton exchange.
 In CFT language, this means that the intercept $j_0 \to 2$ as $\Delta_{g}\to \infty$.
 Convexity of the single-trace leading Regge trajectory would automatically imply this result. However, the convexity property that we prove in appendix \ref{sec:convexity} only applies to the exact leading Regge trajectory of the finite $N$ theory. This is further discussed in our concluding remarks.

Another expected property of tree-level high energy scattering in gravitational theories is that the higher derivative couplings to the graviton are suppressed by the mass scale of higher spinning particles.
In the gravitational context, this follows from causality \cite{Camanho:2014apa}.
Therefore, in the CFT language, we should be able to prove that some OPE coefficients are suppressed by powers of $\Delta_{g}\gg 1$.
Consider for example the three graviton coupling. 
The bulk effective action can be written schematically as
\beq
\frac{1}{16\pi G_N} \int d^{d+1}x \sqrt{g} \left[ \frac{d(d-1)}{\ell^2}+ \mathcal{R} + \alpha_2 \ell^2 \mathcal{R}^2 +\alpha_4 \ell^4 \mathcal{R}^3+\dots \right]  ,
\eeq
where $\ell$ is the radius of the AdS solution when the higher derivative dimensionless couplings $\alpha_2$ and $\alpha_4$ vanish.
The authors of \cite{Camanho:2014apa} showed that causality implies the effective field theory scaling 
\beq
\alpha_2 \ell^2 \sim \frac{1}{m_{g}^2}\,,\qquad
\qquad
\alpha_4 \ell^4 \sim \frac{1}{m_{g}^4}\,,
\eeq
where $m_{g}$ is the mass  of higher spin particles.
In CFT language, this translates into a statement about the three point function of the stress tensor.
In any CFT, this can be written as
\beq
\langle T T T\rangle = \langle T T T\rangle_{\mathcal{R}}+
\alpha_2 \langle T T T\rangle_{\mathcal{R}^2}+
\alpha_4 \langle T T T\rangle_{\mathcal{R}^3}\,,
\eeq
where each term corresponds to a different tensor structure.
We would like to prove that 
\beq
C_{TTT}^{(2)} \sim
\alpha_2   \sim \frac{1}{\Delta_{g}^2}\,,\qquad
\qquad
C_{TTT}^{(3)} \sim
\alpha_4 \sim \frac{1}{\Delta_{g}^4}\,,
\label{eq:alpha_scaling_gap}
\eeq
where $C_{TTT}^{(i)}$ are OPE coefficients.
This has been argued in \cite{Hartman:2016lgu,Afkhami-Jeddi:2016ntf,Kulaxizi:2017ixa,Li:2017lmh}. Here we provide another argument based on unitarity of the bulk phase shift conjectured a while ago in \cite{Cornalba:2008sp} and recently proved in \cite{Kulaxizi:2017ixa}. 

In section \ref{sec:CRT}, we review Conformal Regge Theory and generalize it for the four-point function of two stress tensors and two scalar operators. Section \ref{sec:ads_unitarity_from_cft} reviews the recent proof \cite{Kulaxizi:2017ixa} of the AdS unitarity condition and determines subleading contributions, which allow us to analyze the validity of the condition.
In section \ref{sec:saddle_point_scalar} the AdS unitarity condition is used to derive bounds on 
OPE coefficients of two currents (or two stress tensors) and operators of the leading Regge trajectory.
The phase shift is computed using a saddle point approximation, where the location $\nu_0$ of the saddle depends on the
AdS impact parameter $L$. By varying $L$ one can move the saddle point to different interesting points on the Regge trajectory,
starting from the  intercept at $\nu_0=0$, to the stress-tensor  at $\nu_0=\pm i h$ and to the 
spin 4 operator $\calO_{J=4}$ at $\nu_0=\pm i (\D_g - h)$.
The resulting bounds are summarized in table \ref{tab:summary_of_bounds}. 
In particular, with mild assumptions on the behaviour of  OPE coefficients in the large gap limit, we are able to show (\ref{eq:alpha_scaling_gap}).
We conclude in section \ref{sec:remarks} with some remarks on finite $N$ CFTs.
The appendices contain technical details and a proof of the convexity of the leading Regge trajectory.
\begin{table}[h!]
\centering
{\renewcommand{\arraystretch}{1.5}
\begin{tabular}{ c| c| c| c| c}
$\nu_0$ & $\ J\ $ & $L$ & bounds on $C_{TT j(\nu_0)}^{(i)}$  & $\D_g$\\ \hline
0 & $j_0$ & 0 & $C_{TT j(0)}^{(2)} = C_{TT j(0)}^{(3)} = 0$ & any \\
$\pm i h$ & $2$ &$\sim \ln S$ & conformal collider bounds \cite{Hofman:2008ar} & any \\
$\pm i (\D_g - h)$ & $4$ & $\sim \ln S$ &$\frac{C_{TT \calO_{\!J=4}}^{(2)}}{C_{TT \calO_{\!J=4}}^{(1)}} \lesssim \frac{1}{\D_g^2}, \ \frac{C_{TT \calO_{\!J=4}}^{(3)}}{C_{TT \calO_{\!J=4}}^{(1)}}\lesssim \frac{1}{\D_g^4}$ & 
\begin{tabular}{ c}
$\D_g\gg 1$ and\vspace{-0.3cm}
\\
flat space limit 
\end{tabular}
\end{tabular}
}
\caption[]{Summary of bounds on the leading Regge trajectory.
For theories with a large $\D_g$, we show that the bounds at $\nu_0 = 0$ imply \eqref{eq:alpha_scaling_gap}.
In order to derive the bounds at $\nu_0 =\pm i (\D_g - h)$, given in \eqref{eq:bound_J4_T}, we have to impose in addition to large $\D_g$
the stronger condition that there is a well defined flat space limit.
Analogous bounds are obtained for the OPE coefficient with two conserved currents.}
\label{tab:summary_of_bounds}
\end{table}

%%%%%%%%%%%%%%%%%%%%%%%%%%%%%%%%%%%%%%%%%%%%%%%%%%%%%%%%%%%%%%%%%%%%%%%%%%%%%%%%%%%%%%%%%%%
\section{Conformal Regge theory}
\label{sec:CRT}
%%%%%%%%%%%%%%%%%%%%%%%%%%%%%%%%%%%%%%%%%%%%%%%%%%%%%%%%%%%%%%%%%%%%%%%%%%%%%%%%%%%%%%%%%%%

In this section we will review  the main formulae for the Regge limit of CFT correlators \cite{Cornalba:2007fs,Costa:2012cb}. 
For the sake of clarity, we start with the case of correlators of scalar operators, leaving the complications of 
distinct tensor structures that arise for external spinning operators for subsequent subsections. In order to prepare the ground
to derive non-trivial bounds for OPE coefficients we will finish this section with the case 
of two vector  currents and two scalars already derived in \cite{Cornalba:2009ax}, and then present the extension to the case of 
two stress tensors and two scalars.

%%%%%%%%%%%%%%%%%%%%%%%%%%%%%%%%%%%%%%%%%%%%%%%%%%%%%%%%%%%%%%%%%%%%%%%%%%%%%%%%%%%%%%%%%%%
\subsection{Regge kinematics}
%%%%%%%%%%%%%%%%%%%%%%%%%%%%%%%%%%%%%%%%%%%%%%%%%%%%%%%%%%%%%%%%%%%%%%%%%%%%%%%%%%%%%%%%%%%

We start with the  four-point correlation function 
$A(y_i)= \big\langle  {\cal O}_1(y_1){\cal O}_2(y_2){\cal O}_3(y_3) {\cal O}_4(y_4) \big\rangle$, 
of  four scalar operators of dimension $\Delta_i$ placed at $y_i$.
We will be interested in the case ${\cal O}_1={\cal O}_2$ and ${\cal O}_3={\cal O}_4$. In this case we can write
\begin{equation}
A(y_i)=  \frac{ {\cal A}({z,\bar{z}})}{(y_{12})^{2\Delta_1} (y_{34})^{2\Delta_3}} \,,
\label{eq:4pt-function_scalars_of_zzb}
\end{equation}
where $z,\bar{z}$ are the usual cross ratios
\begin{equation}
z\bar{z} = \frac{y_{12}y_{34}}{y_{13}y_{24}}\,,\ \ \ \ \ \ \ \ \ \
(1-z)(1-\bar{z}) = \frac{y_{14}y_{23}}{y_{13}y_{24}}\,.
\end{equation}
We shall normalize the operators according to $\big\langle {\cal O}_i(y) {\cal O}_i(0) \big\rangle = 1/ y^{2\Delta_i}$.

\begin{figure}[t!]
\begin{centering}
\includegraphics[scale=0.4]{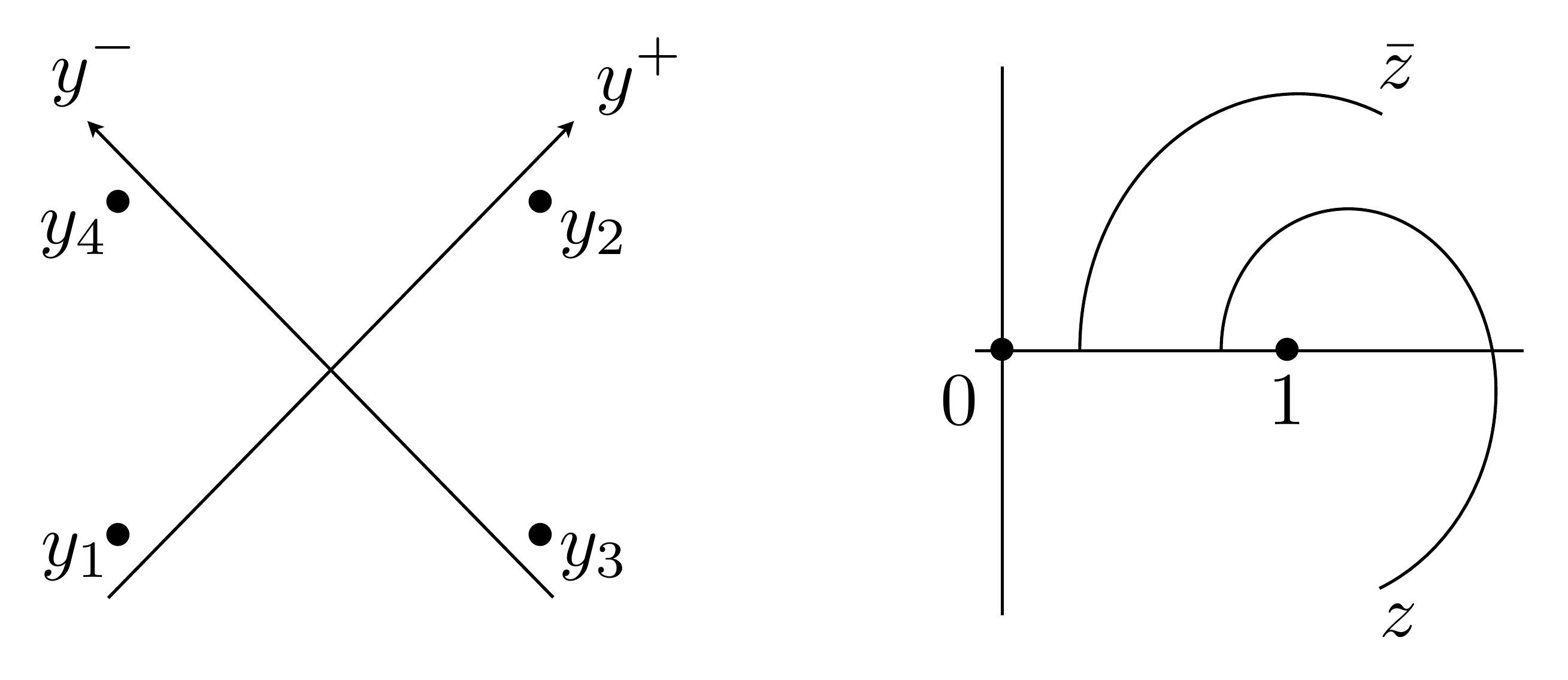}
\par\end{centering}
\caption{\label{fig:ReggeKinematics}
Regge kinematics requires $y_{13}^2,y_{24}^2>0$ and $y_{14}^2,y_{23}^2<0$. The left panel shows the Regge kinematics for 
$y_{12}^2,y_{34}^2>0$, but that is not necessary since we may allow $y_2$ to cross the light cone of $y_1$ and 
$y_4$ the light cone of $y_3$. The right panel shows the path of the cross ratios $z,\bar{z}$ as we
analytically continue from  the Euclidean
region to the Lorentzian one.}
\end{figure}

We wish to consider a CFT in $d$-dimensional Minkowski space  $\mathbb{M}^d$ and study the Regge limit of the above correlation function. 
 In light-cone coordinates $y=(y^+,y^-, y_\perp)$, where 
$y_\perp$ is a point in transverse space $\mathbb{R}^{d-2}$, the Regge limit is defined by 
\begin{equation}
y_1^+\rightarrow -\infty\,,\ \ \ \ \ \ 
y_2^+\rightarrow +\infty\,,\ \ \ \ \ \ 
y_3^-\rightarrow -\infty\,,\ \ \ \ \ \ 
y_4^-\rightarrow +\infty\,,\ \ \ \ \ \ 
\end{equation}
while keeping $y_i^2$ and $y_{i\perp}$ fixed. In particular we shall  keep
the causal relations $y_{14}^2,y_{23}^2<0$ and all the other $y_{ij}^2>0$.
This Lorentzian correlation function, with time ordered operators, is obtained by analytic continuation from the Euclidean one
where $\bar{z}=z^*$ \cite{Cornalba:2006xk}. With the above kinematics, the correct prescription is to fix $\bar{z}$ and rotate $z$ anti-clockwise around the branch point at $z=1$. In the Lorentzian sheet both $z$ and $\bar{z}$ are real. Figure \ref{fig:ReggeKinematics} shows the kinematics and analytic continuation.
  
\begin{figure}[t!]
\begin{centering}
\includegraphics[scale=0.35]{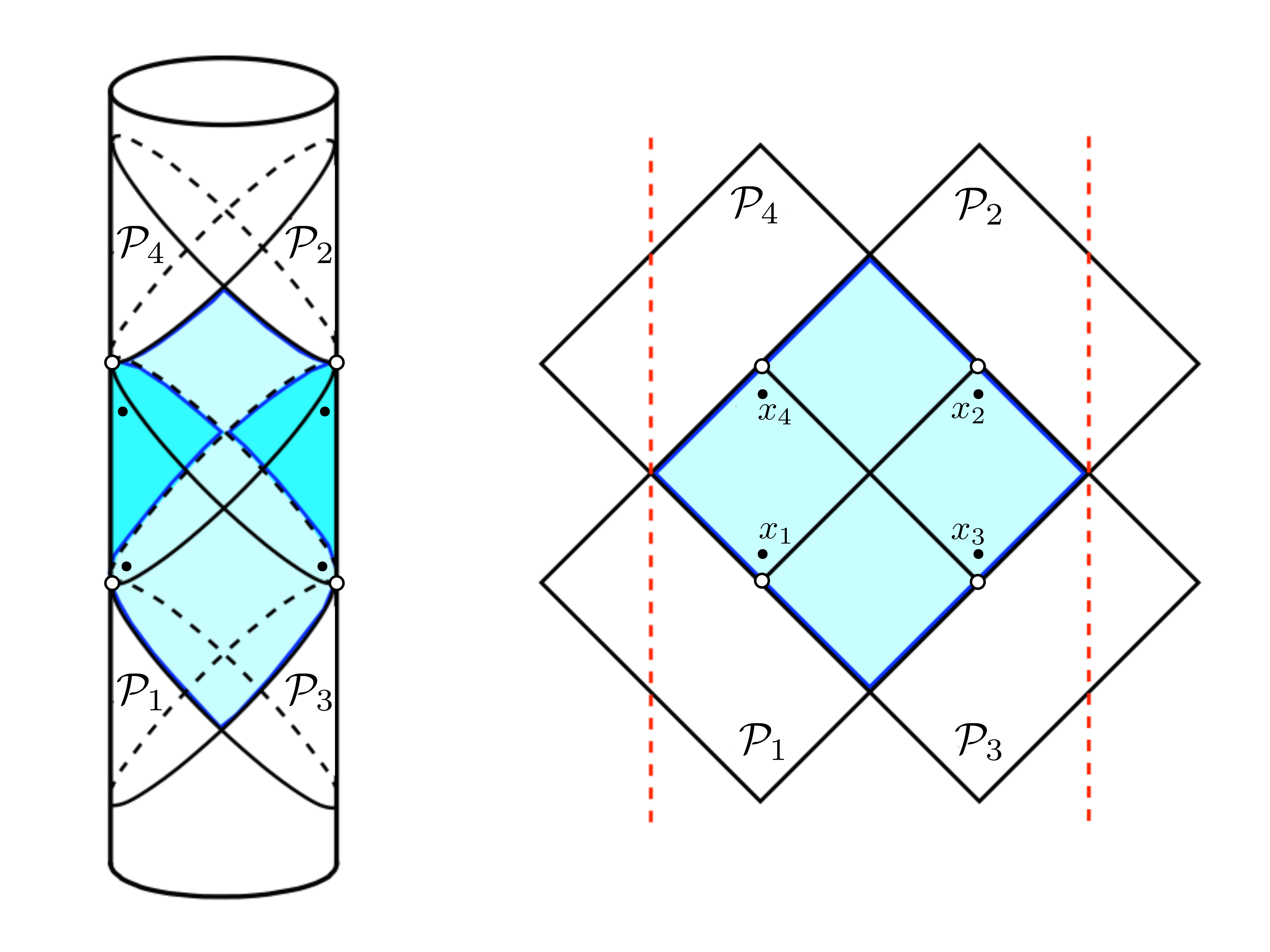}
\par\end{centering}
\caption{\label{fig:PoincarePatches}
The CFT can be defined on the Lorentzian cylinder (left figure). By a conformal transformation one can move to a Poincar\'e  patch, defining the 
theory on Minkowski space, that covers only a portion of the cylinder. The central Poincar\'e patch, where operator insertions are close to null infinity, is shown in blue. 
One may instead consider  Poincar\'e patches whose origins, shown as white dots, are at null infinity of the central Poincar\'e patch. 
The operator ${\cal O}_i$ is then inserted very close to the origin of the  Poincar\'e patch ${\cal P}_i$, where we use coordinates $x_i$.
To visualize the different Poincar\'e patches it is convenient to open the cylinder (right figure). The red lines are identified in this picture.}
\end{figure}

A convenient parameterization of the correlation function considers a   conformal transformation on each point,
 such that each point is close to the origin of a Poincar\'e patch. This can be done with the transformations
\begin{align}
&x_i= (x_i^+,x_i^-, x_{i\perp})  = - \frac{1}{y_i^+} \left(1, y_i^2, y_{i\perp} \right)\,,\ \ \ \ \ \ \ \ \  i=1,2\,,
\label{eq:conformalP1P3}
\\
&x_i= (x_i^+,x_i^-, x_{i\perp})  = - \frac{1}{y_i^-} \left( 1,y_i^2, y_{i\perp} \right)\,,\ \ \ \ \ \ \ \ \ i=3,4\,.
\end{align}
The Regge limit is now the limit $x_i\rightarrow 0$. Notice, however, that 
the operators are not close to each other, since the $x_i$ are close to the origin of distinct Poincar\'e patches.
Figure \ref{fig:PoincarePatches} shows the different Poincar\'e patches, which cover a portion of the Lorentzian cylinder 
$\mathbb{R}\times S^{d-1}$ where we can define our theory. Note that this cylinder
can be thought as belonging to the boundary of global AdS space, although our arguments are purely based in CFT.
Studying the action of the conformal group on the different patches it is possible to show that, in the Regge limit, the correlation function can only depend on 
the combinations \cite{Cornalba:2009ax}
\begin{equation}
x\approx x_1 - x_2\,,\ \ \ \ \ \ \ \ \ \ \ \ 
\bar{x}\approx x_3 - x_4\,.
\label{xxbar}
\end{equation}
Moreover, we can write the only two independent cross ratios 
\begin{equation}
\sigma^2 = x^2 \bar{x}^2 \,,\ \ \ \ \ \ \ \ \  \cosh\rho = -\frac{x\cdot\bar{x}}{|x||\bar{x}|} \,,
\label{SigmaRho}
\end{equation}
so that the Regge limit corresponds to sending $\sigma\rightarrow 0$ with $\rho$ fixed.
Notice that these cross ratios are related to 
$z,\bar{z}$ by
$z\bar{z} = x^2 \bar{x}^2$ and $z+\bar{z} = -2x\cdot\bar{x}$.

Following the standard transformation for conformal primaries 
\begin{equation}
\mathcal{O}(y)= \left| \frac{\partial x}{\partial y}\right|^{\frac{\Delta}{d}} \mathcal{O}(x)\ ,
\end{equation}
the transformed correlation function 
$A(x_i)= \big\langle {\cal O}_1(x_1) {\cal O}_2(x_2) {\cal O}_3(x_3) {\cal O}_4(x_4) \big\rangle$
is related to the original one by
\begin{equation}
A(y_i) = (-y_1^+ y_2^+)^{-\Delta_1}  (-y_3^- y_4^-)^{-\Delta_2} \, A(x_i)\,.
\label{eq:transformAyAx}
\end{equation}
For example, for the disconnected part of the correlation function we have 
\begin{equation}
\langle {\cal O}_1(x_1) {\cal O}_2(x_2) \rangle = \frac{1}{
\left(  -x^{2} + i\epsilon_{x}\right)^{\Delta_1}}\,,\ \ \ \ \ \
\langle {\cal O}_3(x_3) {\cal O}_4(x_4) \rangle = \frac{1}{
\left( - \bar{x}^{2} + i\epsilon_{\bar{x}}\right)^{\Delta_3}}\,,
\label{eq:scalars_normalization}
\end{equation}
where we use the $i\epsilon$-prescription, $ \epsilon_x =  \epsilon \,{\rm sgn}\, x^0 $ \cite{Cornalba:2006xk}.
Notice that each vector $x_i$ is defined with respect to its own Poincar\'e patch ${\cal P}_i$. 
Thus, although both $x$ and $\bar{x}$ are  small vectors, these are two-point functions between points that 
are far apart but approaching the light-cone of each other. For time-like $x$, ${\cal O}_1$ and 
${\cal O}_2$ are space-like related, while for  space-like $x$ they are time-like related.

%%%%%%%%%%%%%%%%%%%%%%%%%%%%%%%%%%%%%%%%%%%%%%%%%%%%%%%%%%%%%%%%%%%%%%%%%%%%%%%%%%%%%%%%%%%
\subsection{Conformal block expansion}
%%%%%%%%%%%%%%%%%%%%%%%%%%%%%%%%%%%%%%%%%%%%%%%%%%%%%%%%%%%%%%%%%%%%%%%%%%%%%%%%%%%%%%%%%%%

We want to  consider the expansion of the correlation function in terms of  $t$-channel  conformal blocks,
\begin{equation}
\mathcal{A}(z,\bar{z})  = \sum_k C_{12k}C_{34k} \, G_{\D_k,J_k}(z,\bar{z})\,,
\label{CBE}
\end{equation}
where $C_{ijk}$ are the OPE coefficients and $G_{\D,J}(z,\bar{z})$ is the conformal block associated to the exchange of a primary of dimension $\Delta$ and spin $J$.
For our purposes, it is more useful to consider the spectral representation \cite{Dobrev:1975ru, Cornalba:2007fs, Costa:2012cb}
\begin{equation}
\mathcal{A}(z,\bar{z})= \sum_J \int_{-\infty}^\infty  d\nu  \, b_J\big(\nu^2\big) \,F_{\nu,J}(z,\bar{z})\,,
\label{CPWEp}
\end{equation}
where 
\begin{equation}
F_{\nu,J}(z,\bar{z})= \kappa_{\nu,J}\, G_{h+i\nu,J}(z,\bar{z})+
\kappa_{-\nu,J}\,G_{h-i\nu,J}(z,\bar{z})\,,
\label{cb+shadow}
\end{equation}
is a sum of two conformal blocks with dimensions $h+i\nu$ and
$h-i\nu$ (we use $h=d/2$ throughout the paper),
with the normalization constant
\begin{equation}
\kappa_{\nu,J} =\frac{i \nu}{2\pi K_{h+i\nu,J}}\,.
\end{equation}
The definition of the function $K_{\D,J}$ is given in equation (\ref{KDeltaJ}) of appendix \ref{sec:discontinuity}.
The two conformal blocks in (\ref{cb+shadow}) satisfy the same differential equation because they have the same Casimir eigenvalue. %$C_{h+i\nu,J}=C_{h-i\nu,J}$.
The second conformal block is usually called the shadow of the first (see for example \cite{SimmonsDuffin:2012uy} for details).
The basis of functions $F_{\nu,J}(z,\bar{z})$ forms a complete basis of single valued functions that satisfy the Casimir
equation \cite{Caron-Huot:2017vep}. They are therefore ideal to expand the Euclidean correlator ${\cal A}(z,\bar{z}=z^*)$.

In order to reproduce, from the spectral representation (\ref{CPWEp}), the contribution in (\ref{CBE}) of a primary of dimension $\D$ and spin $J$ that appears in both OPEs 
$\mathcal{O}_1\mathcal{O}_2$  and
$\mathcal{O}_3\mathcal{O}_4$,
the partial amplitude $b_J(\nu^2)$ must have  poles of the form
\begin{equation}
\nu^2 \to -(\Delta-h)^2\,,\qquad   b_J\big(\nu^2\big)\approx \frac{r(\Delta,J)}{\nu^2+(\Delta-h)^2}\,,
\qquad 
r(\Delta,J)= C_{12J}C_{34J}K_{\D,J}\,.
\label{polesfromCB}
\end{equation}
We remark that the  function $\kappa_{\nu,J}$  has poles corresponding to the  spin $J$ double traces of the type 
$\mathcal{O}_1\partial_{\mu_1} \dots \partial_{\mu_J} \partial^{2m}\mathcal{O}_2$ and 
$\mathcal{O}_3\partial_{\mu_1} \dots \partial_{\mu_J} \partial^{2m}\mathcal{O}_4$. 
This is tailored to the case of large $N$ AdS duals for which a $t$-channel tree level Witten diagram includes 
double trace exchanges \cite{DHoker:1999mic,Cornalba:2006xm}.
For general CFTs these poles are canceled by zeros of the function  $b_J(\nu^2)$.

%%%%%%%%%%%%%%%%%%%%%%%%%%%%%%%%%%%%%%%%%%%%%%%%%%%%%%%%%%%%%%%%%%%%%%%%%%%%%%%%%%%%%%%%%%%
\subsection{Regge theory}
\label{sec:regge_theory}
%%%%%%%%%%%%%%%%%%%%%%%%%%%%%%%%%%%%%%%%%%%%%%%%%%%%%%%%%%%%%%%%%%%%%%%%%%%%%%%%%%%%%%%%%%%

To consider the Regge theory we need to perform an analytic continuation
of the conformal blocks above, corresponding to a Wick rotation from the Euclidean to the Lorentzian correlator.
In terms of the cross ratios $z, \barz$, we need to continue $z$ around $1$ counter clockwise with $\bar{z}$ held fixed. This computation is described in appendix \ref{sec:discontinuity}. 
The result is that after analytic continuation and in the limit $\sigma\rightarrow 0$ at fixed $\rho$, the function $F_{\nu,J}(z,\bar{z})$
is to leading order in $\sigma$ given by
\beq
F_{\nu,J}(z,\bar{z}) \approx -i \pi^h 4^J \sigma^{1-J} \gamma(\nu) \gamma(-\nu)\,
\Omega_{i \nu} (\rho)\,,
\label{eq:ReggelimitofF}
\eeq
where $\gamma(\nu)$, defined in \eqref{eq:gamma},  is a function with poles at the double trace operators 
and $\Omega_{i \nu} (\rho)$ is the harmonic function on hyperbolic space $H_{d-1}$ defined in \eqref{eq:Omega}.
Note that due to the factor $\sigma^{1-J}$ exchanges of large spin contribute most to the Regge
limit. This is the reason why we need to resum all exchanges of the leading Regge trajectory to get a sensible result.
The first step is to rewrite (\ref{CPWEp}) as follows
\begin{equation}
\mathcal{A}(z,\bar{z})= \sum_J \int   d\nu  \, b_J\big(\nu^2\big) 
\,\frac{1}{2}\left[ F_{\nu,J}(z,\bar{z})+
F_{\nu,J}\left(\frac{z}{z-1},\frac{\bar{z}}{\bar{z}-1}\right)
\right]
\,, 
\end{equation}
using the fact that only even spins contribute to the four-point function with $\mathcal{O}_3=\mathcal{O}_4$ and that $F_{\nu,J}\left(\frac{z}{z-1},\frac{\bar{z}}{\bar{z}-1}\right)=(-1)^J F_{\nu,J}(z,\bar{z})$. Notice that this transformation of the cross-ratios corresponds to the  exchange $3\leftrightarrow 4$ (or $1\leftrightarrow 2$).
Next we do the usual Sommerfeld-Watson transform
to replace the sum over $J$ by an integral
\begin{equation}
{\cal A}(z,\bar{z}) = 
\int d\nu \int_C \frac{dJ}{2\pi i} \frac{\pi}{2\sin(\pi J)}\,  b_J\big(\nu^2\big) 
\, \left[ F_{\nu,J}(z,\bar{z})+
F_{\nu,J}\left(\frac{z}{z-1},\frac{\bar{z}}{\bar{z}-1}\right)
\right]\,.
 \label{eq:SWtransform}
\end{equation}
Analytic continuation in the $J$-plane allows us to deform the $J$ contour picking the Regge pole 
with largest ${\rm Re}J$.
In this step, we made the important assumptions that we can drop the contribution from $J=\infty$ and that the leading singularity is a Regge pole. 
More precisely the pole comes from  expanding the denominator of the function (\ref{polesfromCB})
around $J \approx j(\nu)$
\begin{equation}
b_J(\nu^2)\approx -\frac{j'(\nu)\,r(\nu) }{2\nu\big(J-j(\nu)\big)}\,,
\label{eq:Jpane_polesfromCB}
\end{equation}
where $j(\nu)$ is  the inverse function of $\D(J)$ defined by
\begin{equation}
\nu^2+\big(\Delta(j(\nu))-h\big)^2=0\,,
\label{j}
\end{equation}
and we defined the function $r(\nu)$ obtained from the 
analytic continuation of the OPE coefficients that appear in the combination given in (\ref{polesfromCB}),
\begin{equation}
r(\nu) \equiv r \big( h\pm i\nu, j(\nu) \big) = C_{12j(\nu)}C_{34j(\nu)}  K_{h\pm i\nu,j(\nu)} \,.
\label{eq:rtoCC}
\end{equation}
That this analytic continuation is well defined was only recently proved in \cite{Caron-Huot:2017vep}.
Using \eqref{eq:ReggelimitofF}, we conclude that the contribution of this Regge pole is 
\begin{equation}
{\cal A}(\sigma,\rho) =  \int_{-\infty}^\infty  d\nu  \, \alpha(\nu) \,\sigma^{1-j(\nu)} \, \Omega_{i\nu} (\rho)\,,
\label{eq:ARegge}
\end{equation}
where
\begin{equation}
 \alpha(\nu) =-
 \left(  i \cot\!\left( \frac{\pi j(\nu)}{2}\right)  -1 \right)
\pi^{h+1} 4^{j(\nu)}  \gamma(\nu) \gamma(-\nu)   \,\frac{j'(\nu)\,r(\nu) }{4\nu}\,,
\label{eq:alpha}
\end{equation}
and we used that
\begin{equation}
\frac{1+e^{i\pi j(\nu)} }{\sin (\pi j(\nu))} =   \cot\!\left( \frac{\pi j(\nu)}{2}\right)  +i\,.
\end{equation}

The imaginary part of the Regge residue $\alpha(\nu)$ has poles for $j(\nu)$ an even integer corresponding to the elastic exchange of the spin $J$ operators. 
These poles occur on the imaginary $\nu$ axis, given by the condition (\ref{j}) for $j(\nu)=J$.

%%%%%%%%%%%%%%%%%%%%%%%%%%%%%%%%%%%%%%%%%%%%%%%%%%%%%%%%%%%%%%%%%%%%%%%%%%%%%%%%%%%%%%%%%%%
\subsection{AdS physics}
%%%%%%%%%%%%%%%%%%%%%%%%%%%%%%%%%%%%%%%%%%%%%%%%%%%%%%%%%%%%%%%%%%%%%%%%%%%%%%%%%%%%%%%%%%%

Next we describe how to relate the conformal Regge theory to AdS physics. The idea is that,
if a CFT exhibits Regge behavior, there will be a dual
theory for which the Regge behavior arises from the exchange of Regge trajectory of AdS fields. 
The relation to AdS physics can be seen by considering the following transform of the correlation function
\cite{Cornalba:2006xm,Cornalba:2007zb} 
\begin{equation}
A(x,\bar{x})= (-1)^{-\Delta_1-\Delta_3} \int dp\,d\bar{p} \, e^{-2ip\cdot x - 2i\bar{p} \cdot \bar{x}} B(p,\bar{p}) \,,
\label{eq:AtoB}
\end{equation}
where we recall, from (\ref{xxbar}), that $x\approx x_1-x_2$ and $\bar{x}\approx x_3-x_4$. 
Notice that, because of the $i\epsilon$-prescription in (\ref{eq:scalars_normalization}), the correlator $B(p,\bar{p})$ only has support in the future Milne wedge (\emph{i.e.}  for $p^2<0$ and $p^0>0$) . Conformal
symmetry implies it  can be written
in the form
\begin{equation}
B(p,\bar{p}) = \frac{{\cal B}(S,L)}{(-p^2)^{h-\Delta_1} (-\bar{p}^2)^{h-\Delta_3}}\,,
\end{equation}
where
\begin{equation}
S= 4|p| |\bar{p}|\,, \qquad\qquad
\cosh L = -\frac{p\cdot \bar{p}}{|p| |\bar{p}|}\,.
\label{eq:S&L}
\end{equation}
The Regge limit is now $S\rightarrow \infty$ at fixed $L$.

The connection to AdS physics appears when we consider the Fourier transform to momentum space of the original correlation function
\begin{align}
(2\pi)^d\, \delta\left( \sum k_j \right) i \,T(k_j) = \int \prod_{j=1}^4 dy_j \,e^{i  k_j \cdot y_j} \, A(y_j)\ .
\label{Tamp}
\end{align}
Then, writing the amplitude $A(y_j)$ in terms of the transform $\mathcal{B}(S,L)$ and considering the standard
Regge limit of the external momenta $k_i$,
one arrives at the expression \cite{Cornalba:2009ax}
\begin{align}
 T(k_j) \approx- 2is\int dl_\perp
 e^{iq_\perp\cdot l_\perp}
 \int \frac{dr}{r^3}\, \frac{d\bar{r}}{\bar{r}^3} \,
F_{1}(r)\,F_{2}(r)\,F_3(\bar{r})\,F_4(\bar{r})  \,
\mathcal{B}(S,L)  \,,
\label{finalScat}
\end{align}
where $k_1-k_2=q_\perp$ is the transferred momentum,  $l_\perp$ is the impact parameter $l_\perp=y_{1\perp}-y_{3\perp}$ and
the cross ratios $S$ and $L$ defined in (\ref{eq:S&L}) become 
\begin{equation}
S=r\bar{r} s \ ,\ \ \ \ \ \ \ \ \ \ 
\cosh L=\frac{r^2+\bar{r}^2+l_\perp^2}{2r\bar{r}}\,.
\end{equation}
The functions $F_i(r)$ are expressed in terms of Bessel functions and depend on the virtuality $k_i^2$. These functions are given
precisely by the radial dependence of the boundary-bulk propagators of the dual fields of the scalar operators ${\cal O}_i$ that one would obtain
from computing the Witten diagram for the correlation function in the Regge limit.
Thus, (\ref{finalScat}) is mostly fixed by kinematics and acquires the standard AdS form due to conformal symmetry. 
In the AdS language $S$ is related to the total energy of the process with respect to global AdS time, and $L$ is the geodesic distance
in the impact parameter space, which in this case is the $(d-1)$-dimensional hyperboloid $H_{d-1}$.

All the dynamical  information in the AdS impact parameter representation (\ref{finalScat}) is encoded in the function $\mathcal{B}(S,L)$. It is determined by the  
propagator of the exchanged state and the coupling between this state and the external fields. This coupling
is dual to the  OPE coefficient in the CFT side. 
For example, one could consider the exchange 
of a single graviton, or instead the exchange of the entire graviton Regge trajectory. 
For the exchange of a Regge trajectory, the 
transform of (\ref{eq:ARegge}) yields
\begin{equation}
{\cal B}(S,L) =  \int_{-\infty}^\infty  d\nu  \, \beta(\nu) \,S^{j(\nu)-1} \, \Omega_{i\nu} (L)\,,
\label{eq:BRegge}
\end{equation}
with 
\begin{equation}
\alpha(\nu) = \gamma(\nu) \gamma(-\nu) \beta(\nu)\,.
\label{eq:AlphaBeta}
\end{equation}
The function $\beta(\nu)$ can then be read from (\ref{eq:alpha}). 
Notice that the Fourier transform (\ref{eq:AtoB}) automatically takes care of the double trace exchanges that appear
explicitly in $\alpha(\nu) $ but not in $ \beta(\nu)$. Thus,
for AdS physics $\beta(\nu)$ only has poles associated to the exchange of bulk fields.

%%%%%%%%%%%%%%%%%%%%%%%%%%%%%%%%%%%%%%%%%%%%%%%%%%%%%%%%%%%%%%%%%%%%%%%%%%%%%%%%%%%%%%%%%%%
\subsection{Correlators with conserved currents or stress-tensors}
%%%%%%%%%%%%%%%%%%%%%%%%%%%%%%%%%%%%%%%%%%%%%%%%%%%%%%%%%%%%%%%%%%%%%%%%%%%%%%%%%%%%%%%%%%%

In the remainder of this section we consider two cases of four-point correlation functions of operators with spin,
for which non-trivial bounds for OPE coefficients can be derived.
The first is the correlator of two conserved currents ${\cal J}^a$ of dimension  $\D_1=d-1$ and two scalar operators  ${\cal O}$ of dimension 
$\Delta_3$,
\begin{equation}
A^{ab}(y_i)= \big\langle {\cal J}^a(y_1) {\cal J}^b(y_2) {\cal O}(y_3)  {\cal O}(y_4) \big\rangle\,.
\label{eq:4pt-function_j}
\end{equation}
The second case is the correlator of two stress-tensors $T^{ab}$ of dimension  $\D_1=d$ and two scalars,
\begin{equation}
A^{abcd}(y_i)= \big\langle T^{ab}(y_1) T^{cd}(y_2)  {\cal O}(y_3)   {\cal O}(y_4) \big\rangle\,.
\label{eq:4pt-function_st}
\end{equation}
We will use the same conventions as in \cite{Cornalba:2009ax}, except that we exchange $y_2$ and $y_3$.

Following the standard transformation for conformal primaries 
\begin{equation}
{\cal O}(y)= \left| \frac{\partial x}{\partial y}\right|^{\frac{\Delta_3}{d}} {\cal O}(x)\ ,\ \ \ \ \ \ \ \ \ \ \ \  
{\cal J}^a(y) =  \left| \frac{\partial x}{\partial y}\right|^{\frac{\D_1+1}{d}} \frac{\partial y^a}{\partial x^m}\, {\cal J}^m(x)\, ,
\end{equation}
the transformed correlation function
\begin{equation}
A^{mn}(x_i)= \big\langle {\cal J}^m(x_1) {\cal J}^n(x_2) {\cal O}(x_3)  {\cal O}(x_4) \big\rangle\, ,
\label{eq:Aofx_j}
\end{equation}
is related to the original correlation function (\ref{eq:4pt-function_j}) by
\begin{equation}
A^{ab}(y_i) = (-y_1^+ y_2^+)^{-1-\D_1}  (-y_3^- y_4^-)^{-\Delta_3} \,\frac{\partial y_1^a}{\partial x_1^m}\,\frac{\partial y_2^b}{\partial x_2^n}\, A^{mn}(x_i)\,.
\label{eq:transformAyAx_spin}
\end{equation}
The corresponding relation for the transformed correlator of stress-tensors
\beq
A^{mnpq}(x_i)= \langle T^{mn}(x_1) T^{pq}(x_2) {\cal O}(x_3)  {\cal O}(x_4) \rangle\, ,
\label{eq:Aofx_st}
\eeq
is analogous.

We also introduce free-index notation, writing both correlators as polynomials
\beq
A(x,\bar{x}, z_1, z_2) = z_{1m} z_{2n} A^{mn}(x, \barx) \quad \text{ or } \quad z_{1m} z_{1n} z_{2p} z_{2q} A^{mnpq}(x, \barx)\,,
\eeq
where $z_1,z_2$ are polarizations satisfying $z_i^2=0$.

%%%%%%%%%%%%%%%%%%%%%%%%%%%%%%%%%%%%%%%%%%%%%%%%%%%%%%%%%%%%%%%%%%%%%%%%%%%%%%%%%%%%%%%%%%%
\subsubsection{Regge theory}
%%%%%%%%%%%%%%%%%%%%%%%%%%%%%%%%%%%%%%%%%%%%%%%%%%%%%%%%%%%%%%%%%%%%%%%%%%%%%%%%%%%%%%%%%%%

We can run a similar argument as for the correlation function of scalars to obtain the
 contribution of a Regge pole $j(\nu)$ to the four point functions with vectors or with stress-tensors. The result is
\begin{align}
A(x,\bar{x}, z_1, z_2)  \approx  \int 
  d\nu~ \frac{\sum_{k}\alpha_k (  \nu) \,\mathcal{D}_k \,\Omega_{i\nu} ( \rho)}{
( -x^2 + i\epsilon_{x})^{\D_1 +\frac{j(\nu)-1}{2}}  ( - \bar{x}^2+i\epsilon_{\bar{x}})^{\Delta_3+\frac{j(\nu)-1}{2}}     }\,.
\label{ReggeA}
\end{align}
The coefficients $\alpha_k(\nu)$ and the spin $j(\nu)$ encode the dynamical information of the correlation function.
The index $k$ labels the tensor structures that appear in correlators of spinning operators,
which are generated by the differential operators ${\cal D}_{k}$. 
It is natural to construct operators that are homogeneous in $x$ (i.e.\ they only depend only on ${\hat x} \equiv x/|x|$) out of
covariant derivatives on the space $H_{d-1}$. As we explain in appendix \ref{sec:casimir_equation}, in this way each of the operators
generates a solution to the Casimir equation to leading order in the Regge limit $\sigma \to 0$.
For the correlation function with two vectors we have\footnote{These four operators are the same as given in \cite{Cornalba:2009ax} up to terms containing $x \cdot \partial$ which vanish when acting on a function of $\rho$.}
\bea
\mathcal{D}_{1}  &=  z_1 \cdot z_2  + (z_1 \cdot \hat x) (z_2 \cdot \hat x)\,,\\
\mathcal{D}_{2}  &=  -(z_1 \cdot \hat x) (z_2 \cdot \hat x)\,,\\
\mathcal{D}_{3}  &=  (z_1 \cdot \hatx) (z_2 \cdot \nabla) + (z_2 \cdot \hatx) (z_1 \cdot \nabla)\,,\\
\mathcal{D}_{4}  &=  - (z_1 \cdot \nabla) (z_2 \cdot \nabla) + \frac{1}{d-1} \mathcal{D}_{1} \nabla^2 \,.
\eea{DTensors}
For the correlation function with two stress-tensors we choose the operators
\bea
\mathcal{D}_{1}  ={}& P_{m_1 n_1}(z_1) P_{m_2 n_2}(z_2) \big(\eta^{m_1 m_2} + \hatx^{m_1} \hatx^{m_2}\big)
 \big(\eta^{n_1 n_2} + \hatx^{n_1} \hatx^{n_2}\big)\,,\\
\mathcal{D}_{2}  ={}& P_{m_1 n_1}(z_1) P_{m_2 n_2}(z_2) \big(\eta^{m_1 m_2} + \hatx^{m_1} \hatx^{m_2}\big) \nabla^{n_1} \nabla^{n_2} 
-\frac{1}{d-1} \mathcal{D}_{1} \nabla^2 \,,\\
\mathcal{D}_{3}  ={}& \frac{1}{2} P_{m_1 n_1}(z_1) P_{m_2 n_2}(z_2) \big( \nabla^{m_1} \nabla^{m_2} \nabla^{n_1} \nabla^{n_2} +  \nabla^{m_2} \nabla^{m_1} \nabla^{n_2} \nabla^{n_1} \big)\\
&- \frac{1}{d+1} \left( \mathcal{D}_{2} +\frac{1}{d-1} \mathcal{D}_{1} \nabla^2 \right) \Big(3-d +2 \nabla^2\Big) \,,\\
\mathcal{D}_{4}  ={}& (z_1 \cdot \hat x) (z_2 \cdot \hat x) (z_1 \cdot \nabla) (z_2 \cdot \nabla)\,,\\
\mathcal{D}_{5}  ={}& (z_1 \cdot \hat x)^2 (z_2 \cdot \hat x)^2 \,,\\
\mathcal{D}_{6}  ={}& (z_1 \cdot \hat x) (z_2 \cdot \hat x) (z_1 \cdot \nabla) (z_2 \cdot \nabla) \,,\\
\mathcal{D}_{7}  ={}& \Big(z_1 \cdot z_2  + (z_1 \cdot \hat x) (z_2 \cdot \hat x)\Big) \Big( (z_1 \cdot \hat x) (z_2 \cdot \nabla) + (z_2 \cdot \hat x) (z_1 \cdot \nabla) \Big)\,,\\
\mathcal{D}_{8}  ={}& (z_1 \cdot \hat x) (z_2 \cdot \hat x) \Big( (z_1 \cdot \hat x) (z_2 \cdot \nabla) + (z_2 \cdot \hat x) (z_1 \cdot \nabla) \Big)\,,\\
\mathcal{D}_{9}  ={}&  (z_1 \cdot \hat x)^2 (z_2 \cdot \nabla)^2 + (z_2 \cdot \hat x)^2 (z_1 \cdot \nabla)^2\,,\\
\mathcal{D}_{10} ={}& \Big( (z_1 \cdot \hat x) (z_2 \cdot \nabla) + (z_2 \cdot \hat x) (z_1 \cdot \nabla) \Big) (z_1 \cdot \nabla) (z_2 \cdot \nabla) \,,
\eea{DTensors_st}
where we introduced the object
\beq
P_{mn}(z_i) = \Big(z_i^p + (z_i \cdot \hatx) \hatx^p\Big) \Big(z_i^q + (z_i \cdot \hatx) \hatx^q\Big)
 \left(\eta_{pm} \eta_{qn} - \frac{1}{d-1} \eta_{pq} \eta_{mn} \right)\,,
\eeq
to make the first three operators transverse and traceless. This will be a convenient 
choice when we use the same basis of operators in the impact parameter representation.
Furthermore, the subtractions of $\calD_1$ in  $\calD_2$ and of $\calD_1$ and  $\calD_2$ in
$\calD_3$ were chosen such that
\beq
\left( \frac{\partial}{\partial z_1} \cdot \frac{\partial}{\partial z_2} \right)^2 \mathcal{D}_{2} =
\left( \frac{\partial}{\partial z_1} \cdot \frac{\partial}{\partial z_2} \right)^2 \mathcal{D}_{3} = 0\,,
\eeq
to match the convention chosen in \cite{Camanho:2014apa}.

%%%%%%%%%%%%%%%%%%%%%%%%%%%%%%%%%%%%%%%%%%%%%%%%%%%%%%%%%%%%%%%%%%%%%%%%%%%%%%%%%%%%%%%%%%%
\subsubsection{AdS physics}
%%%%%%%%%%%%%%%%%%%%%%%%%%%%%%%%%%%%%%%%%%%%%%%%%%%%%%%%%%%%%%%%%%%%%%%%%%%%%%%%%%%%%%%%%%%

As for the scalar case we wish to relate the Regge behavior of the correlation function $A(x,\bar{x},z_1,z_2)$  to the  phase shift computed in the dual AdS scattering process. 
To that end we introduce the Fourier transform
\begin{align}
A(x,\bar{x},z_1,z_2)=  (-1)^{-\D_1-\D_3} \int dp\,  d\bar{p} \,  e^{ -2 i p\cdot x -2 i \bar{p} \cdot \bar{x}}
B(p,\bar{p},z_1,z_2)\ .\label{AB}
\end{align}
The $i\epsilon$-prescription in (\ref{ReggeA}) implies that $B(p,\bar{p},z_1,z_2)$ only has support   on the 
future light-cones.
The result for future directed timelike vectors $p$ and $\bar{p}$ can be written as
\begin{align}
B(p,\bar{p},z_1,z_2)  = \frac{\mathcal{B}(p,\bar{p},z_1,z_2) }{
( -p^2 )^{h-\D_1 }  ( -\bar{p}^2 )^{h-\D_3 }     }~,
\label{eq:BtoCalB}
\end{align}
with $\mathcal{B}$ given by
\begin{equation}
\mathcal{B}(p,\bar{p},z_1,z_2)  \approx \int  
  d\nu~ S^{j(\nu)-1}\, \sum_{k}\beta_k (  \nu)\,\hat{\mathcal{D}}_k\, \Omega_{i\nu}(L) \,.
  \label{calB}
\end{equation}
The differential operators $ \hat{{\cal D}}_{k}$ have the same form as in (\ref{DTensors}) and (\ref{DTensors_st}) but with $\hatx$ replaced by $\hat p \equiv p/|p|$ (and derivatives now also taken with respect to $\hat p$).
The coefficients $\beta_k(\nu)$ can be written as linear combinations of the $\alpha_k(\nu)$
by computing the Fourier transform \eqref{AB}. These relations are derived in appendix \ref{sec:fourier_transformation}.

%%%%%%%%%%%%%%%%%%%%%%%%%%%%%%%%%%%%%%%%%%%%%%%%%%%%%%%%%%%%%%%%%%%%%%%%%%%%%%%%%%%%%%%%%%%
\subsubsection{Differential operators for conserved operators}
%%%%%%%%%%%%%%%%%%%%%%%%%%%%%%%%%%%%%%%%%%%%%%%%%%%%%%%%%%%%%%%%%%%%%%%%%%%%%%%%%%%%%%%%%%%

In the impact parameter representation the conservation condition for the currents or stress-tensors simply becomes
\beq
p \cdot \frac{\partial}{\partial z_i} \mathcal{B} (p,\bar{p},z_1,z_2)= 0\,.
\label{eq:conservation_B}
\eeq
In the case of the correlation function with stress-tensors, 
this was the reason for the choice of the form of the first three operators in \eqref{DTensors_st},
since  they automatically satisfy the conservation condition
by themselves, and the others do not. For the case of vectors,  the structures built out of 
$\mathcal{D}_{1}$ and $\mathcal{D}_{4}$ satisfy the conservation condition.
Thus for both correlation functions the conservation condition \eqref{eq:conservation_B} becomes
\bea
0 &= \beta_2 = \beta_3 \,, \qquad \quad && \text{for }\langle {\cal J}{\cal J}{\cal O}{\cal O} \rangle\,,\\
0 &= \beta_4 = \beta_5 = \beta_6 = \beta_7 = \beta_8 = \beta_9 = \beta_{10} \,, \qquad \quad &&\text{for }\langle TT{\cal O}{\cal O}\rangle\,.
\eea{eq:conservation_cond_beta}
It is therefore possible to define the amplitude directly in  $H_{d-1}$
by performing a coordinate transformation where we write
 $p=E\,e$ and $\bar{p}=\bar{E}\,\bar{e}$ with $E$ and $\bar{E}$ positive
and $e$ and $\bar{e}$ points in $H_{d-1}$,
\begin{equation}
e= \frac{1}{r}\left(1,r^2+e_\perp^2,e_\perp\right) \,,\ \ \ \ \ \ \ 
\bar{e}=\frac{1}{\bar{r}}\left(1,\bar{r}^2+\bar{e}_\perp^2,\bar{e}_\perp\right)\,.
\label{e}
\end{equation}
We can think of $(r,e_\perp)$ and $(\bar{r},\bar{e}_\perp)$ as the coordinates in impact parameter space (defined by the locus of the dual AdS null geodesics) 
associated to the boundary sources of the vector currents (or stress tensors) and scalar operators, respectively.
In the new coordinate system $p^\mu=(E,r,e_\perp)$, the reduced amplitude  $ \mathcal{B}$ has components
\begin{align}
 \mathcal{B}^{\mu\nu}  =  \frac{\partial p^\mu }{\partial p^m}\frac{\partial p^\nu}{\partial p^n}\,\mathcal{B}^{mn}\,,
 \end{align}
and the metric element is
\begin{equation}
ds^2=-dE^2 + \frac{E^2}{r^2} \left(  dr^2 + de_\perp^2 \right).
\end{equation}
The conservation condition \eqref{eq:conservation_B} now becomes $\mathcal{B}^{\mu E} = \mathcal{B}^{E \nu} = 0$ for the case of vectors, or $\mathcal{B}^{\mu\nu\alpha E} =\dots = \mathcal{B}^{E \mu\nu\alpha} = 0$ for stress-tensors. Thus 
 it is natural to work directly in $H_{d-1}$ at a fixed value of $E$, where we have coordinates $p^{\hat \mu}=(r,e_\perp)$ and metric
\beq
ds^2=\frac{1}{r^2} \left(  dr^2 + de_\perp^2 \right).
\eeq
For two external vector currents the remaining differential operators become
\bea
\hat \calD_{1\ \nuh}^{\muh} &= \delta^{\muh}_{\ \nuh}\,,\\
\hat \calD_{4\ \nuh}^{\muh} &= - \nabla^{\muh}\nabla_{\nuh} - \frac{\nu^2 + \left( \frac{d-2}{2} \right)^2}{d-1}  \delta^{\muh}_{\ \nuh}\,.
\eea{eq:DH3}
Similarly, the operators for two external stress-tensors are
\bea
\hat \calD_{1\  \rhoh \sigmah}^{\muh \nuh} &= \delta^{\{\muh}_{\ \{\rhoh} \delta^{\nuh\}}_{\ \sigmah\}}\,,\\
\hat \calD_{2\  \rhoh \sigmah}^{\muh \nuh} &= \delta^{\{\muh}_{\ \{\rhoh} \nabla^{\nuh\}} \nabla_{\sigmah\}}  + \frac{\nu^2 + \left( \frac{d-2}{2} \right)^2}{d-1}  \delta^{\{\muh}_{\ \{\rhoh} \delta^{\nuh\}}_{\ \sigmah\}}\,,\\
\hat \calD_{3\  \rhoh \sigmah}^{\muh \nuh} &= \frac{1}{2} \left( \nabla^{\{\muh} \nabla_{\{\rhoh} \nabla^{\nuh\}} \nabla_{\sigmah\}}
+ \nabla_{\{\rhoh} \nabla^{\{\muh} \nabla_{\sigmah\}} \nabla^{\nuh\}} \right)
+ \frac{4\nu^2 +d^2-2d-2}{2(d+1)}
\delta^{\{\muh}_{\ \{\rhoh} \nabla^{\nuh\}} \nabla_{\sigmah\}}  \,.
\eea{eq:DH3_grav}
The indices $\muh \nuh$ belong to one stress-tensor and $\rhoh \sigmah$ to the other. 
The curly brackets indicate symmetrization and subtraction of the trace, as appropriate for stress tensors. 

For later convenience we define the following shorthand for the differential operators,
including also the simplest case of the correlation function between two scalar operators $\phi$ and two other scalars ${\cal O}$
\beq 
\mathfrak{D} (\nu) =
\begin{cases}
\beta(\nu)\,, & \text{for } \langle \phi\phi{\cal O}{\cal O} \rangle\,,\\
\epsilon_\muh \epsilon^{* \nuh} \sum\limits_{k=1,4} \beta_k(\nu)  \,\hat  \calD_{k\ \nuh}^{\muh}\,, & \text{for }
\langle {\cal J}{\cal J}{\cal O}{\cal O} \rangle\,,\\
\epsilon_\muh \epsilon_\nuh \epsilon^{* \rhoh} \epsilon^{* \sigmah} \sum\limits_{k=1,2,3} \beta_k(\nu)\, \hat \calD_{k\  \rhoh \sigmah}^{\muh \nuh}\,, & \text{for }\langle TT{\cal O}{\cal O} \rangle\,,
\end{cases}
\label{eq:cases}
\eeq
where $\epsilon_\muh$ is a (complex) polarization which we take to satisfy $\epsilon_\muh \epsilon^\muh =0$ due to tracelessness of the stress-tensor.
The polarizations are chosen to be complex conjugate $z_1 = z_2^* \equiv \epsilon$ because it is only in this
configuration that we can relate the correlator to the norm of a state in section  \ref{sec:proof_ads_unitarity_spin} below.
The functions $\beta_k(\nu)$ are the same as in the scalar case 
(\eqref{eq:AlphaBeta} with \eqref{eq:alpha}), only the piece $r(\nu)$ depends on the index $k$ and will be denoted
$r_k(\nu)$. We can define a basis of t-channel OPE coefficients by generalizing \eqref{eq:rtoCC}
\beq
r_k(\nu) \equiv C^{(k)}_{12j(\nu)}C_{34j(\nu)}  K_{h\pm i\nu,j(\nu)} \,.
\label{eq:rtoCC_spin}
\eeq
In appendices 
\ref{sec:embedding_space} and \ref{sec:fourier_transformation} we derive a  set of linear relations
that relates these OPE coefficients to more conventional bases (the basis of conformal blocks constructed with
derivative operators \cite{Costa:2011dw} and a basis of three-point functions constructed using 
the embedding space formalism \cite{Costa:2011mg}).
For ratios we obviously have
\beq
\frac{\beta_k(\nu)}{\beta_j(\nu)} = \frac{r_k(\nu)}{r_j(\nu)}
=\frac{C^{(k)}_{12 j(\nu)}}{C^{(j)}_{12 j(\nu)}}\,,
\eeq
where $12$ denote the operators $\phi\phi$, ${\cal J}{\cal J}$ or $TT$, inserted at $y_1$ and $y_2$, that couple to the operators
in the leading Regge trajectory $J=j(\nu)$.

%%%%%%%%%%%%%%%%%%%%%%%%%%%%%%%%%%%%%%%%%%%%%%%%%%%%%%%%%%%%%%%%%%%%%%%%%%%%%%%%%%%%%%%%%%%
\section{AdS unitarity from CFT}
\label{sec:ads_unitarity_from_cft}
%%%%%%%%%%%%%%%%%%%%%%%%%%%%%%%%%%%%%%%%%%%%%%%%%%%%%%%%%%%%%%%%%%%%%%%%%%%%%%%%%%%%%%%%%%%

\subsection{Scalars}
Let us define an AdS phase shift $\chi (S,L)$ by expressing
 $\mathcal{B}(S,L)$, including the disconnected piece, as
\begin{equation}
{\cal B}(S,L) =  {\cal N} \,e^{i\chi (S,L)} \,,
\label{eq:PhaseShift}
\end{equation}
where the real constant ${\cal N}$ is fixed by the disconnected term. In analogy with the standard impact parameter representation for the phase shift 
$\delta(s,l_\perp)$, for which S-matrix unitarity implies that  ${\rm Im}\big(\delta(s,l_\perp)\big)<0$, we conjectured in 
\cite{Cornalba:2008sp}  that AdS unitarity would imply
\begin{equation}
{\rm Im}\big(\chi(S,L)\big)\ge 0\,.
\label{eq:AdSunitarity}
\end{equation}
At the time this inequality was conjectured on the  basis of the analogy with the S-matrix unitarity condition. It was at the heart of the AdS black disk model
for deep inelastic scattering that reproduces data at low Bjorken $x$ in the so-called saturation region. 

Recently the AdS unitarity condition (\ref{eq:AdSunitarity}) 
was proved using first principle CFT arguments \cite{Kulaxizi:2017ixa}. 
Let us start by reviewing their arguments.
The basic idea is to consider
the state formed by the operators ${\cal O}_1\equiv \phi$ and ${\cal O}_3\equiv {\cal O}$ placed at the center of its own Poincar\'e patch
\begin{equation}
| \Psi \rangle  = \int dx_1dx_3 f(x_1,x_3) \, \phi_{{\cal P}_1} (x_1)  {\cal O}_{{\cal P}_3} (x_3) | 0 \rangle\,.
\end{equation}
The wave function $f(x_1,x_3)$ is localized at small values of $x_1$ and $x_3$ and defines our incoming scattering states.
In global coordinates they are  placed at antipodal points in the boundary $S^{d-1}$
sphere and close to time $\tau=-\pi/2$. The norm of this state is then given by
\begin{equation}
\langle  \Psi  | \Psi \rangle  =
\int dx_2dx_4\int dx_1dx_3  f^*(x_2,x_4) f(x_1,x_3)  
\big\langle  \phi_{{\cal P}_1} (x_2) {\cal O}_{{\cal P}_3} (x_4) \phi_{{\cal P}_1} (x_1)  {\cal O}_{{\cal P}_3} (x_3) \big\rangle\,.
\label{eq:norm}
\end{equation}
In this equation $ \phi_{{\cal P}_1}  (x_2)$ is inserted in the Poincar\'e patch  ${\cal P}_1$ of $ \phi_{{\cal P}_1} (x_1)$, 
the operators are therefore close to each other with 
separation $x\approx x_1-x_2$ and ordered as shown in (\ref{eq:norm}). 
Similarly  $ {\cal O}_{{\cal P}_3}(x_4)$  is in the same Poincar\'e patch as $ {\cal O}_{{\cal P}_3}(x_3)$.
Thus we can use the OPE expansion to write
\begin{align}
\langle  \Psi  | \Psi \rangle  =&
\int dx_2dx_4\int dx_1dx_3 \, f^*(x_2,x_4) f(x_1,x_3) 
\nonumber\\
 & \sum_k C_k(x_2-x_1,\partial_{x_1} ) C_k(x_4-x_3,\partial_{x_3} )
 \big\langle {\cal O}_{k{\cal P}_1}  (x_1) {\cal O}_{k{\cal P}_3} (x_3)  \big\rangle\,.
\end{align}
Here $\langle{\cal O}_{k{\cal P}_1}  (x_1) {\cal O}_{k{\cal P}_3} (x_3)  \rangle$ is the 2-point function between Euclidean separated points. 
In global coordinates this is just the correlation function for operators placed at antipodal points in the boundary $S^{d-1}$
sphere and  close to time $\tau=-\pi/2$. 

To construct the above two-point function,
we use global embedding space coordinates
\beq
X = \left(X^0, X^a, X^{d+1}\right) = \big(\cos \tau, \Omega^a, -\sin \tau \big)\,,
\eeq
where $X^0$ and $X^{d+1}$ are time-like coordinates and $\Omega^a$ is a unit vector on $\mathbb{R}^{d}$.
In our case we can fix the points $x_1$  and $x_3$ at the embedding points
\beq
X_1 = (0,1,0,\ldots,0,1)\,,\qquad \qquad
X_3 = (0,-1,0,\ldots,0,1)\,,
\eeq
with polarization vectors
\beq
Z_1 = (z_1^0,z_1^0,\overrightarrow{z_1},z_1^0)\,,\qquad \qquad
Z_3 = (z_3^0,z_3^0,\overrightarrow{z_3},-z_3^0)\,,
\eeq
where the $Z_i$ are defined in terms of null vectors $z_i = (z_i^0, \overrightarrow{z_i}) \in \mathbb{R}^{1,d-1}$ and
were chosen to satisfy $Z_i^2 = Z_i \cdot X_i = 0$.
Using this one finds the two-point function
\bea
&z_{1 \mu_1} \ldots z_{1 \mu_J}
z_{3 \nu_1} \ldots z_{3 \nu_J}
\langle {\cal O}_{k{\cal P}_1}^{\mu_1 \ldots \mu_J}  (x_1) {\cal O}_{k{\cal P}_3}^{\nu_1 \ldots \nu_J} (x_3)  \rangle\\
={}&\frac{\big(\left( Z_1 \cdot Z_3 \right)\left( X_1 \cdot X_3 \right) - \left( Z_1 \cdot X_3 \right)\left( Z_3 \cdot X_1 \right)\big)^J}{\left( -2 X_1 \cdot X_3 \right)^{\D + J}}
= \frac{(-2 z_1 \cdot z_3)^J}{4^{\D+J}}\,,
\eea{eq:2point_antipodal}
where $z_1 \cdot z_3$ is the Lorentzian scalar product for $ \mathbb{R}^{1,d-1}$ vectors.
Thus, without the contraction to null vectors the two-point function is
\beq
\langle {\cal O}_{k{\cal P}_1}^{\mu_1 \ldots \mu_J}  (x_1) {\cal O}_{k{\cal P}_3 \nu_1 \ldots \nu_J} (x_3)  \rangle \propto \delta^{(\mu_1}_{(\nu_1} \ldots \delta^{\mu_J)}_{\nu_J)} - \text{traces}\,.
\label{eq:two-point_traceless}
\eeq

We are now in position to write the norm of the state $| \Psi \rangle$ as follows 
\begin{align}
\langle  \Psi  | \Psi \rangle  =&
\int dx_2dx_4\int dx_1dx_3 \, f^*(x_2,x_4) f(x_1,x_3) 
\nonumber\\
 &%\frac{1}{(x^2-i\epsilon_x)^{2\Delta_1} (\bar{x}^2-i\epsilon_{\bar{x}})^{2\Delta_3}} 
  \sum_k\frac{1}{(-x^2+i\epsilon_x)^{\Delta_1-\frac{\Delta_k}{2}} (-\bar{x}^2+i\epsilon_{\bar{x}})^{\Delta_3-\frac{\Delta_k}{2}}} 
  \,C_{J_k}^{(h-1)}\left(\frac{x\cdot \bar{x}}{|x||\bar{x}|}\right)
+\dots \,,
\label{eq:normpsi}
\end{align}
where $C_{J_k}^{(h-1)}$ is a Gegenbauer polynomial (which encodes the combination of contractions in \eqref{eq:two-point_traceless}) and the dots represent sub-leading contributions (at small $x$ and $\bar{x}$) coming from descendants.
The sum over $k$ is dominated by the identity operator,
corresponding to the disconnected term in the correlation function. Let us consider first the transform
$B_0(p,\bar{p})$ of this disconnected term
\begin{equation}
B_0(p,\bar{p})= 
\int \frac{dx d\bar{x}}{\pi^{2d}}\, \frac{e^{2ip\cdot x+2i\bar{p}\cdot \bar{x}}}{(x^2-i\epsilon_x)^{\Delta_1} (\bar{x}^2-i\epsilon_{\bar{x}})^{\Delta_3}} 
=
 \frac{ \theta(-p^2)\theta(p^0)\theta(-\bar{p}^2)\theta(\bar{p}^0)}{(-p^2)^{h-\Delta_1} (-\bar{p}^2)^{h-\Delta_3}} \, {\cal N} (\Delta_1,\Delta_3)\,,
\end{equation}
where $ {\cal N} (\Delta_1,\Delta_3)$ is a constant that depends on the external dimensions.
Then the contribution of the other operators can be obtained by acting with $\partial_p$ and $\partial_{\bar{p}}$,
\begin{equation}
\int dx d\bar{x} \,\frac{e^{2ip\cdot x+2i\bar{p}\cdot \bar{x}}}{(x^2-i\epsilon_x)^{2\Delta_1} (\bar{x}^2-i\epsilon_{\bar{x}})^{2\Delta_3}} 
\,F(x\cdot \bar{x})
=
F\!\left(- \frac{1}{4}\partial_p\cdot \partial_{\bar{p}} \right)
 B_0(p,\bar{p})\,.
\end{equation}
Using these equations, expression \eqref{eq:normpsi} simplifies to
\begin{align}
\langle  \Psi  | \Psi \rangle  =&
\int dx_2dx_4\int dx_1dx_3 \, f^*(x_2,x_4) f(x_1,x_3) 
\\
 & \int_{M} dp d\bar{p} \, e^{-2ip\cdot x-2i\bar{p}\cdot \bar{x}}
  \frac{ {\cal N} (\Delta_1,\Delta_3)
}{(-p^2)^{h-\Delta_1} (-\bar{p}^2)^{h-\Delta_3}} 
\left(
1+\sum_k \sum_{w=0}^{J_k}c_w\frac{(p\cdot \bar{p})^{w}}{(p^2\bar{p}^2)^{
\frac{1}{2}(\Delta_k+w)}}+
\dots
\right).
\nonumber
\end{align}
where $k$ denotes traceless symmetric primary operators in the theory with spin $J_k$ and dimension $\D_k$, and where 
the specific form of the constants $c_w$ will not be needed in what follows.
Finally, using $x\approx x_1-x_2$ and $\bar{x}=x_3-x_4$, we can integrate over $x_i$ to find
\begin{equation}
\langle  \Psi  | \Psi \rangle  
=\int_M dp d\bar{p} \,  \frac{ | \hat{f}(p,\bar{p})|^2 {\cal N} (\Delta_1,\Delta_3)}{(-p^2)^{h-\Delta_1} (-\bar{p}^2)^{h-\Delta_3}} \, 
 \left(1 + \sum_k\sum_{w=0}^{J_k}c_w\frac{(\cosh L)^w}{S^{\Delta_k}}  +\dots\right) .
 \label{eq:Norm}
\end{equation}
where
\beq
 \hat{f}(p,\bar{p})=\int dx_1dx_3 \,  f(x_1,x_3) e^{-2ip\cdot x_1-2i\bar{p}\cdot x_3}\,.
\eeq

One can consider, on the other hand, a final state $| \Psi' \rangle$ constructed by the unitary evolution of  $| \Psi \rangle$ by $\pi$
in global time together with inversion on the $S^{d-1}$ sphere. This transformation places the operator $\phi$ close to the 
center of the  Poincar\'e patch ${\cal P}_2$ and  ${\cal O}$ close to the  center of the  Poincar\'e patch ${\cal P}_4$. It defines
the outgoing scattering state.  This transition amplitude is computed precisely from the correlation function considered in this paper
\begin{equation}
\langle  \Psi'  | \Psi \rangle  =
\int dx_2dx_4\int dx_1dx_3  f^*(x_2,x_4) f(x_1,x_3)  \langle  \phi_{{\cal P}_2} (x_2) {\cal O}_{{\cal P}_4} (x_4) \phi_{{\cal P}_1} (x_1) {\cal O}_{{\cal P}_3} (x_3)  \rangle\,.
\end{equation}
Using the transform (\ref{eq:AtoB}) we can perform the $x_i$ integration obtaining 
\begin{equation}
\langle  \Psi'  | \Psi \rangle  = \int dp d\bar{p}  \, \frac{ | f(p,\bar{p})|^2 }{(-p^2)^{h-\Delta} (-\bar{p}^2)^{h-\Delta'}} \,{\cal B}(S,L) \,.
\end{equation}

We may now make use of  the Cauchy-Schwarz inequality
\begin{equation}
|\langle  \Psi'  | \Psi \rangle  | \le \sqrt{\langle  \Psi  | \Psi \rangle\langle  \Psi'  | \Psi' \rangle} = \langle  \Psi  | \Psi \rangle\,,
\end{equation}
and consider external wave functions $ f(x_1,x_3)$ such that its Fourier transform  is localized at high momenta  $p$ and $\bar{p}$. 
Using the definition of the phase shift introduced in (\ref{eq:PhaseShift}) one concludes that \cite{Kulaxizi:2017ixa}
\begin{equation}
\left| e^{i\chi (S,L)}   \right|  \le
1 + \epsilon(S,L)  +\dots \,,
\label{eq:AdSunitarityFinal_appendix}
\end{equation}
where the leading contribution of each operator to the error function $\epsilon(S,L)$ can be read from (\ref{eq:Norm})
\beq
\epsilon(S,L) = \sum_k \sum_{w=0}^{J_k}c_w\frac{(\cosh L)^w}{S^{\Delta_k}}\,.
\label{eq:epsilon_S_L_appendix}
\eeq
This justifies (\ref{eq:AdSunitarity}) in the large $S$ limit.
For theories with a small parameter such that $\chi (S,L)$ is small (large $N$ theories), and
if
\beq
|\epsilon(S,L)| \ll |  \chi(S,L)| \,,
\label{eq:error_function_suppressed}
\eeq
we also obtain the conjectured result (\ref{eq:AdSunitarity}).
For fixed $L$, condition \eqref{eq:error_function_suppressed} is true provided the phase shift 
$\chi(S,L)$ grows with $S$ faster than $\epsilon(S,L)$. In fact, $\epsilon(S,L)$ decays with a power of $S$   determined by the 
smallest dimension in the theory, denoted $\D_{min}$. If on the other hand we consider kinematics
with $L \propto \ln S$, the decay of $\epsilon(S,L)$ will be determined by the minimal
twist $\tau_{min} = (\D_k - J_k)_{min}$.

%%%%%%%%%%%%%%%%%%%%%%%%%%%%%%%%%%%%%%%%%%%%%%%%%%%%%%%%%%%%%%%%%%%%%%%%%%%%%%%%%%%%%%%%%%%
\subsection{Vector currents and stress-tensors}
\label{sec:proof_ads_unitarity_spin}
%%%%%%%%%%%%%%%%%%%%%%%%%%%%%%%%%%%%%%%%%%%%%%%%%%%%%%%%%%%%%%%%%%%%%%%%%%%%%%%%%%%%%%%%%%%

Let us briefly generalize the above argument to the case where the external operators have spin. We will just point out the main differences to the scalar case.
Let $| \Psi \ket$ be defined as before, just let the operator at $x_1$ have spin.
For the norm of the state we get a formula like \eqref{eq:normpsi}, but now with differential
operators which generate the external spins at positions $x_1$ and $x_2$
\begin{align}
\langle  \Psi  | \Psi \rangle  =&
\int dx_2dx_4\int dx_1dx_3 \, f^*(x_2,x_4) f(x_1,x_3) 
\nonumber\\
 & \sum_k\frac{1}{(x^2-i\epsilon_x)^{2\Delta_1-\Delta_k} (\bar{x}^2-i\epsilon_{\bar{x}})^{2\Delta_3-\Delta_k}} 
\sum_i a_i \calD_i \,
  C_{J_k}\!\left(\frac{x\cdot \bar{x}}{|x||\bar{x}|}\right)
+\dots \,,
\label{eq:normpsi_spin}
\end{align}
where $a_i$ are some constants and $\calD_i$ are the differential operators \eqref{DTensors}
or \eqref{DTensors_st} with polarizations $z_1 = z_2^*$.
Now we can perform a Fourier transformation and arrive
at a result analogous to \eqref{eq:Norm}
\begin{equation}
\langle  \Psi  | \Psi \rangle  =\int_M dp d\bar{p} \,  \frac{ | f(p,\bar{p})|^2 }{(-p^2)^{h-\Delta} (-\bar{p}^2)^{h-\Delta'}} \, 
{\cal N} (\Delta_1,\Delta_3) \sum_k
\sum_i b_i \hat \calD_i
 \left(1 + \sum_{w=0}^{J_k}c_w\frac{(\cosh L)^w}{S^{\Delta_k}}  +\dots\right) .
 \label{eq:Norm_spin}
\end{equation}
Here $b_i$ are some new constants which are linearly related to the $a_i$, similarly to what happens in appendix \ref{sec:fourier_transformation} between $\alpha_i(\nu)$ and $\beta_i(\nu)$. We do not need the exact form of this relation since the discussion below only relies on the asymptotic behavior of the subleading term in \eqref{eq:Norm_spin} at large or small $L$. This behavior is not changed by the differential operators since the covariant derivatives do not act on $S$ and the exponent of $(\cosh L)^w = (-\hatp \cdot {\hat {\bar p}})^w$ does not change. To see this one can compute
\beq
(\epsilon \cdot \nabla)^{2k} (-\hatp \cdot {\hat {\bar p}})^w
= (-\hatp \cdot {\hat {\bar p}})^w \sum\limits_{i=0}^k f_i(w) \left( \tanh(L) (n \cdot \epsilon)\right)^{2i} \epsilon^{2(k-i)}\,,
\eeq
where $\epsilon$ is a polarization vector satisfying $\epsilon \cdot \hatp = 0$, $n$ is the unit vector defined below in \eqref{eq:n}, $k$ is an integer and
$f_i(w)$ are polynomials in $w$.

%%%%%%%%%%%%%%%%%%%%%%%%%%%%%%%%%%%%%%%%%%%%%%%%%%%%%%%%%%%%%%%%%%%%%%%%%%%%%%%%%%%%%%%%%%%
\section{Bounds for OPE coefficients from Reggeon exchange}
\label{sec:saddle_point_scalar}
%%%%%%%%%%%%%%%%%%%%%%%%%%%%%%%%%%%%%%%%%%%%%%%%%%%%%%%%%%%%%%%%%%%%%%%%%%%%%%%%%%%%%%%%%%%
In this section we  explore  consequences of the unitarity condition on the imaginary part of the AdS phase shift. 
The results are more interesting in the case of correlation functions with operators with spin, so we shall consider 
simultaneously the computation of the phase shift for the three cases
$\langle \phi\phi{\cal O}{\cal O} \rangle$, $\langle {\cal J}{\cal J}{\cal O}{\cal O} \rangle$ and $\langle TT{\cal O}{\cal O} \rangle$.

To compute the phase shift we need to recall some properties of the function $j(\nu)$. 
It is an even function of $\nu$ and it must pass
through the protected point $J=2$ at $\Delta=h\pm i\nu=d$, thus $j(\pm h i) = 2$. From the chaos bound \cite{Maldacena:2015waa} we also know that, 
at the symmetric   point $\nu=0$, the intercept value  $j(0)$ can not be larger than  2. 
Along the spectral region where 
$i\nu$ is a positive real number and $\Delta=h+ i\nu$,  we expect the function $j(\nu)$  to be a growing convex function. \footnote{
Convexity of $j(\nu)$ has not been proven but we shall assume it in this section.}
Hence $j'(\nu)$ is a positive imaginary number 
for $i\nu$  real and positive. Moreover, due to the unitarity bound $\Delta \ge J+d-2$,
for $i\nu$ real and positive, we must have that  ${\rm Im}  \big(j'(\nu) \big)<1$. 
For large $\Delta_g$, $j(\nu)$  can be written has an infinite series of the form 
\begin{equation}
j(\nu)=2- \sum_{n=1}^\infty {\cal D}^n  j_n(\nu^2) = 2 -   {\cal D} (h^2+\nu^2)  \left( 1 +   \sum_{n = 2}^\infty  {\cal D}^{n-1} \tilde{j}_n(\nu^2) \right)\,,
\label{ReggeSpinStrong}
\end{equation}
where the expansion parameter ${\cal D} = 1/\Delta_g^2$ and we will use the definition $\Delta_g=\Delta(J=4) $. For theories with a well defined
large gap limit we must be able to take ${\cal D}\rightarrow 0$ and $\nu\rightarrow \infty$ keeping ${\cal D} \nu^2$ fixed. Otherwise in this limit
we would never be able to obtain $j\big (\pm i(\Delta_g-h)  \big)=4$. This requirement imposes that $\tilde{j}_n(\nu^2)$ is a polynomial of maximal degree 
$n-1$. Notice that we are only imposing that there is a large $\Delta_g$ limit. This is, however, very similar to the flat space limit of the dual AdS
physics for which in the limit the fixed quantity is ${\cal D} \nu^2 \equiv - \alpha' t$. In this case, 
since in the region of large $s$ and $t$ the Regge trajectory must be linear with $t$, 
as shown recently in \cite{Caron-Huot:2016icg},  the degree of  $\tilde{j}_n(\nu^2)$ is at most  $n-2$.

The integral in (\ref{calB}) can be computed using a saddle point approximation. 
Let us write the phase shift as
\beq
\chi(S,L) =  -\frac{i}{\calN} \int_{-\infty}^\infty d\nu S^{1-j(\nu)} \mathfrak{D} (\nu) \Omega_{i\nu} (L)\,,
\eeq
where the operator  $\mathfrak{D}(\nu)$ is given in (\ref{eq:cases}).
 Since both $j(\nu)$ and $\mathfrak{D}(\nu)$ are even functions of $\nu$, using the explicit form of the function 
$\Omega_{i\nu} (L)$ given in (\ref{eq:Omega}), 
we have that
\begin{equation}
 {\cal \chi}(S,L)  = \frac{1}{ {\cal N} \pi}  \frac{1}{ S} \int_{-\infty}^\infty  d\nu  \, \nu\, e^{ j(\nu) \ln S-i\nu L} 
 \left(e^{i\nu L} \mathfrak{D} (\nu)  \Pi_{i \nu}(L)  \right)\,,
\label{eq:BReggeSaddleSpin}
\end{equation}
where $\Pi_{i \nu}(L)$ is the scalar propagator on $H_{d-1}$ defined in \eqref{eq:propagator}.
Note that the combination $e^{i \nu L} \Pi_{i \nu}(L)$ does not depend exponentially on $\nu$,
and that the action of the operator $\mathfrak{D}(\nu)$ on the propagator does not change that fact.
Thus the integral in (\ref{eq:BReggeSaddleSpin}) has 
a saddle point at $\nu=\nu_0$ defined by
\begin{equation}
j'(\nu_0) \ln S - i L=0\,,
\label{eq:SaddlePoint}
\end{equation}
located along the negative imaginary $\nu$ axis. We remark 
that the saddle point is itself a function of $L$. It is at the symmetric point $\nu_0=0$
if we fix $L$ and let $\ln S\gg 1$. On the other hand, if we scale  $L$ with $\ln S$, we can vary 
the location of the saddle point  from $0$ to $-i\infty$, as $L$ grows. 

To do the integral in (\ref{eq:BReggeSaddleSpin}) we expand around the saddle point
\begin{equation}
{\cal \chi}(S,L) =  \frac{S^{j(\nu_0)-1}}{{\cal N}\pi}  \,e^{-  i\nu_0 L}  \int_{-\infty}^\infty  d\nu  \, \nu\,  e^{ \,j''(\nu_0) \ln S\, \frac{(\nu-\nu_0)^2}{2}+\dots} 
\left( e^{i \nu L} 
\mathfrak{D} (\nu)\Pi_{i \nu}(L)\right).
\label{eq:AReggeSaddle}
\end{equation}
It is now clear that the saddle point approximation is valid provided $ |j''(\nu_0)| \ln S\gg 1$. This can always be achieved for large enough $\ln S$, even if $ |j''(\nu_0)|\ll 1$, as 
it is the case for large gap. Thus we have
\begin{equation}
{\cal \chi}(S,L) =  \frac{S^{j(\nu_0)-1}  }{{\cal N} \pi}  \,\nu_0   \,  \sqrt{\frac{2\pi}{ -j''(\nu_0) \ln S}} \, \mathfrak{D} (\nu_0)  \Pi_{i \nu_0} (L)\,,
\label{eq:ChiReggeonSpin}
\end{equation}
where we recall that the functions $ \beta_k(\nu)$ that appear in $ \mathfrak{D} (\nu) $ can be read 
from (\ref{eq:alpha}) and (\ref{eq:AlphaBeta}), and are given by
\begin{equation}
\beta_k(\nu) =
 \left(  i \cot\!\left( \frac{\pi j(\nu)}{2}\right)  -1 \right)
\pi^{h+1} 4^{j(\nu)-1}  \,\frac{-ij'(\nu)}{i\nu}\, r_k(\nu) \,.
\label{eq:beta}
\end{equation}

%%%%%%%%%%%%%%%%%%%%%%%%%%%%%%%%%%%%%%%%%%%%%%%%%%%%%%%%%%%%%%%%%%%%%%%%%%%%%%%%%%%%%%%%%%%
\subsection{Unitarity condition along Regge trajectory}
%%%%%%%%%%%%%%%%%%%%%%%%%%%%%%%%%%%%%%%%%%%%%%%%%%%%%%%%%%%%%%%%%%%%%%%%%%%%%%%%%%%%%%%%%%%

Since $i\nu_0$ is a positive real number,
the unitarity condition $\Im \big({\cal \chi}(S,L)\big) \geq 0$ becomes
\beq
\Re \big(  \mathfrak{D} (\nu_0)  \Pi_{i \nu_0} (L) \big) \leq 0 \,.
\label{eq:unitarity_cond_spin}
\eeq
In particular, this becomes a condition on the real part of the $\beta_k(\nu_0)$, which is given by
\begin{equation}
{\rm Re} \big( \beta_k(\nu_0) \big)=-
\pi^{h+1} 4^{j(\nu_0)-1}  \,\frac{-ij'(\nu_0)}{i\nu_0}\, r_k(\nu_0) \,,
\label{eq:Real_beta}
\end{equation}
where we recall that $i j'(\nu_0) $ is real. Thus
we do not need to worry about the poles of the imaginary part of $\beta_k(\nu_0)$, which are related to the elastic 
exchanges of the spin $J$ fields in the Regge trajectory. In fact, when the saddle point collides with such points, the above computation of  ${\rm Re}\big(\chi(S,L)\big)$ is no longer valid.
On the other hand, we focus on the absorptive part of the correlation function.
Thus, for scalar operators the unitarity condition becomes simply 
\begin{equation}
r(\nu_0) = C_{\phi\phi j(\nu_0)}C_{{\cal O}{\cal O}j(\nu_0)}  K_{h +  i\nu_0,j(\nu_0)}\ge 0\,.
\label{eq:ConditionScalars}
\end{equation}
This is a condition on the analytic continued OPE coefficients at the saddle point value. However, 
as we shall see below, we still need to
make sure that the AdS unitarity condition applies, due to the restriction (\ref{eq:error_function_suppressed})
from corrections to the unitarity condition.

To implement condition (\ref{eq:unitarity_cond_spin}) in the case of operators with spin, we need to act
with $H_{d-1}$ covariant derivatives on functions of $L$. 
The results can be written in terms of a unit vector $n$ that is tangent to the geodesic on 
$H_{d-1}$ that connects the impact points $(r,e_\perp)$  and $(\bar{r},\bar{e}_\perp)$ of the external operators, computed at the impact point $(r,e_\perp)$ 
of the external vector (or stress tensor). 
Solving  $\nabla_n n = 0$ one obtains
\beq
n_\muh = \frac{1}{r \bar r \sinh(L)} \big(r- \bar r \cosh(L), e_\perp - \bar{e}_\perp \big)\,.
\eeq
This can also be expressed in terms of the coordinates $\hat p$ and $\hat {\bar p}$ as
\beq
n^m = - \frac{{\hat {\bar p}}^m + (\hat p \cdot \hat {\bar p}) \,\hatp^m}{\sqrt{(\hat p \cdot \hat {\bar p})^2-1}}\,.
\label{eq:n}
\eeq
Next we consider the two cases of $\langle {\cal J}{\cal J}{\cal O}{\cal O} \rangle$ and $\langle TT{\cal O}{\cal O} \rangle$.

%%%%%%%%%%%%%%%%%%%%%%%%%%%%%%%%%%%%%%%%%%%%%%%%%%%%%%%%%%%%%%%%%%%%%%%%%%%%%%%%%%%%%%%%%%%
\subsubsection{Correlator $\langle {\cal J}{\cal J}{\cal O}{\cal O} \rangle$ }
%%%%%%%%%%%%%%%%%%%%%%%%%%%%%%%%%%%%%%%%%%%%%%%%%%%%%%%%%%%%%%%%%%%%%%%%%%%%%%%%%%%%%%%%%%%

In this case, acting with the operator that generates two conserved currents given in (\ref{eq:cases}), one obtains
\begin{equation}
\mathfrak{D} (\nu_0)\, \Pi_{i \nu_0} (L)  =
 \Pi_{i \nu_0} (L) |\epsilon|^2 \beta_1(\nu_0)
 \left( 1 + \frac{ \beta_4(\nu_0)}{\beta_1(\nu_0)}
\,a(\nu_0,L) \left(\frac{|n \cdot \epsilon|^2}{|\epsilon|^2} - \frac{1}{d-1} \right)
\right),
\label{eq:DonH_currents}
\end{equation}
where
\bea
a(\nu_0,L) ={}&
\left(h+i \nu _0-1\right) \Bigg(
(1-2 h) \coth (L)+h-i \nu _0-1\\
&-\frac{2 (h-1) (2 h-1) e^{-2 L} \coth (L) \, _2F_1\big(h,h+i \nu _0;i \nu _0+2;e^{-2 L}\big)}{\left(1+i \nu _0\right) 
\, _2F_1\big(h-1,h+i \nu _0-1;i \nu _0+1;e^{-2 L}\big)}\Bigg)\,.
\eea{eq:currents_n}
We assume that the condition \eqref{eq:unitarity_cond_spin} is satisfied if $\beta_4(\nu_0) =0$ 
(i.e.\ $\Re\big( \beta_1(\nu_0) \big) \leq 0$). 
In this case the above discussion for scalar operators 
applies without change, since the only difference here is an additional factor of $|\epsilon|^2$.
When we turn on the $\beta_4(\nu_0)$ term in \eqref{eq:DonH_currents},
we have to make sure that it does not change the sign of the expression for any choice of polarization $\epsilon_{\hat{\mu}}$. 
This gives the condition
\beq
-\frac{d-1}{d-2} \leq  \frac{ r_4(\nu_0)}{ r_1(\nu_0)}\, a(\nu_0,L) \leq d-1\,,
\label{eq:bound_currents}
\eeq
where we note that $\beta_4(\nu_0) /  \beta_1(\nu_0) =r_4(\nu_0) /  r_1(\nu_0) = C^{(4)}_{{\cal J}{\cal J} j(\nu_0)} /C^{(1)}_{{\cal J}{\cal J} j(\nu_0)}$.

%%%%%%%%%%%%%%%%%%%%%%%%%%%%%%%%%%%%%%%%%%%%%%%%%%%%%%%%%%%%%%%%%%%%%%%%%%%%%%%%%%%%%%%%%%%
\subsubsection{Correlator $\langle TT{\cal O}{\cal O} \rangle$ }
%%%%%%%%%%%%%%%%%%%%%%%%%%%%%%%%%%%%%%%%%%%%%%%%%%%%%%%%%%%%%%%%%%%%%%%%%%%%%%%%%%%%%%%%%%%

Acting with the operator that generates the tensor structures of two stress-tensors, we get
\bea
\mathfrak{D} (\nu_0) \,\Pi_{i \nu_0} (L)  =
   \Pi_{i \nu_0} (L)|\epsilon|^4 \beta_1(\nu_0) &\Bigg[
   1 +
t_2(\nu_0,L)   \left(\frac{|n \cdot \epsilon|^2}{|\epsilon|^2} - \frac{1}{d-1} \right)
\\&\quad
+ t_4(\nu_0,L) \left(  \frac{|n \cdot \epsilon|^4}{|\epsilon|^4} - \frac{2}{(d+1)(d-1)} 
\right)\Bigg]\,,
\eea{eq:tensors_n}
where 
\beq
t_2(\nu_0,L) = -\frac{r_2(\nu_0)}{r_1(\nu_0)} \,a(\nu_0,L)
+ \frac{r_3(\nu_0)}{r_1(\nu_0)} \,f_2(\nu_0,L)\,, \qquad
t_4(\nu_0,L) = \frac{r_3(\nu_0)}{r_1(\nu_0)} \,f_4(\nu_0,L)\,.
\label{eq:t2t4}
\eeq
Since $f_2$ and $f_4$ are slightly lengthy functions, their full expressions can
be found in appendix \ref{sec:t2_and_t4}.
Similar to the previous case, when $\beta_2(\nu_0) = \beta_3(\nu_0) =0$  the
discussion reduces to that of the scalar correlator.
Requiring that the terms proportional to $\beta_2(\nu_0)$ and $\beta_3(\nu_0)$ do not
change the sign of the expression \eqref{eq:tensors_n} leads to bounds on $t_2(\nu_0,L)$ and $t_4(\nu_0,L)$ of the same form as in
\cite{Hofman:2008ar,Camanho:2014apa}
\bea
0 &\leq 1-\frac{t_2(\nu_0,L)}{d-1} - \frac{2 t_4(\nu_0,L)}{(d+1)(d-1)}\,, \\
0 &\leq \left(1-\frac{t_2(\nu_0,L)}{d-1} - \frac{2 t_4(\nu_0,L)}{(d+1)(d-1)} \right) +\frac{t_2(\nu_0,L)}{2}\,, \\
0 &\leq \left(1-\frac{t_2(\nu_0,L)}{d-1} - \frac{2 t_4(\nu_0,L)}{(d+1)(d-1)} \right) +\frac{d-2}{d-1} \big(t_2(\nu_0,L) + t_4(\nu_0,L)\big) \,.
\eea{eq:t2t4_bounds}
Recall that these bounds can be derived by decomposing the polarization tensor appearing in  \eqref{eq:tensors_n}
into irreducible representations of the $O(d-2)$ invariant
space orthogonal to $n$ given in (\ref{eq:n}) (traceless symmetric two-tensor, vector and scalar). In practice this can be achieved by restoring explicit tracelessness of the polarization tensor
\beq
\epsilon^{\hat \mu}_{\ \hat \nu} = \frac{1}{2}
\left( \epsilon^{\hat \mu}_{1} \epsilon_{2 \hat \nu} +
\epsilon^{\hat \mu}_{2} \epsilon_{1 \hat \nu} \right)
- \frac{1}{d-1}\, \delta^{\hat \mu}_{\ \hat \nu} \,\epsilon_1 \cdot \epsilon_2 \,,
\eeq
and inserting the three choices
\bea
n \perp \epsilon_1\,, n \perp \epsilon_2\,, \epsilon_1 \perp \epsilon_2 \,,
\qquad n \parallel \epsilon_1\,, n \perp \epsilon_2\,, \epsilon_1 \perp \epsilon_2\,,
\qquad \text{or} \qquad n \parallel \epsilon_1 \parallel \epsilon_2 \,.
\eea{eq:polarization_tensor_choices}
Note that the first choice is only possible if $d \geq 4$ and the second for $d \geq 3$.
This corresponds to the fact that the three-point functions have less tensor
structures in two or three dimensions. The three-point function of three stress-tensors,
for example, has only two parity even tensor structures in $d=3$ dimensions \cite{Maldacena:2011jn}.

%%%%%%%%%%%%%%%%%%%%%%%%%%%%%%%%%%%%%%%%%%%%%%%%%%%%%%%%%%%%%%%%%%%%%%%%%%%%%%%%%%%%%%%%%%%
\subsection{Unitarity condition at the intercept}
\label{sec:intercept}
%%%%%%%%%%%%%%%%%%%%%%%%%%%%%%%%%%%%%%%%%%%%%%%%%%%%%%%%%%%%%%%%%%%%%%%%%%%%%%%%%%%%%%%%%%%

We wish  to understand the region of validity of condition  (\ref{eq:unitarity_cond_spin})
for the OPE coefficients 
as we vary the impact parameter $L$, since we can think of (\ref{eq:SaddlePoint}) as
$\nu_0=\nu_0(L)$.  We must guarantee that corrections to the AdS unitarity condition are suppressed, that is  
(\ref{eq:error_function_suppressed}) holds.

First we consider the case where $L$ is kept fixed, that is $L\ll \ln S$,
which requires ${\rm Im} \big( j'(\nu_0) \big) \ll 1$. 
In this case we can expand the saddle point equation (\ref{eq:SaddlePoint}) around the symmetric point $\nu=0$ where $j'(0)=0$,
obtaining 
\begin{equation}
\nu_0 = \frac{iL}{ j''(0) \ln S}\,.
\label{eq:SaddlePointExpansion}
\end{equation}
Using this approximation, for the particular case of scalar operators the phase shift becomes
\begin{equation}
{\rm Im} \big({\cal \chi}(S,L) \big)  = 
 \frac{ \Gamma(h-1) }{ 2{\cal N}}   \sqrt{\frac{2\pi}{ -j''(0) }}    \,  r(0)\,  \frac{  (4S)^{j(0)-1} }{( \ln S)^{3/2}}   
L\,e^{(1-h)L} \,_2F_1\big( h-1,h-1;1  ;e^{-2L}\big)\,.
\end{equation}
In particular, we can obtain the small $L$ behavior
\begin{equation}
{\rm Im} \big({\cal \chi}(S,L) \big)  = \frac{1}{ {\cal N}}    \sqrt{\frac{1}{ -2 j''(0) }}     \frac{  (4S)^{j(0)-1} }{( \ln S)^{3/2}}   \,  r(0)\,
\times
\begin{dcases}
L \ln (8)  \big(1+O(L) \big)\,, & d = 3\,, \\
\frac{\Gamma(h-\frac{3}{2})}{2  L^{2h-4}} \big(1+O(L) \big)\,, \qquad & d \geq 4\,.
\end{dcases}
\label{eq:ImChiIntercept}
\end{equation}
Hence, we conclude that  OPE coefficients between the same two scalar operator and the Regge trajectory
at the intercept value, $C_{{\cal O}{\cal O}j(0)}$, must all have the same sign, 
\begin{equation}
r(0) \ge 0 \quad \Rightarrow \quad
C_{\phi\phi j(0)}  C_{{\cal O}{\cal O}j(0)} \ge 0\,.
\label{eq:ResultsOPEScalarsLfixed}
\end{equation}
This condition holds for any value of the gap and is valid provided we can neglect the $1/S$ corrections in the AdS unitarity condition (\ref{eq:error_function_suppressed}). That is, provided
\begin{equation}
j(0) > 1- \Delta_{min}\,,
\label{eq:ConditionScalarsLfixed}
\end{equation}
where $ \Delta_{min}$ is the operator with smallest dimension in the theory. This gives $j(0)>2-h$ if one uses the scalar unitarity bound.

%%%%%%%%%%%%%%%%%%%%%%%%%%%%%%%%%%%%%%%%%%%%%%%%%%%%%%%%%%%%%%%%%%%%%%%%%%%%%%%%%%%%%%%%%%%
\subsubsection{Correlator $\langle {\cal J}{\cal J}{\cal O}{\cal O} \rangle$ }
%%%%%%%%%%%%%%%%%%%%%%%%%%%%%%%%%%%%%%%%%%%%%%%%%%%%%%%%%%%%%%%%%%%%%%%%%%%%%%%%%%%%%%%%%%%

Next we want to extend the previous discussion to the case of operators with spin. 
For the case of vector currents we need to analyze the function 
$a(\nu_0,L)$ in  (\ref{eq:currents_n}) for $\nu_0=0$. 
Since $\ln S$ is large, we could still keep the impact parameter $L$ fixed. However the optimal 
bound is obtained when we take $L$ to be very small.  In this limit we have
\beq
a_2(0,L) = 
\begin{dcases} 
\frac{2}{L^2 \ln (L)} \big(1+O(1/\ln L) \big)\,, & d=3\,, \\
-\frac{3 + 4 (h-2) h}{L^2} \big(1+O(L) \big)\,, & d \geq 4\,.
\end{dcases}
\eeq
We conclude that the only way to satisfy the bounds 
(\ref{eq:bound_currents}) is to have 
\begin{equation}
r_4(0)=0\quad \Rightarrow \quad
C^{(4)}_{{\cal J}{\cal J} j(0)} = 0\,.
\label{eq:r4(0)=0}
\end{equation}
The argument is similar to that obtained for the physical OPE, $C^{(4)}_{{\cal J}{\cal J} T}=0$, in the large gap limit. However,
here we derive the condition for the analytically continued OPE coefficient at the intercept value  for any value of the 
gap. 
Moreover, let us write $r_4(\nu)$ as a polynomial in $\nu^2$. 
The $1/L^2$ behavior of the function $a_2(0,L)$ as $L\rightarrow 0$, together  with the bounds (\ref{eq:bound_currents}), 
require the 
polynomial  $r_4(\nu)$  to start with a power of  $\nu^2$, that is
\begin{equation}
r_4(\nu)= \sum_{n=1}^\infty a_n \nu^{2n}\,.
\label{eq:r4Prediction}
\end{equation}
It would be interesting to bound the coefficient $a_1$. However, if we analyze  \eqref{eq:bound_currents} using \eqref{eq:SaddlePointExpansion} and \eqref{eq:r4Prediction} in the limit $L\to 0$, we obtain a very weak bound on $|a_1|$ of order $(\ln S)^2$. 
\footnote{We thank Alexander Zhiboedov for suggesting this calculation.}

Condition (\ref{eq:r4(0)=0}) was derived for fixed impact parameter $L\ll \ln S$, in the limit
 $L\rightarrow 0$. 
 From the  dual AdS physics view point, one could worry that the computation will not be valid for $L$ sufficiently small. 
 However, Regge theory is resumming all tree level exchanges of string states in the leading $S$ approximation. Of course 
 there string loops effects that  become important for smaller $L$, but those are suppressed at large $N$.

%%%%%%%%%%%%%%%%%%%%%%%%%%%%%%%%%%%%%%%%%%%%%%%%%%%%%%%%%%%%%%%%%%%%%%%%%%%%%%%%%%%%%%%%%%%
\subsubsection{Correlator $\langle TT{\cal O}{\cal O} \rangle$ }
%%%%%%%%%%%%%%%%%%%%%%%%%%%%%%%%%%%%%%%%%%%%%%%%%%%%%%%%%%%%%%%%%%%%%%%%%%%%%%%%%%%%%%%%%%%

For the case of stress tensors the computation is similar. We need to 
analyze the functions 
$t_2(\nu_0,L)$ and $t_4(\nu_0,L)$
in (\ref{eq:t2t4}) for $\nu_0\ll1$ and in the limit $L\rightarrow 0$. This yields the following behavior
\bea
 t_2(0, L) ={}&    
\begin{dcases} 
\left( \frac{1}{L^2}\frac{r_2(0)}{r_1(0)} - \frac{16}{L^4} \frac{r_3(0)}{r_1(0)}\right)
\frac{-2}{ \ln (L)} 
\big(1+O(1/\ln L) \big), & d=3\,, \\
\left( \frac{1}{L^2} \frac{r_2(0)}{r_1(0)} - \frac{4(2h+1)}{L^4} \frac{r_3(0)}{r_1(0)}  \right) \big(3 + 4 (h-2) h\big) \big(1+O(L) \big), & d \geq 4\,,
\end{dcases}\\
t_4(0,L) ={}& \begin{dcases} 
\frac{-48}{L^4 \ln (L)} \frac{r_3(0)}{r_1(0)}
\big(1+O(1/\ln L) \big), \qquad \qquad \qquad \qquad \qquad \qquad \qquad & d=3\,, \\
\frac{1}{L^4} \frac{r_3(0)}{r_1(0)} \big(9 - 40 h^2 + 16 h^4\big) \big(1+O(L) \big), & d \geq 4\,.
\end{dcases}\\
\eea{eq:t2_t4_small_L}
Thus, the only way to satisfy the bounds  (\ref{eq:t2t4_bounds})
is to have 
\begin{equation}
r_2(0)=r_3(0)=0\quad \Rightarrow \quad
C^{(2)}_{TT j(0)} = C^{(3)}_{TT j(0)} =0\,.
\label{eq:r2(0)=0}
\end{equation}
Moreover, now in order to satisfy the bounds   (\ref{eq:t2t4_bounds}) as $L\rightarrow 0$, the polynomial 
$r_2(\nu)$ must start with a power of  $\nu^2$, while 
$r_3(\nu)$ must start with a power of  $\nu^4$, 
that is
\begin{equation}
r_2(\nu)= \sum_{n=1}^\infty b_n \nu^{2n}\,,\qquad \qquad
r_3(\nu)= \sum_{n=2}^\infty c_n \nu^{2n}\,.
\label{eq:r2r3Prediction}
\end{equation}

It would be  nice to check the predictions (\ref{eq:r4Prediction}) and (\ref{eq:r2r3Prediction}) in a concrete example at finite gap.
One such example could the Banks-Zaks fixed points \cite{Banks:1981nn}.

%%%%%%%%%%%%%%%%%%%%%%%%%%%%%%%%%%%%%%%%%%%%%%%%%%%%%%%%%%%%%%%%%%%%%%%%%%%%%%%%%%%%%%%%%%%
\subsection{Unitarity condition for coupling to stress tensor  at large gap}
\label{sec:intercept_to_J2}
%%%%%%%%%%%%%%%%%%%%%%%%%%%%%%%%%%%%%%%%%%%%%%%%%%%%%%%%%%%%%%%%%%%%%%%%%%%%%%%%%%%%%%%%%%%

Let us show that for a theory with a large gap, the unitarity condition obtained at the intercept can also be made useful for the physical OPE coefficients 
with the stress tensor $C_{{\cal J}{\cal J}T}$  and $C_{TTT}$. This happens because,
just like the spin $j(\nu)$, the function $r(\nu)$ can also be written as an expansion in ${\cal D}=1/\Delta_g^2$,
\begin{equation}
r(\nu) = r(0) + \sum_{n=1}^\infty   {\cal D}^n   f_n(\nu^{2}) \,,
\end{equation}
where $f_n(0)=0$. The assumption that there is a well defined  $\Delta_g\rightarrow \infty$
limit, namely ${\cal D}\rightarrow 0$, $\nu\rightarrow\infty$ keeping ${\cal D} \nu^2$ fixed, implies that 
 the function $f_n(\nu^{2})$ is a polynomial of maximal degree $2n$ given by
\begin{equation}
f_n(\nu^{2})= \sum_{k=1}^n  \nu^{2k} a_{k,n}\,,
\end{equation}
where $a_{k,n}$ are coefficients that depend on the specific theory and on the OPE coefficient. 
In other words, we are assuming that the functions $r(\nu)$ remain finite in the large gap limit.

The point now is that we can consider the protected point $\nu=\pm i h$, corresponding to  
spin $J=2$, and take the large gap limit ${\cal D}\ll 1$. In this limit
\begin{equation}
r(\pm i h ) = r(0)  - h^2 a_{1,1} {\cal D} +   \left( h^4 a_{2,2} -h^2 a_{1,2}   \right)  {\cal D}^2 + 
O({\cal D}^3)\,,
\end{equation}
where the $a_{k,n}$ are assumed to be of order unit.
This means we can use the unitarity condition imposed on $r(0)$ to impose a condition on the physical OPE coefficients $r(\pm h i)$,
that describes the coupling to the stress tensor of the two external vector currents or stress tensors.

For the case of the correlator $\langle {\cal J}{\cal J}{\cal O}{\cal O}\rangle$ and using (\ref{eq:r4Prediction}) we conclude  that
\begin{equation}
r_4(\pm i h ) = - h^2 a_{1,1}^{(4)}\calD + \calO(\calD^2)\,.
\label{eq:r4grav}
\end{equation}
Thus, we confirm the result of \cite{Camanho:2014apa} that the OPE coefficient associated with the non-minimal coupling in the dual AdS theory is related to the gap as
$C^{(4)}_{{\cal J}{\cal J}T} \sim 1/ \Delta_g^2$.

For the case of the correlator $\langle TT{\cal O}{\cal O}\rangle$ and using (\ref{eq:r2r3Prediction}) we conclude  that
\begin{equation}
r_2(\pm i h ) = - h^2 a_{1,1}^{(2)}\calD + \calO(\calD^2)\,,
\qquad\qquad
r_3(\pm i h ) = - h^4 a_{2,2}^{(3)}\calD^2 + \calO(\calD^3)\,,
\label{eq:r2r3grav}
\end{equation}
Again we obtained the non-trivial behavior predicted by the non-minimal AdS couplings
$C^{(2)}_{TTT} \sim 1/ \Delta_g^2$ and $C^{(3)}_{TTT} \sim 1/ \Delta_g^4$.

%%%%%%%%%%%%%%%%%%%%%%%%%%%%%%%%%%%%%%%%%%%%%%%%%%%%%%%%%%%%%%%%%%%%%%%%%%%%%%%%%%%%%%%%%%%
\subsection{Conformal collider bounds}
%%%%%%%%%%%%%%%%%%%%%%%%%%%%%%%%%%%%%%%%%%%%%%%%%%%%%%%%%%%%%%%%%%%%%%%%%%%%%%%%%%%%%%%%%%%

Let us  consider the unitarity condition   (\ref{eq:unitarity_cond_spin}) at the  stress tensor protected point $\nu_0=-ih $. In the next section
we shall analyze the validity of this condition using  (\ref{eq:AdSunitarityFinal_appendix}).
For the saddle point to be at a finite value along the  imaginary $\nu$ axis we need to scale the impact 
parameter $L$ with $\ln S$.
For scalar operators this gives simply 
\begin{equation}
r(\pm  i h ) \ge 0 \quad \Rightarrow \quad
C_{\phi\phi T}  C_{{\cal O}{\cal O}T}  \ge 0\,,
\label{eq:ConditionScalarsLargegraviton}
\end{equation}
which follows from the conformal Ward identities.

Next we consider the more interesting  case of operators with spin. 
In the case of vector currents, since  $L \sim \ln S$, we need to analyze the behavior of the function 
$a(\nu_0,L)$ in (\ref{eq:currents_n}) for large $L$,
\beq
\lim_{L\to \infty} a(\nu_0,L) = - (h+i\nu_0)(h+i\nu_0-1)\,.
\label{eq:a_2_large_L}
\eeq
In order to make contact to known bounds, we use the three basis changes
\eqref{eq:alpha_beta_vectors}, \eqref{eq:alpha_to_C_vectors}, \eqref{eq:OPE_relation_vectors} and the
conservation condition \eqref{eq:conservation_embedding_vectors} to change to the basis of OPE coefficients $\beta$ and 
$\eta$ defined in \eqref{eq:JJL_3point}. Evaluated at $\nu_0$, this change of basis reads
\bea
&
\frac{r_4(\nu_0)}{r_1(\nu_0)} =
 \bigg(
\beta  (2 h-1) \left(4 h j+2 (h-3) h+j^2-5 j+2 \nu _0^2-2 i \nu _0+4\right)\\
&-\eta  (2 h-1) \left(3 h+j-i \nu _0-4\right) \left(2 h+j+2 i \nu _0-2\right)
\bigg) \Big/\\
&\bigg( \Big( \beta  \left(-2 h^3+h^2 \left(7+4 i \nu _0\right)-
2 h \left(j-\nu _0 \left(\nu _0-5 i\right)+3\right)-(j-4) j-3 \nu _0 \left(\nu _0-2 i\right)\right)+\\
&\eta  \left(3 h+j-i \nu
   _0-4\right) \left(h \left(2 h-2 i \nu _0-3\right)+j+3 i \nu _0\right) \Big)
(h + i \nu_0) (h + i \nu_0 - 1)
\bigg).
\eea{eq:b4_over_b1}
In particular, for  $\nu_0=-ih $ we have
\beq
\left. \frac{r_4(- h i)}{r_1(- h i)} \right|_{j=2} =
\frac{\beta +4(h-1)h \eta }{ 4 (h-1)h(\beta - \eta) }\,.
\eeq
Finally, we relate the OPE basis $\beta,\eta$  to the normalization of the two-point function of the conserved current $C_{\cal J}$, which is positive, and
to the OPE coefficient $\lambda$ (see \cite{arXiv:1603.03771} for details)
\beq
\beta = -2 \lambda \,, \qquad \eta = \frac{2h C_{\cal J}}{S_{2h}} - 2 \lambda \,, \qquad S_{2h} = \frac{2 \pi^h}{\Gamma (h)}\,.
\label{eq:Ward_id_vectors}
\eeq
Using the large $L$ limit of  \eqref{eq:a_2_large_L} for $a(-ih,L)$,  the bound \eqref{eq:bound_currents} turns into the conformal collider bounds 
\beq
\frac{ (h-1) \Gamma(h+1)}{(2h-1)\pi^h}  C_{\cal J}
\leq \lambda 
\leq
\frac{\Gamma(h+1)}{2\pi^h} C_{\cal J}\,.
\eeq

For stress-tensors the conformal collider bounds are found similarly. 
For large $L$ the functions $t_2(\nu_0,L)$ and $t_4(\nu_0,L)$ satisfy
\begin{align}
\lim_{L \to \infty} t_2(\nu_0,L) ={}& \left(\frac{r_2(\nu_0)}{r_1(\nu_0)} - \frac{2}{2h+1} \frac{r_3(\nu_0)}{r_1(\nu_0)} (1 + h + i \nu_0) (1 + 3 h + i \nu_0) \right) (h+i\nu_0)(h+i\nu_0-1) \,, 
\nonumber\\
\lim_{L \to \infty} t_4(\nu_0,L) ={}& \frac{r_3(\nu_0)}{r_1(\nu_0)}
\left(h+i \nu _0-1\right) \left(h+i \nu _0\right) \left(h+i \nu _0+1\right) \left(h+i \nu _0+2\right)\,.
\label{eq:t2_t4_large_L}
\end{align}
Using the basis changes \eqref{eq:beta_to_alpha_st}, \eqref{eq:alpha_to_C_st} and \eqref{eq:OPE_relation_st}, together with the conservation condition
\eqref{eq:conservation_embedding_st}, and finally the relation \eqref{eq:OPE_to_t2_t4} to the quantities $t_2$ and $t_4$ from the conformal collider literature, we find
\beq
\lim_{L \to \infty} \left. t_2(-ih,L) \right|_{ j=2} = t_2, \qquad
\lim_{L \to \infty} \left. t_4(-ih,L) \right|_{ j=2} = t_4.
\eeq
For these $t_2$ and $t_4$ equation \eqref{eq:t2t4_bounds} are precisely the conformal collider bounds.

%%%%%%%%%%%%%%%%%%%%%%%%%%%%%%%%%%%%%%%%%%%%%%%%%%%%%%%%%%%%%%%%%%%%%%%%%%%%%%%%%%%%%%%%%%%
\subsection{Validity of unitarity condition along  Regge trajectory}
\label{sec:cond_along_trajectory}
%%%%%%%%%%%%%%%%%%%%%%%%%%%%%%%%%%%%%%%%%%%%%%%%%%%%%%%%%%%%%%%%%%%%%%%%%%%%%%%%%%%%%%%%%%%

Finally we consider the validity of the unitarity condition when  the saddle point moves along the negative imaginary $\nu$ axis. As explained before,
this corresponds to 
scaling  $L$ with $\ln S$.
We need to compare the decay with $S$ of  the phase shift with that  of the error function (\ref{eq:epsilon_S_L_appendix}) 
in the unitarity condition (\ref{eq:AdSunitarityFinal_appendix}).
Inserting the saddle point condition \eqref{eq:SaddlePoint} one sees that the phase shift (\ref{eq:ChiReggeonSpin}) 
scales as
\beq
\chi(S,L) \propto 
S^{j(\nu_0)-1} \Pi_{i \nu_0} (L) \propto
S^{j(\nu_0)-1} S^{- i j'(\nu_0) (1-i \nu_0 -h)} \,.
\eeq
Let us define the exponent $\eta_k$  for the leading contribution of the error function of an operator ${\cal O}_k$ as 
\beq
\epsilon_k(S,L) \propto S^{-\eta_k}\,.
\eeq
Thus, condition (\ref{eq:unitarity_cond_spin}) will only be true provided 
\begin{equation}
j(\nu_0)-1 + (1 -i \nu_0-h) \big(-i j'(\nu_0)\big)   >  -\eta_k  \,,
\label{eq:ConditionScalarsLlarge2}
\end{equation}
is satisfied for all operators ${\cal O}_k$ in the theory. 
As one can read off from \eqref{eq:epsilon_S_L_appendix}, we have
\beq
\eta_k =  \Delta_k+ij'(\nu_0) J_k\,.
\eeq
Since $0<-i j'(\nu_0) < 1$, this exponent can take values between the dimension $  \Delta_k$ and the twist $\tau_k = (\Delta_k-J_k)$. 
If ${\cal O}_k$ is a scalar, $\eta_k$ is bounded from below by the unitarity bound
\beq
\eta_k = \Delta_k \geq  (h-1)\,.
\label{eq:eta_bound_scalars}
\eeq
For operators with spin we have a similar condition from the unitarity bound
\beq
\eta_k \geq \tau_k \geq (2h -2)\,.
\label{eq:eta_bound_spin}
\eeq

%%%%%%%%%%%%%%%%%%%%%%%%%%%%%%%%%%%%%%%%%%%%%%%%%%%%%%%%%%%%%%%%%%%%%%%%%%%%%%%%%%%%%%%%%%%
\subsubsection{Large gap}
%%%%%%%%%%%%%%%%%%%%%%%%%%%%%%%%%%%%%%%%%%%%%%%%%%%%%%%%%%%%%%%%%%%%%%%%%%%%%%%%%%%%%%%%%%%

Next we analyze the LHS of (\ref{eq:ConditionScalarsLlarge2}). 
In the case of theories with a large gap, we can estimate the  LHS of (\ref{eq:ConditionScalarsLlarge2}) using 
the spin function (\ref{ReggeSpinStrong}) in the large gap limit.
For the stress tensor protected point where $\nu_0=-ih $, and taking the limit ${\cal D}\ll 1$, we obtain the condition
\begin{equation}
1-(2h-1)2h{\cal D} > - \eta_k \,.
\end{equation}
Thus, if we are conservative and set $\eta_k=(h-1)$ from the scalar unitary bound,
the unitarity condition clearly works for fixed $\nu_0=-i h $, corresponding to the physical OPE with the stress-tensor.

We also wish to consider the region $\nu_0^2\sim 1/{\cal D}$. In this case we can use 
\begin{equation}
j(\nu) \approx  2  -  {\cal D} (h^2+\nu^2)  \left( 1 +   \sum_{n = 1}^\infty b_n {\cal D}^{n} \nu^{2n}  \right)\,,
\label{eq:ReggeSpinLargeGap}
\end{equation}
from  which we conclude that the unitarity condition holds provided 
\begin{equation}
1+ {\cal D} \nu_0^2  \left( 1 + \sum_{n=1}^\infty  (2n+1) b_n {\cal D}^n \nu_0^{2n}\right) > -\eta_k \,.
\label{eq:ConditionScalarsLargeGap}
\end{equation}
We look at $\nu_0$ such that $\nu_0^2=- \epsilon^2/{\cal D}$, with   $ {\cal D}\ll \epsilon^2\ll 1$. 
That is, $\nu_0$ is at a small fraction of $\Delta_g$ away from the energy-momentum tensor, but still 
very far from it, since $|\nu_0|\gg1$. Then the condition becomes $1- \epsilon^2 >  -\eta_k$,
which is in general  satisfied.
For scalar operators this gives the usual condition 
$r(-i\epsilon/{\cal D} )\ge0$. In the case of vector currents, we consider the large $\nu_0$ limit of
(\ref{eq:a_2_large_L}),
\beq
\lim_{L\to \infty} a(\nu_0,L) = \nu_0^2\,, \qquad\qquad    (|\nu_0|\gg 1) \,.
\label{eq:a_large_L_nu}
\eeq
This means that for large $\nu_0$ we must have that $ r_4(\nu_0) / r_1(\nu_0)\sim 1/\nu_0^2$.
In the case  of stress tensors, the large $\nu_0$ limit of
(\ref{eq:t2_t4_large_L}) gives
\begin{align}
\lim_{L \to \infty} t_2(\nu_0,L) ={}&  -  \nu_0^2\, \frac{r_2(\nu_0)}{r_1(\nu_0)} - \frac{2\nu_0^4}{2h+1} \frac{r_3(\nu_0)}{r_1(\nu_0)}   \,, 
\nonumber\\
\lim_{L \to \infty} t_4(\nu_0,L) ={}& \nu_0^4\,\frac{r_3(\nu_0)}{r_1(\nu_0)}\,,
 \qquad\qquad   \qquad\qquad  \qquad   (|\nu_0|\gg 1) \,,
 \label{eq:t2_t4_large_L_nu}
\end{align}
implying that for large $\nu_0$ we must have that $ r_2(\nu_0) / r_1(\nu_0)\sim 1/\nu_0^2$ and
$ r_3(\nu_0) / r_1(\nu_0)\sim 1/\nu_0^4$.

We may wish, however, to reach the next operator at $J=4$,
which happens for ${\cal D} \nu_0^2$ of order unit. 
In that case  condition  (\ref{eq:ConditionScalarsLargeGap}) takes the form  $1- g( {\cal D} \nu_0^2 ) < - \eta_k$, for some unknown function $g$, and we can not make any general statement. 
On the other hand, if we restrict to theories that have a well defined flat space limit, then all the coefficients $b_n$ in (\ref{eq:ReggeSpinLargeGap}) are zero.  In that case
$J=4$ is reached for  $i\nu_0=\sqrt{2/{\cal D}}$ and the condition \eqref{eq:ConditionScalarsLlarge2} becomes $-1>-\eta_k$. Using \eqref{eq:eta_bound_scalars} and \eqref{eq:eta_bound_spin}, this is satisfied for $d>4$ and can also be achieved in $d=4$ by requiring that there is no scalar operator saturating the unitarity bound.
Thus, for theories with a well defined flat space limit, the ratios of OPE coefficients are also suppressed by $1/\Delta_g^2$, for example in the vector case
\begin{equation}
- \frac{d-1}{ \Delta_g^2} \leq \frac{C^{(4)}_{{\cal J}{\cal J}{\cal O}_{\!J=4}}}{C^{(1)}_{{\cal J}{\cal J}{\cal O}_{\!J=4}}} \leq \frac{d-1}{(d-2) \Delta_g^2} \,.
\end{equation}
And for the stress tensor OPE coefficients 
\bea
0 &\leq \frac{1}{\D_g^2} - \frac{1}{d-1} 
\frac{C^{(2)}_{TT{\cal O}_{\!J=4}}}{C^{(1)}_{TT{\cal O}_{\!J=4}}} \,,\\
0 &\leq \frac{1}{\D_g^4} + \frac{d-3}{2(d-1)\D_g^2} 
\frac{C^{(2)}_{TT{\cal O}_{\!J=4}}}{C^{(1)}_{TT{\cal O}_{\!J=4}}}
-\frac{1}{d+1}  \frac{C^{(3)}_{TT{\cal O}_{\!J=4}}}{C^{(1)}_{TT{\cal O}_{\!J=4}}} \,,\\
0 &\leq \frac{1}{\D_g^4} + \frac{d-3}{(d-1)\D_g^2} 
\frac{C^{(2)}_{TT{\cal O}_{\!J=4}}}{C^{(1)}_{TT{\cal O}_{\!J=4}}}
+\frac{d-2}{d+1}  \frac{C^{(3)}_{TT{\cal O}_{\!J=4}}}{C^{(1)}_{TT{\cal O}_{\!J=4}}} \,.
\eea{eq:bound_J4_T}
It would be nice to check this prediction for the case of ${\cal N}=4$ SYM.

%%%%%%%%%%%%%%%%%%%%%%%%%%%%%%%%%%%%%%%%%%%%%%%%%%%%%%%%%%%%%%%%%%%%%%%%%%%%%%%%%%%%%%%%%%%
\subsubsection{Weakly coupled CFTs}
%%%%%%%%%%%%%%%%%%%%%%%%%%%%%%%%%%%%%%%%%%%%%%%%%%%%%%%%%%%%%%%%%%%%%%%%%%%%%%%%%%%%%%%%%%%

Finally let us consider a weakly interacting CFT with a small coupling. In this case we can
 write $\Delta= d-2 + J + \gamma(J)$, which implies that $j'(\nu)= i (1 + d\gamma/dJ)^{-1}$. Thus the unitarity condition is satisfied provided  
 \begin{equation}
 -\frac{d}{2}- \gamma\big(j(\nu_0)\big)  +  \frac{d\gamma}{dJ} \left(j(\nu_0)-1 +\frac{d}{2} \right) > - \eta_k =
\left\{
\begin{array}{ll}
-\Delta_{min}\,, & J_k=0\\
- \tau_k  - \frac{d\gamma}{dJ}\,, & J_k>0
\end{array}
\right.
\,.
\end{equation}
For ${\cal O}_k$ a scalar operator, and if we impose $\Delta_{min}=2h-2$ as in large $N$ gauge theories, we have
\begin{equation}
- \gamma\big(j(\nu_0)\big)  +  \frac{d\gamma}{dJ} \big(j(\nu_0)-1 +h \big) > -h+2\,.
\end{equation}
For ${\cal O}_k$ a spin $J_k$ operator we consider a twist gap of $2h-2$ to obtain
\begin{equation}
- \gamma\big(j(\nu_0)\big)  +  \frac{d\gamma}{dJ} \big(j(\nu_0) +h \big) > -h+2\,.
\end{equation}
We conclude that
for $h>2$ the AdS unitarity condition is always satisfied at weak coupling, while for 
$h<2$ it is never satisfied at weak coupling. 
For $h=2$ the condition is only satisfied for low spin. 
For instance, it is clearly satisfied at $j(\nu_0)=2$ because $\gamma(2)=0$ and $\frac{d\gamma}{dJ}>0$. On the other hand, for large spin the condition is not satisfied neither in gauge theories where $\gamma(J)\sim \log J$ nor in other CFTs where typically $\gamma(J)\sim c_1- c_2/J^{\tau_0}$ for some positive constants $c_1, c_2, \tau_0$.

%%%%%%%%%%%%%%%%%%%%%%%%%%%%%%%%%%%%%%%%%%%%%%%%%%%%%%%%%%%%%%%%%%%%%%%%%%%%%%%%%%%%%%%%%%%
\section{Concluding remarks}
\label{sec:remarks}
%%%%%%%%%%%%%%%%%%%%%%%%%%%%%%%%%%%%%%%%%%%%%%%%%%%%%%%%%%%%%%%%%%%%%%%%%%%%%%%%%%%%%%%%%%%

In this paper, we argued that it is natural to define a phase shift associated to the Regge limit of large $N$ CFT four-point functions. Unitarity implies that the imaginary part of the phase shift must be positive, as usual in scattering theory.
Using this condition, we derived bounds on the analytic continuation of OPE coefficients to complex angular momentum. In particular, we showed that OPE coefficients associated to higher derivative couplings in the dual AdS theory, must vanish when continued to the intercept $J \to j(0)$.
It would be very interesting to test this result with explicit calculations. Given that the argument does not involve large gap, it should be testable in weakly coupled large $N$ conformal gauge theories.
This result also allowed us to give a new argument for the expected effective field theory suppression of the  OPE coefficients of the stress tensor operator ($J=2$) when all single-trace higher spin ($J>2$) operators have parametrically large dimension.

Conformal Regge theory was constructed by analogy with Regge theory for scattering amplitudes in flat space \cite{Cornalba:2007fs,Costa:2012cb}. Its validity rests on the assumption that one can drop the contribution from infinity in the complex angular momentum plane when deforming the Sommerfeld-Watson contour (as reviewed around equation \eqref{eq:SWtransform}).
It is not known if this assumption is valid in general or only in special cases like the planar limit of large $N$ gauge theories. If the assumption is valid in general, then it implies that the Regge trajectories at large but finite $N$ are quite different from the planar limit Regge trajectories.
In the planar limit, it is natural to define a leading Regge trajectory of single-trace operators $j (\nu)$. It is unclear to us, what is the relation between $j (\nu)$ and the true leading Regge trajectory $j_{\rm exact}(\nu)$ of a theory with large but finite $N$. Notice that for large gap $\Delta_{g}$, the single-trace operators with $J\ge 4$ have dimension much larger than some double-trace operators with the same spin. Therefore, $j (\nu)$ and $j_{\rm exact}(\nu)$ are very different curves in the region $J\ge 4$. Moreover, we expect that the single-trace intercept $j(0) \to 2$ when $\Delta_{g} \to \infty$ but the exact intercept $j_{\rm exact}(0) \le 1$ because the correlator is bounded by 1 in the Regge limit at finite $N$. \footnote{This follows from analyzing the OPE channel $(13)(24)$  which is convergent in the Regge limit. 
This is also consistent with  a naive exponentiation  of the planar level phase shift (eikonalization).
If the (imaginary part of the) planar level phase shift grows with $S$ then we expect the correlator to vanish in the Regge limit at finite $N$; if it decreases then we expect the correlator to approach 1. These two possibilities have been discussed recently  in \cite{Murugan:2017eto} from the chaos point of view.
}
Therefore, the two curves $j(\nu)$ and $j_{\rm exact}(\nu)$ must also be different in the region $\nu \sim 0$. In appendix \ref{sec:convexity}, we prove convexity of $j_{\rm exact}(\nu)$. 
It is unclear if the argument can be applied or generalized to the single-trace leading Regge trajectory $j(\nu)$ which plays a central role in this paper.
We leave these important questions for future investigations.

%%%%%%%%%%%%%%%%%%%%%%%%%%
\section*{Acknowledgments}
%%%%%%%%%%%%%%%%%%%%%%%%%%
The authors  benefited from discussions with Simon Caron-Huot,
 Thomas Hartman, Jared Kaplan, Manuela Kulaxizi, Daliang Li, David Meltzer, Andrei Parnachev, David Poland, Alexander Zhiboedov, 
 and are grateful to ICTP-SAIFR for hosting a great 
Bootstrap 2017 meeting.
MSC thanks Universidade de Santiago de Compostela for the hospitality in the initial and final stages of this work.
This research received funding from the [European Union] 7th Framework Programme (Marie Curie Actions) under grant agreement 317089 (GATIS), from the grant CERN/FIS-NUC/0045/2015 and from the 
Simons Foundation grants 488637 and 488649 (Simons collaboration on the Non-perturbative bootstrap). 
Centro de F\'\i sica do Porto is partially funded by the Foundation for Science and Technology of Portugal (FCT). 
 JP is supported
by the National Centre of Competence in Research SwissMAP funded by the Swiss National
Science Foundation

%%%%%%%%%%%%%%%%%%%%%%%%%%%
\appendix
% Don't show subsections of appendices in toc
\addtocontents{toc}{\protect\setcounter{tocdepth}{1}}
%%%%%%%%%%%%%%%%%%%%%%%%%%%

%%%%%%%%%%%%%%%%%%%%%%%%%%%
\section{Discontinuity of scalar conformal block}
\label{sec:discontinuity}
%%%%%%%%%%%%%%%%%%%%%%%%%%%

In this appendix we derive the discontinuity of the scalar conformal block, normalized as
\beq
\lim_{z \to 0} \lim_{\barz \to 0} G_{\D,J}(z,\bar{z}) \sim z^{\frac{\Delta+J}{2}} \barz^{\frac{\Delta-J}{2}}\,,
\label{eq:block_boundary}
\eeq
with cuts on the real $z$ and $\bar{z}$ axis for $z,\bar{z}<0$ and $z,\bar{z}>1$.
We will compute the analytic continuation of this block, as $z$ goes around $1$ counter clockwise with $\bar{z}$ held fixed,
generalizing the derivation of section (4.3) in \cite{Cornalba:2006xm} for the case that $\D_{12}$ or $\D_{34}$ are nonzero ($\Delta_{ij}=\Delta_i-\Delta_j$). Although these parameters are
ultimately zero in our setup, this generalization is needed in order to
use the result with the spin generating differential operators of \cite{Costa:2011dw}, because these operators include shifts in $\D_{12}$ or $\D_{34}$
of the scalar blocks.

In this appendix we use the definitions
\beq
\tau = \frac{\D + J}{2}\,, \quad \bar{\tau} = \frac{\D - J}{2} \,, \quad a =-\frac{\D_{12}}{2}\,, \quad b =\frac{\D_{34}}{2}\,.
\eeq
The conformal Casimir operator in the limit $\barz \to 0$ is (see e.g.\ \cite{Dolan:2011dv})
\beq
z^2 (1-z) \partial^2 - (1+a+b)z^2 \partial - a b z +\barz^2 \bar{\partial}^2 - (d-2)\barz \bar{\partial}\,.
\eeq
In this limit and for the boundary condition \eqref{eq:block_boundary}
the Casimir equation is solved by 
\beq
 \lim_{\barz \to 0} G_{\D,J}(z,\bar{z}) \sim z^{\tau} \barz^{\bar{\tau}} {}_2 F_1 (\tau+a,\tau+b,2\tau,z)\,.
\eeq

%%%%%%%%%%%%%%%%%%%%%%%%%%%
\subsection{Analytic continuation around 1}
%%%%%%%%%%%%%%%%%%%%%%%%%%%
To derive the monodromy around one can use the following expansion of the hypergeometric function around $1$, which is valid for $(a+b) \in \mathbb{Z}$ \cite{MR0167642}
\begin{align}
&
{}_2 F_1 (\tau+a,\tau+b,2\tau,z) =  \text{ terms without branch cut} \;- 
\label{eq:hypergeo_id_1}\\
&- \frac{(-1)^{a+b}\Gamma(2\tau)}{\Gamma(\tau-a) \Gamma(\tau-b) \Gamma(1+a+b)}\,
 {}_2 F_1 (\tau+a, \tau+b, 1+a+b, 1-z) \ln(1-z)\,.
\nonumber
\end{align}
From analytically continuing counter-clockwise around one the logarithm picks up a factor of $2 \pi i$ and we arrive at
\beq
\text{Disc } G_{\D,J}(z,\bar{z}) \sim 
- \barz^{\bar{\tau}}z^{\tau} \frac{2\pi i (-1)^{a+b}\Gamma(2\tau)}{\Gamma(\tau-a) \Gamma(\tau-b) \Gamma(1+a+b)}\,
{}_2 F_1 (\tau+a, \tau+b, 1+a+b, 1-z)\,.
\eeq
The leading behavior for small $z$ is
\bea
\text{Disc } G_{\D,J}(z,\bar{z}) &\sim 
- 2\pi i\,
\frac{\Gamma(2\tau) \Gamma(2\tau-1) (-1)^{a+b}}{\Gamma(\tau-a) \Gamma(\tau+a) \Gamma(\tau-b) \Gamma(\tau+b) }
\barz^{\bar{\tau}}z^{1-\tau} \,.
\eea{eq:disc_one_expansion}
This is the result for $\barz \to 0$, however the result can in general be a function of $\barz/z$ in the region $z, \barz \sim 0$, i.e.\
\beq
\text{Disc } G_{\D,J}(z,\bar{z}) \sim \barz^{\bar{\tau}}z^{1-\tau}  g(\barz/z)\,.
\label{eq:disc_ansatz}
\eeq
The function $g(\barz/z)$ can be found by solving the Casimir equation near $z, \barz \sim 0$, where the Casimir operator becomes
\beq
z^2 \partial^2 +\barz^2 \bar{\partial}^2 + (d-2) \frac{z \barz}{z-\barz} (\partial - \bar{\partial})\,.
\eeq
Inserting \eqref{eq:disc_ansatz} in the corresponding Casimir equation one finds a hypergeometric differential equation with  solution
\beq
g(\barz/z) \propto {}_2 F_1 \left(\frac{d}{2}-1,\tau+\bar{\tau}-1,\tau+\bar{\tau}+1 -\frac{d}{2}, \frac{\barz}{z}\right).
\eeq
We arrive at the final result
\bea
\text{Disc } G_{\D,J}(z,\bar{z}) ={}& 
- 2\pi i\,
\frac{\Gamma(2\tau) \Gamma(2\tau-1) (-1)^{a+b}}{\Gamma(\tau-a) \Gamma(\tau+a) \Gamma(\tau-b) \Gamma(\tau+b) }\\
& \times \barz^{\bar{\tau}}z^{1-\tau} 
{}_2 F_1 \left(\frac{d}{2}-1,\tau+\bar{\tau}-1,\tau+\bar{\tau}+1 -\frac{d}{2}, \frac{\barz}{z}\right).
\eea{eq:disc_one}

%%%%%%%%%%%%%%%%%%%%%%%%%%%
\subsection{Relation to harmonic functions on hyperbolic space}
%%%%%%%%%%%%%%%%%%%%%%%%%%%

Restoring the dependence on the conformal dimensions and spin, we can rewrite the result
\eqref{eq:disc_one} as
\bea
\text{Disc } G_{\D,J}(z,\bar{z}) \approx{}&
-4 i \pi^{h} \sigma^{1-J}
 \frac{\Gamma(\D-h+1)}{\Gamma(\D-1)}\, \Pi_{\Delta-h}(\rho) \\
&\frac{\Gamma(\D+J) \Gamma(\D+J-1) (-1)^{\frac{-\D_{12}+\D_{34}}{2}}}{
\Gamma\left(\frac{\D+J+\D_{12}}{2}\right) \Gamma\left(\frac{\D+J-\D_{12}}{2}\right) \Gamma\left(\frac{\D+J+\D_{34}}{2}\right) \Gamma\left(\frac{\D+J-\D_{34}}{2}\right)}\,,
\eea{eq:disc_with_propagator}
where $\Pi_{\Delta-h}(\rho)$ is the scalar propagator on hyperbolic space $H_{d-1}$,
\beq
\Pi_{\Delta-h}(\rho)
=  \frac{\pi^{1-h}}{2}\frac{\Gamma(\D-1)}{\Gamma(\D-h+1)}\,
\e^{(1-\D)\rho}
{}_2 F_1 \Big(h-1, \D-1, \D-h+1,\e^{-2\rho} \Big)\,.
\label{eq:propagator}
\eeq
We can use this to write down the analytic continuation of $F_{\nu,J}(z,\bar{z})$ defined in \eqref{cb+shadow}.
To do that we need the definition of
\begin{align}
K_{\D,J}=\ & \frac{\Gamma(\Delta+J) \,\Gamma(\Delta-h+1) \, (\Delta-1)_J   }{ 
4^{J-1} 
\Gamma\!\left( \frac{\Delta +J+\Delta_{1 2}}{2}\right)
\Gamma\!\left( \frac{\Delta +J-\Delta_{1 2}}{2}\right) 
\Gamma\!\left( \frac{\Delta +J+\Delta_{3 4}}{2}\right)
\Gamma\!\left( \frac{\Delta +J-\Delta_{3 4}}{2}\right)}
\label{KDeltaJ}\\&
\frac{1}{
 \Gamma\!\left( \frac{\Delta_1 +\Delta_{2} -\D+J}{2}\right)
\Gamma\!\left( \frac{\Delta_3 +\Delta_{4} -\D+J}{2}\right)
\Gamma\!\left( \frac{\Delta_1 +\Delta_{2} +\D+J-d}{2}\right)
\Gamma\!\left( \frac{\Delta_3 +\Delta_{4} +\D+J-d}{2}\right)}\,.
 \nonumber
\end{align}
Setting
$\Delta_1=\Delta_2$ and $\Delta_3=\Delta_4$ we obtain 
\beq
\text{Disc }  F_{\nu,J}(z,\bar{z}) \approx -i \pi^h 4^J \sigma^{1-J} \gamma(\nu) \gamma(-\nu)\,
\Omega_{i \nu} (\rho)\,,
\eeq
where
\begin{equation}
\gamma(\nu) = \Gamma\!\left(  \frac{2\Delta_1 +J +i\nu-h}{2} \right)  \Gamma\!\left(  \frac{2\Delta_3 +J +i\nu-h}{2} \right) ,
\label{eq:gamma}
\end{equation}
and
\beq
\Omega_{i \nu} (\rho) = \frac{i \nu}{2 \pi}\, \big(\Pi_{i\nu} - \Pi_{-i\nu}\big)\,,
\label{eq:Omega}
\eeq
is the harmonic function on $H_{d-1}$, which satisfies
\beq
\left( \nabla_{H_{d-1}}^2 + \nu^2 + (h-1)^2 \right) \Omega_{i \nu} (\rho) = 0 \,.
\label{eq:nabla}
\eeq

%%%%%%%%%%%%%%%%%%%%%%%%%%%
\section{Casimir equation in the Regge limit}
\label{sec:casimir_equation}
%%%%%%%%%%%%%%%%%%%%%%%%%%%

In this appendix we discuss the differential operators that are used to give spin to the harmonic functions on
hyperbolic space $H_{d-1}$. It will turn out that spherical tensor harmonics on $H_{d-1}$ automatically solve the Casimir
equation in the Regge limit.
Note that  in  \cite{Costa:2016other}
we already exploited an analogous construction in the lightcone limit 
(with tensor harmonics on the sphere $S^{d-1}$).

We begin by deriving the leading term of the Casimir equation in the Regge limit.
To this end we introduce the embedding space coordinates $P^M \in \mathbb{R}^{2,d}$
\beq
P^M = \big(P^+,P^-,P^m\big)\,, \qquad P\cdot P = -P^+ P^- + \eta_{m n} P^m P^n\,.
\eeq
They are related to the coordinates $y^m \in \mathbb{R}$ of physical Minkowski space by \cite{Cornalba:2009ax}
\beq
P^M = \big(y^+, y^-, 1, y^2,y_\perp\big)\,,
\eeq
and to the coordinates $x_i$ by
\bea
P_1^M &= \left(-1,-x_1^2,x_1^m\right)\,, \qquad &&
P_3^M &=& \left(-x_3^2,-1,x_3^m\right)\,,\\
P_2^M &= \left(1,x_2^2,-x_2^m\right) \,,\qquad &&
P_4^M &=& \left(x_4^2,1, -x_4^m\right)\,.
\eea{eq:xiPatches}
For the external polarizations $z_1$ and $z_2$ (satisfying $z_i^2 = 0$) the corresponding polarizations in embedding space are
\beq
Z_i^M = \big(0,-2 x_i \cdot z_i,z_i^m\big)\,.
\label{eq:ziPatches}
\eeq
Using these relations one can derive the leading term  in $x = x_1 - x_2$ of the Casimir operator
\begin{align}
\frac{1}{2} \left( J_{M N}^{(1)} + J_{M N}^{(2)} \right)^2
={}& \frac{1}{2} \left( J_{m n}^{(x)} + J_{m n}^{(z_1)} + J_{m n}^{(z_2)} \right)^2
+ \big( x \cdot \partial_x + \Delta_1 + \Delta_2 \big)
\big( x \cdot \partial_x + \Delta_1 + \Delta_2 - d\big)
\nonumber\\
& + \text{ terms that increase homogeneity in }x\,,
\label{eq:casimir_x_expansion}
\end{align}
where
\bea
J_{M N}^{(i)} &= - i \left( P_{iM} \frac{\partial}{\partial P_i^N} - P_{iN} \frac{\partial}{\partial P_i^M} + Z_{iM} \frac{\partial}{\partial Z_i^N} - Z_{iN} \frac{\partial}{\partial Z_i^M} \right),\\
J_{m n}^{(x)} &= - i \left( x_m \frac{\partial}{\partial x_n} - x_n \frac{\partial}{\partial x^m} \right).
\eea{eq:J}
The leading term of the conformal partial wave  is given in terms of a function $f$ of 
the variables $\hat x = x/|x|, \hat \barx = \barx/|\barx|, z_1$ and $z_2$, that is
\beq
\mathcal{W} (x, \barx, z_1, z_2) = 
\frac{\sigma^{1-l}}{(x^2)^\frac{\D_1+\D_2}{2}(\barx^2)^\frac{\D_3+\D_4}{2}} \,f\!\left( \hat x, \hat \barx , z_1, z_2\right).
\label{eq:W_ansatz}
\eeq
Inserting \eqref{eq:casimir_x_expansion}, \eqref{eq:W_ansatz} and $\D = h \pm i \nu $
into the Casimir equation
\beq
\left(
\frac{1}{2} \left( J_{M N}^{(1)} + J_{M N}^{(2)} \right)^2 - c_{\Delta, l}
\right) \mathcal{W} (P_i,Z_i)= 0\,, \qquad c_{\Delta, l} = \Delta(\Delta-d) + l (l + d - 2)\,,
\eeq
one finds the Casimir equation in the Regge limit
\beq
\left(
\frac{1}{2} \left( J_{m n}^{(\hat{x})} + J_{m n}^{(z_1)} + J_{m n}^{(z_2)} \right)^2  + \nu^2 + \left( h-1 \right)^2
\right)
\!f\left( \hat x, \hat \barx , z_1, z_2 \right) = 0\,.
\label{eq:casimir_eq_regge}
\eeq
For the case where the operators ${\cal O}_1$ and ${\cal O}_2$ are scalars, this equation becomes \eqref{eq:nabla} and is solved by $\Omega_{i \nu} (\rho)$, the harmonic function on $H_{d-1}$ (note that $\frac{1}{2} \big( J_{m n}^{(\hat x)} \big)^2 = \nabla_{H_{d-1}}^2$).
In the general case, solutions can be easily constructed by noting that $\left( J_{m n}^{(\hat x)} + J_{m n}^{(z_1)} + J_{m n}^{(z_2)} \right)^2$ commutes with $(z_i \cdot \hat x)$, $(z_i \cdot \nabla)$ and $(z_1 \cdot z_2)$, where $\nabla$ is the covariant derivative on $H_{d-1}$.
Hence \eqref{eq:casimir_eq_regge} is solved by
\beq
f\big( \hat x, \hat \barx , z_1, z_2 \big) = \calD\, \Omega_{i \nu} (\rho) \,,
\eeq
where $\calD$ is any operator constructed from $(z_i \cdot \hat x)$, $(z_i \cdot \nabla)$ and $(z_1 \cdot z_2)$. This is enough to generate all the independent tensor structures, as the examples \eqref{DTensors} and \eqref{DTensors_st} shown in the main text.
In practice the covariant derivatives can be computed without doing a coordinate change by taking the usual derivative and then projecting all indices to $H_{d-1}$, for example
\beq
\nabla^m\nabla^n \Omega_{i\nu} ( \rho) = \big(\delta^{m}_p+\hat x^{m} \hat x_{p}\big) \big(\delta^{n}_q+\hat x^{n} \hat x_{q}\big) \frac{\partial}{\partial \hat x_p} \big(\delta^{q}_r+\hat x^{q} \hat x_{r}\big) \frac{\partial}{\partial \hat x_r} \,\Omega_{i\nu} ( \rho)\,.
\eeq

%%%%%%%%%%%%%%%%%%%%%%%%%%%
\section{Spinning conformal blocks in the embedding space formalism}
\label{sec:embedding_space}
%%%%%%%%%%%%%%%%%%%%%%%%%%%
In this appendix we perform the change of basis necessary to write the derived bounds on OPE coefficients in terms of
a more conventional basis.

Firstly let us note that 
the discontinuity of the conformal partial wave in the Regge limit can be obtained by acting 
on the discontinuity of the scalar conformal partial wave
with the differential operators $D_i$ introduced in  \cite{Costa:2011dw}
\beq
\text{Disc} \ \mathcal{W} = \frac{1}{(P_{12})^{\frac{\D_1+\D_2+2}{2}} (P_{34})^{\frac{\D_3+\D_4}{2}}}
\sum_{i} c_{12J}^{i} C_{34J}
\, D_i
\left( \frac{P_{24}}{P_{14}} \right)^{\frac{\D_{12}}{2}}
\!\!\!\left( \frac{P_{14}}{P_{13}} \right)^{\frac{\D_{34}}{2}}
\!\!\text{Disc} \ G_{\D, J} (z, \barz)\,,
\label{eq:spinning_discontinuity}
\eeq
where, in terms of $\mathbb{R}^{2,d}$ embedding space vectors,  $P_{ij} = - 2 P_i \cdot P_j $
and $c_{12J}^{i}$ is a basis of OPE coefficients different from the one defined in \eqref{eq:rtoCC_spin}.

%%%%%%%%%%%%%%%%%%%%%%%%%%%
\subsection{Correlator \texorpdfstring{$\bra {\cal J} {\cal J} {\cal O} {\cal O} \ket$}{<J J phi phi>}}
%%%%%%%%%%%%%%%%%%%%%%%%%%%
In the case of two currents the differential operators $D_i$ in \eqref{eq:spinning_discontinuity} are
\beq
D_1 = D_{11} D_{22} \,, \ D_2 = H_{12} \,, \ D_3= D_{12} D_{22} \Sigma^{-2} + D_{21} D_{11} \Sigma^{2}\,, \ D_4 = D_{12} D_{21} \,,
\label{eq:Dij_vectors}
\eeq
where  $D_{ij}$ and $H_{ij}$ are defined as in \cite{Costa:2011dw} and $\Sigma^n$ is an operator shifting $\D_{12} \to \D_{12} + n$.
We want to relate these structures to our basis of differential operators \eqref{DTensors},
which was defined at leading order in $\sigma$.
We need to make the dependence of $\text{Disc} \ G_{\D, J} (z, \barz)$ on the cross ratios and $\D_{12}$ explicit in order to read off
differential operators which act on a function of $\rho$. This dependence can be read off from \eqref{eq:disc_with_propagator}
and has the form
\beq
\text{Disc} \ G_{\D, J} (z, \barz) = \sigma^{1-J} \frac{(-1)^{\frac{-\D_{12}}{2}}}{\Gamma(\frac{\D+J+\D_{12}}{2}) \Gamma(\frac{\D+J-\D_{12}}{2})} 
\,f(\rho)\,.
\eeq
In order to map the resulting tensor structures to the $x, \barx$ coordinates we use the embedding introduced above in \eqref{eq:xiPatches} and \eqref{eq:ziPatches},
and make the choice  $x_1=x_4=0$, $x_2=-x$ and $x_3=\barx$,
\bea
P_1^M &= \left(-1,0,0\right)\,, \qquad &&
P_3^M &=& \left(-\barx^2,-1,\barx^m\right)\,,\\
P_2^M &= \left(1,x^2,x^m\right) \,,\qquad &&
P_4^M &=& \left(0,1, 0\right)\,,\\
Z_1^M &=\left(0,0,z_1^m\right)\,, \qquad &&
Z_2^M &=& \left(0,2 x \cdot z_2,z_2^m\right)\,.
\eea{eq:xPatches}
To leading order in $\sigma$ we find (now setting $\D_{12} = \D_{34} = 0$)
\beq
\text{Disc} \ \mathcal{W} = \frac{\sigma^{1-J}}{(-x^2)^{\D_1} (-\barx^2)^{\D_3}}
\frac{1}{\Gamma(\frac{\D+J}{2})^2}
\sum_{i=1}^4 c_{12J}^{i} C_{34J} \tilde{D}_i f(\rho)\,,
\label{eq:Dtilde_embedding}
\eeq
where the operators $\tilde D_i$ can be expressed in terms of the operators defined in \eqref{DTensors}
\bea
\tilde D_1 &= \frac{1}{2} \big( (1-J+\omega) \calD_1 + J(J-1) \calD_2 + J \calD_3 + \calD_4 \big)\,,
\\
\tilde D_2 &= \calD_1 - \calD_2\,, 
\\
\tilde D_3 &= \frac{\D+J-2}{\D+J} \big( (1+J+\omega) \calD_1 - J(J+1) \calD_2  + \calD_4 \big)\,,
\\
\tilde D_4 &= \frac{1}{2} \big( (1-J+\omega) \calD_1 + J(J-1) \calD_2 - J \calD_3 + \calD_4 \big)\,,
\eea{eq:Dleft_to_DTensors}
where we used the notation
\beq
\omega \equiv \frac{\nu^2 + \left(h-1\right)^2}{2h-1}\,.
\label{eq:omega}
\eeq
The relation between the differential operators can also be written in matrix form
\beq
\tilde D_i = \sum_k M_{ik} \calD_k\,.
\eeq
By comparing \eqref{ReggeA} and \eqref{eq:Dtilde_embedding} one sees that the same matrix also relates the OPE coefficients
\beq
\alpha_k(\nu) \propto \sum_i c_{{\cal J}{\cal J} j(\nu)}^{i} M_{ik}\,.
\eeq
The overall factor is the same as in the scalar case \eqref{eq:alpha}
\beq
\chi(\nu) \equiv-
 \left(  i \cot\!\left( \frac{\pi j(\nu)}{2}\right)  -1 \right)
\pi^{h+1} 4^{j(\nu)}  \gamma(\nu) \gamma(-\nu)  \frac{j'(\nu)\,c_{{\cal O}{\cal O}j(\nu)}  K_{h\pm i\nu,j(\nu)} }{4\nu}\,,
\eeq
so that the relations are
\bea
\frac{\alpha_1 (\nu)}{\chi(\nu)} &= \frac{1}{2} (1-j+\omega )\, c_1+c_2+\frac{(-2+j+\Delta ) (1+j+\omega )}{j+\Delta }\, c_3+\frac{1}{2} (1-j+\omega ) \,c_4\,,\\
\frac{\alpha_2 (\nu)}{\chi(\nu)} &= \frac{1}{2} (-1+j) j \,c_1-c_2-\frac{j (1+j) (-2+j+\Delta )}{j+\Delta }\, c_3+\frac{1}{2} (-1+j) j \,c_4\,,\\
\frac{\alpha_3 (\nu)}{\chi(\nu)} &= \frac{j}{2} (c_1 - c_4)\,,\\
\frac{\alpha_4 (\nu)}{\chi(\nu)} &= \frac{c_1}{2}+\left(1-\frac{2}{j+\Delta }\right) c_3+\frac{c_4}{2}\,,
\eea{eq:alpha_to_C_vectors}
where   $\D = h\pm i \nu $ and we used the notation $c_i \equiv c^i_{{\cal J}{\cal J}j(\nu)}$.

The OPE coefficients are further related to the coefficients $\alpha, \beta, \gamma, \eta$
 appearing in the three-point function basis defined by 
(see \cite{Costa:2011dw} for details)
\beq
\bra {\cal J}_1^{\D_1} {\cal J}_2^{\D_1} \calO_3^{\D,J} \ket = V_3^{J-2} \,
\frac{\alpha V_1 V_2 V_3^2 + \beta (H_{13} V_2 + H_{23} V_1) V_3 + \gamma H_{12} V_3^2 + \eta H_{13} H_{23}}{(P_1 \cdot P_2)^{\frac{\D - 2\D_1}{2}} (P_2 \cdot P_3)^{-\frac{\D}{2}} (P_3 \cdot P_1)^{-\frac{\D}{2}}  }\,.
\label{eq:JJL_3point}
\eeq
This three-point function is given in terms of the building blocks\footnote{Note that the normalization of $H_{ij}$ is different than in \eqref{eq:Dij_vectors}.}
\beq
H_{i,j} = Z_i \cdot Z_j - \frac{(P_i \cdot Z_j) (P_j \cdot Z_i)}{(P_i \cdot P_j)} \,,
\ \ \ \ 
V_{i,jk} = - \frac{i}{\sqrt{2}}\frac{(Z_i \cdot P_j) (P_k \cdot P_i) - (Z_i \cdot P_k) (P_j \cdot P_i)}{\sqrt{P_i \cdot P_j} \sqrt{P_j \cdot P_k} \sqrt{P_k \cdot P_i}}\,,
\eeq
and we used the shorthands
\beq
V_1 = V_{1,23}\,, \qquad V_2 = V_{2,31}, \qquad V_3 = V_{3,12}\,.
\eeq
The relation of the OPE coefficients in these two bases is
\bea
c_{{\cal J}{\cal J}j(\nu)}^1 &=
\frac{(-1+J) \big(J (\alpha -2 \beta )+2 \beta  \Delta \big)+(J-\Delta )^2 \eta }{2 (-1+J) J (-1+\Delta ) \Delta }\,,\\
c_{{\cal J}{\cal J}j(\nu)}^2 &=
\frac{-\alpha +2 \beta +\gamma  \Delta +\frac{\Delta-J  }{J-1}\,\eta }{\Delta }\,,\\
c_{{\cal J}{\cal J}j(\nu)}^3&=
\frac{(-1+J) J (\alpha -2 \beta )+(J-\Delta ) (J+\Delta ) \eta }{2 (-1+J) J (-1+\Delta ) \Delta }\,,\\
c_{{\cal J}{\cal J}j(\nu)}^4&=
\frac{(-1+J) \big(J (\alpha -2 \beta )-2 \beta  \Delta \big)+\big(-4 \Delta +(J+\Delta )^2\big) \eta }{2 (-1+J) J (-1+\Delta ) \Delta }\,.
\eea{eq:OPE_relation_vectors}
In this basis it is easy to compute the conservation conditions for currents in which case
$\D_1 =d-1$.
These  conditions are
\bea
0 ={}& \beta  \Big(\Delta  (d-\Delta -2)+J \big(3 d-2 (\Delta +2)\big)+J^2\Big)\\
&+\alpha  J (-d+\Delta +1)-\eta  (\Delta +J) (2 d-\Delta +J-4)\,,\\
0 ={}& \beta  (d-\Delta -2)+\eta  (-2 d+\Delta -J+4)+\gamma  J\,.
\eea{eq:conservation_embedding_vectors}
It is a nontrivial consistency check that using the three basis changes \eqref{eq:OPE_relation_vectors}, \eqref{eq:alpha_to_C_vectors}
and \eqref{eq:alpha_beta_vectors} given in appendix \ref{sec:fourier_transformation} below, 
these conservation conditions are related to the one in the $\beta_k (\nu)$ basis, $\beta_2 (\nu) = \beta_3 (\nu) = 0$,
stated  in \eqref{eq:conservation_cond_beta}.

%%%%%%%%%%%%%%%%%%%%%%%%%%%
\subsection{Correlator \texorpdfstring{$\bra T T {\cal O}{\cal O}\ket$}{<T T phi phi>}}
%%%%%%%%%%%%%%%%%%%%%%%%%%%

In the case two of the external operators are stress-tensors, the differential operators in \eqref{eq:spinning_discontinuity}
are
\begin{equation}
\begin{aligned}
D_1 &= D^2_{11} D^2_{22} \,,\\
D_2 &= H_{12} D_{11} D_{22}\,,\\
D_3 &= H_{12}^2\,,\\
D_4 &= D_{12} D_{11} D^2_{22} \Sigma^{-2} + D_{21} D_{22} D^2_{11} \Sigma^{2}\,,\\
D_5 &= H_{12} \left( D_{12} D_{22} \Sigma^{-2} + D_{21} D_{11} \Sigma^{2} \right),
\end{aligned}
\quad
\begin{aligned}
D_6 &= D^2_{12} D^2_{22} \Sigma^{-4} + D^2_{21} D^2_{11} \Sigma^{4}\,,\\
D_7 &= D_{12} D_{21} D_{11} D_{22} \,, \\
D_8 &= H_{12} D_{12} D_{21}\,, \\
D_9 &= D^2_{12} D_{21} D_{22} \Sigma^{-2} + D^2_{21} D_{12} D_{11} \Sigma^{2}\,, \\
D_{10} &= D^2_{12} D^2_{21}\,.
\end{aligned}
\label{eq:Dij_st}
\end{equation}
After commuting past the $\D_{12}$ dependent factor of the scalar block in the Regge limit, 
they can be written in terms of the basis of operators \eqref{DTensors_st}.
As in the case of currents one can read off the relations
\begin{align}
  \frac{\alpha_{k=1,\ldots,10} (\nu)}{\chi(\nu)} &= \sum\limits_{i=1}^{10} \#_i \, c^i_{TTj(\nu)} \,, \qquad
(\mathtt{\alpha to C} \text{ in Mathematica file})\,.
\label{eq:alpha_to_C_st}
\end{align}
We refrain from printing this and other lengthy relations here and provide them in
a Mathematica notebook which is included in the arXiv submission of this paper.

Another basis of tensor structures is defined by the three-point function
\beq
\bra T_1^{\D_1} T_2^{\D_1} \calO_{3}^{\D,J} \ket = \frac{  \sum_i \lambda_i Q_i}{(P_1 \cdot P_2)^{\frac{\D - 2\D_1}{2}} (P_2 \cdot P_3)^{-\frac{\D}{2}} (P_3 \cdot P_1)^{-\frac{\D}{2}}  }\,,
\eeq
where $Q_i$ are the tensor structures
\begin{equation}
\begin{aligned}
&Q_{1}=V_{1}^{2}V_{2}^{2}V_{3}^{J}\,,  \\
&Q_{2}=H_{23}V_{1}^{2}V_{2}V_{3}^{J-1}+H_{13}V_{1}V_{2}^{2}V_{3}^{J-1}\,,\\
&Q_{3}=H_{12}V_{1}V_{2}V_{3}^{J}\,,\\
&Q_{4}=H_{12}H_{13}V_{2}V_{3}^{J-1}+H_{12}H_{23}V_{1}V_{3}^{J-1}\,,\\
&Q_{5}=H_{13}H_{23}V_{1}V_{2}V_{3}^{J-2}\,,
\end{aligned}
\quad
\begin{aligned}
&Q_{6}=H_{12}^{2}V_{3}^{J}\,,\\
&Q_{7}=H_{13}^{2}V_{2}^{2}V_{3}^{J-2}+H_{23}^{2}V_{1}^{2}V_{3}^{J-2}\,,\\
&Q_{8}=H_{12}H_{13}H_{23}V_{3}^{J-2}\,,\\
&Q_{9}=H_{13}^{2}H_{23}V_{2}V_{3}^{J-3}+H_{13}H_{23}^{2}V_{1}V_{3}^{J-3}\,,\\
&Q_{10}=H_{13}^{2}H_{23}^2V_{3}^{J-4}\,.
\end{aligned}
\label{eq:3pt_tensor_structures_st}
\end{equation}
The relation between this basis and the one defined in \eqref{eq:spinning_discontinuity} is
\begin{align}
c^{k=1,\ldots,10}_{TTj(\nu)}
&= \sum\limits_{i=1}^{10} \#_i \, \lambda_i  \,, \qquad
(\mathtt{Cto \lambda} \text{ in Mathematica file})\,.
\label{eq:OPE_relation_st}
\end{align}
The conservation condition for the case $\D_1 =d$  is
\begin{align}
\lambda_{k=1,3,5,6,8,9,10}
= \sum\limits_{i=2,4,7} \#_i \, \lambda_i  \,, \qquad
(\mathtt{conservation \lambda} \text{ in Mathematica file})\,,
\label{eq:conservation_embedding_st}
\end{align}
and we checked that the three basis changes \eqref{eq:beta_to_alpha_st}, \eqref{eq:alpha_to_C_st} and
\eqref{eq:OPE_relation_st} relate this condition to \eqref{eq:conservation_cond_beta}.

Finally note the relation to the quantities $t_2, t_4$ which appear in the literature on 
conformal collider bounds. In appendix C.3 of \cite{arXiv:1603.03771} one can find the following relations, valid at 
$\nu_0 = - i h, j=2$,
\begin{align}
\lambda_2 ={}&
\frac{C_T  \Gamma(\frac{d}{2}+2)}{(d-1)^3 (d+1)^2 (d+2) \pi^\frac{d}{2}}
\Big(
2 \left(d^4+3 d^3-10 d^2+6 d+4\right) t_4
\nonumber\\
&-2 (d-1) (d+1) \left(3 d^3-5 d^2+d+2\right)+(d+1) \left(d^4+3 d^3-9 d^2+3 d+6\right) t_2
\Big)\,,
\nonumber
\\
\lambda_4 ={}&
\frac{C_T  \Gamma(\frac{d}{2}+1)}{(d-1)^3 (d+1)^2  \pi^\frac{d}{2}}
\Big(
\left(d^2-2 d+3\right) (d+1)^2 t_2+2 \left(d^3-d^2+2 d+2\right) t_4
\label{eq:OPE_to_t2_t4}
\\
&-2 (d-1) \left(d^3-d^2+1\right) (d+1)
\Big)\,,
\nonumber
\end{align}
\begin{align}
\lambda_7 ={}&
\frac{C_T  \Gamma(\frac{d}{2}+2)}{(d-1)^3 (d+1)^2 (d+2) \pi^\frac{d}{2}}
\Big(
(d+1) \left(d^3-3\right) t_2-(d-1) d (d+1) \left(2 d^2-2 d-1\right)
\nonumber\\
&+\left(3 d^3-d^2-2 d-4\right) t_4
\Big)\,.
\nonumber
\end{align}

%%%%%%%%%%%%%%%%%%%%%%%%%%%
\section{Fourier transformation}
\label{sec:fourier_transformation}
%%%%%%%%%%%%%%%%%%%%%%%%%%%

In this appendix we compute the Fourier transformation that relates the phase shift to the 
conformal correlator of a Regge pole  generalizing appendix A of \cite{Cornalba:2009ax} to any dimension $d$
and also to the correlator with two external stress-tensors.
%%%%%%%%%%%%%%%%%%%%%%%%%%%
\subsection{Correlator \texorpdfstring{$\bra {\cal J} {\cal J} {\cal O} {\cal O} \ket$}{<J J O O>}}
%%%%%%%%%%%%%%%%%%%%%%%%%%%
We start by
\begin{align}
B(p,\bar{p},z_1,z_2)= \frac{(-1)^{\D_1+\D_3}}{\pi^{2d}}   \int dx  d\bar{x} \,  e^{ 2 i p\cdot x +2 i \bar{p} \cdot \bar{x}}
A(x,\bar{x},z_1,z_2) \ , \label{Bft}
\end{align}
with  $A(x,\bar{x},z_1,z_2)$ given by (\ref{ReggeA}).
It is convenient to rewrite (\ref{ReggeA}) as 
\begin{align}
A(x,\bar{x},z_1,z_2)  \approx \int 
  d\nu~\sum_{k=0}^4\alpha_k (  \nu)\,x^2\mathcal{D}_k  \frac{ (-1)^{1-\D_1-\D_3} \Omega_{i\nu} ( \rho)}{
( x^2-i\epsilon_{x})^{\D_1 +\frac{j(\nu)+1}{2}}  ( \bar{x}^2-i\epsilon_{\bar{x}})^{\D_3+\frac{j(\nu)-1}{2}}     }~,
\end{align}
where in this equation we can write the operators $\mathcal{D}_k$ from \eqref{DTensors} explicitly in terms of $x$
\begin{align}
x^2\mathcal{D}_1=&\; (z_1 \cdot z_2) x^2- (z_1 \cdot x) (z_2 \cdot x) \,,
\nonumber\\ 
x^2\mathcal{D}_2=&\; (z_1 \cdot x) (z_2 \cdot x) \,,
\label{eq:op_fourier_1}
\\ 
x^2\mathcal{D}_3=&\;x^2\big((z_1 \cdot x) (z_2 \cdot \partial)+(z_2 \cdot x) (z_1 \cdot \partial)\big) -2 (z_1 \cdot x) (z_2 \cdot x) x\cdot \partial\,,
\nonumber\\ 
x^2\mathcal{D}_4=&\;x^4 (z_1 \cdot \partial) (z_2 \cdot \partial) -x^2 x^q \big((z_1 \cdot x) (z_2 \cdot \partial)+(z_2 \cdot x) (z_1 \cdot \partial)\big) \partial_q +  (z_1 \cdot x) (z_2 \cdot x)x^qx^s \partial_q \partial_s
\nonumber\\
&-\frac{1}{d-1} \left( (z_1 \cdot z_2) x^2- (z_1 \cdot x) (z_2 \cdot x)\right) \left( x^2 \partial^2 - x^qx^s \partial_q \partial_s \right)\,.
\nonumber
\end{align}
 One can then use integration by parts in (\ref{Bft}) to write
\begin{equation}
B(p,\bar{p},z_1,z_2) = -\frac{1}{\pi^{2d}}  \int
d\nu~\sum_{k=0}^4 \alpha_k(  \nu) \tilde{\mathcal{D}}_k 
\int 
\frac{ dx d\bar{x}\, e^{ 2i x\cdot p +2i \bar{x} \cdot \bar{p}}\, \Omega_{i\nu} ( \rho )}{
( x^2-i\epsilon_{x})^{\D_1 +\frac{j(\nu)+1}{2}}  ( \bar{x}^2-i\epsilon_{\bar{x}})^{\D_3+\frac{j(\nu)-1}{2}}    } \,,
 \label{Bint} 
\end{equation}
where 
\begin{align}
-4\tilde{\mathcal{D}}_1=&\; (z_1 \cdot z_2)\hat{\partial}^2- (z_1 \cdot \hat{\partial})(z_2 \cdot \hat{\partial}) \,,
\nonumber\\
 -4\tilde{\mathcal{D}}_2=&\;  (z_1 \cdot \hat{\partial})(z_2 \cdot \hat{\partial}) \,,
\nonumber\\
 -4\tilde{\mathcal{D}}_3=&\; -\hat{\partial}^2 \left( (z_1 \cdot \hat{\partial})(z_2 \cdot p)+(z_2 \cdot \hat{\partial})(z_1 \cdot p)\right)
 +2 (z_1 \cdot \hat{\partial})(z_2 \cdot \hat{\partial}) \hat{\partial}   \cdot p\,,
\\ 
-4\tilde{\mathcal{D}}_4 =&\;  \hat{\partial}^4(z_1 \cdot p)(z_2 \cdot p)  -  
\hat{\partial}^2 \hat{\partial}_s \left((z_1 \cdot \hat{\partial})(z_2 \cdot p)+(z_2 \cdot \hat{\partial})(z_1 \cdot p)\right) p^s +
\nonumber\\
&+ (z_1 \cdot \hat{\partial})(z_2 \cdot \hat{\partial}) \hat{\partial}_s\hat{\partial}_q p^s p^q
-\frac{1}{d-1} \left( (z_1 \cdot z_2)\hat{\partial}^2- (z_1 \cdot \hat{\partial})(z_2 \cdot \hat{\partial}) \right) \left(\hat{\partial}^2 p^2 - \hat{\partial}_s\hat{\partial}_q p^s p^q \right),
\nonumber
\end{align}
and $\displaystyle{\hat{\partial}_n =\frac{\partial\ }{\partial p^n}}$.
The scalar integral in the second line of (\ref{Bint}) can be done explicitly.
First notice that the $i\epsilon$-prescription implies that the integral vanishes if either $p$ or $\bar{p}$ is spacelike or past-directed.
We can then write
\begin{equation}
\frac{1}{\pi^{2d}}
 \int 
\frac{ dx d\bar{x} \,e^{ 2i x\cdot p +2i \bar{x} \cdot \bar{p}}\,\Omega_{i\nu} ( \rho )}{
( x^2-i\epsilon_{x})^{\D_1 +\frac{j(\nu)+1}{2}}  ( \bar{x}^2-i\epsilon_{\bar{x}})^{\D_3+\frac{j(\nu)-1}{2}}    }=
\frac{ \theta(p^0) \theta(-p^2) \theta(\bar{p}^0) \theta(-\bar{p}^2) G \left( e\cdot \bar{e} \right)}{
(-p^2)^{h -\D_1-\frac{j(\nu)+1}{2}}  (-\bar{p}^2)^{h-\D_3-\frac{j(\nu)-1}{2}}     }\,,
\end{equation}
just using Lorentz invariance and scaling.
Performing a Fourier transform we have 
\begin{equation}
\frac{  \Omega_{i\nu} ( \rho )}{
( x^2-i\epsilon_{x})^{\D_1 +\frac{j(\nu)+1}{2}}  ( \bar{x}^2-i\epsilon_{\bar{x}})^{\D_3+\frac{j(\nu)-1}{2}}    }=
\int_{\rm M} 
\frac{ dp d\bar{p}  \,e^{ -2i x\cdot p -2i \bar{x} \cdot \bar{p}}\, G \left( e\cdot \bar{e} \right)}{
(-p^2)^{h -\D_1-\frac{j(\nu)+1}{2}}  (-\bar{p}^2)^{h-\D_3-\frac{j(\nu)-1}{2}}     }\,,
\end{equation}
where we denote by ${\rm M}$ the future light-cone or Milne wedge.
To determine the function $G$ it is sufficient to consider future directed $x$ and $\bar{x}$.
In this case, after integrating over $E$ and $\bar{E}$ (recall that $p=E \,e$) we find
\begin{equation}
 \Omega_{i\nu} ( \rho ) 
 =\int_{H_{d-1}} de d\bar{e}\,  
\frac{ \Gamma(2\D_1+j(\nu)+1) \Gamma(2\D_3+j(\nu)-1)  \,G \left( e\cdot \bar{e} \right)}{
\left(-2 e \cdot x /|x|  \right)^{2\D_1+j(\nu)+1}  
\left(-2\bar{e}\cdot \bar{x}/|\bar{x}|\right)^{2\D_3+j(\nu)-1}     }\,.
\end{equation}
Each integral is a convolution of radial functions on $H_{d-1}$ that is easily done using the harmonic basis \cite{Penedones:2007ns}. This gives
$
G \left( e\cdot \bar{e} \right)= \zeta(\nu,1) \,\Omega_{i\nu}(L) 
$, with
\begin{equation}
\zeta(\nu,n)
= \frac{4 \pi^{2-d}   }
{
\Gamma\!\left( \frac{2\D_1+j(\nu) -h +i\nu}{2} +n \right)\Gamma\!\left( \frac{2\D_1+j(\nu)-h-i\nu}{2} +n \right)
\Gamma\!\left( \frac{2\D_3+j(\nu)-h+i\nu}{2} \right)\Gamma\!\left( \frac{2\D_3+j(\nu)-h-i\nu}{2} \right)
}\,,
\end{equation}
where the parameter $n$ was introduced for later convenience.

We may now return to (\ref{Bint}) to find 
\begin{equation}
B(p,\bar{p},z_1,z_2)\approx -  \int
d\nu~\sum_{k=0}^4 \alpha_k(  \nu) \, \tilde{\mathcal{D}}_k
\frac{\zeta(\nu,1) \,\Omega_{i\nu}(L) }{
(-p^2)^{h -\D_1-\frac{j(\nu)+1}{2}}  (-\bar{p}^2)^{h-\D_3-\frac{j(\nu)-1}{2}}     }\,.
\end{equation}
With long but trivial manipulations we can rewrite the operators $\tilde{\mathcal{D}}_k$ in the following convenient form
\begin{align}
-4p^2\tilde{\mathcal{D}}_1= &\;(z_1 \cdot z_2)p^2\hat{\partial}^2- p^2(z_1 \cdot \hat{\partial})(z_2 \cdot \hat{\partial}) \,,
\nonumber\\ -4p^2\tilde{\mathcal{D}}_2= &\; p^2(z_1 \cdot \hat{\partial})(z_2 \cdot \hat{\partial}) \,,
\nonumber\\ -4p^2\tilde{\mathcal{D}}_3=&\;- \left((z_1 \cdot p)(z_2 \cdot \hat{\partial})+(z_2 \cdot p)(z_1 \cdot \hat{\partial})\right) p^2\hat{\partial}^2
-2\left( (z_1 \cdot z_2)-2\frac{(z_1 \cdot p)(z_2 \cdot p)}{p^2}\right) p^2\hat{\partial}^2
\nonumber\\& +2  p^2 (z_1 \cdot \hat{\partial})(z_2 \cdot \hat{\partial}) \left(p\cdot \hat{\partial} +d-2\right)\,,
\nonumber
%\\
\end{align}
\begin{align}
 -4p^2\tilde{\mathcal{D}}_4 =&\;
\frac{(z_1 \cdot p)(z_2 \cdot p)}{p^2} \left( p^2\hat{\partial}^2 - \frac{2(d-4)}{d-1} p \cdot \hat{\partial} +\frac{12}{d-1}  \right)  p^2\hat{\partial}^2\\&
- \left((z_1 \cdot p)(z_2 \cdot \hat{\partial})+(z_2 \cdot p)(z_1 \cdot \hat{\partial})\right)  p^2\hat{\partial}^2\left( \frac{d+2}{2(d-1)} p\cdot \hat{\partial} + \frac{d^2-5 d +10}{2(d-1)} \right)
\nonumber\\&
 +\frac{1}{d-1}  p^2 (z_1 \cdot \hat{\partial})(z_2 \cdot \hat{\partial})
 \left(  p^2 \hat{\partial}^2+2(p\cdot \hat{\partial})^2  +2(d+1) p\cdot \hat{\partial}+6 (d-2)\right)
\nonumber\\
&-\frac{1}{d-1}  (z_1 \cdot z_2)p^2 \hat{\partial}^2
\left(  p^2 \hat{\partial}^2 -  (p\cdot \hat{\partial})^2 + (5-d)p\cdot \hat{\partial} +6\right)\,,
\nonumber
\end{align}
so that the commuting operators $p\cdot \hat{\partial}$ and $p^2 \hat{\partial}^2$ can be traded by their eigenvalues,
\begin{align}
p\cdot \hat{\partial}&\to 2\D_1+j(\nu)+1-2h \ ,
\nonumber\\
p^2 \hat{\partial}^2
&= \left(p\cdot \hat{\partial}\right)^2 +(2h-2) p\cdot \hat{\partial} -\nabla^2  
\\
& \to \big(2\D_1+j(\nu)+1-2h\big)\big(2\D_1+j(\nu)-1\big) +\left(h-1 \right)^2+\nu^2\ ,
 \nonumber
\end{align}
where $ \nabla^2$ is the Laplacian on the $(d-1)$-dimensional hyperboloid $p^2=-1$.
It is then a trivial computation to obtain the form (\ref{calB}) with\footnote{We suppressed the argument $\nu$ of 
the functions $j(\nu)$ and $\alpha_k(\nu)$ on the right-hand-side to reduce the size of the expressions. Also recall the definition of $\omega$ in \eqref{eq:omega}.}  
\begin{align}
\frac{4\beta_1(\nu)}{\zeta(\nu,1)} ={}&
\left(2+j-j^2+d (-2+j-\omega )+2 \omega +2 \left(1+d-2 j-2 \Delta _1\right) \Delta _1\right) \alpha _1
\nonumber\\
&+\left(-1+d-j-\omega -2 \Delta _1\right) \alpha _2+2 \omega  \left(d-j-2 \Delta _1\right) \alpha _3+(-2+d) \omega  (1+\omega ) \alpha _4\,,
\nonumber\\
\frac{4\beta_2(\nu)}{\zeta(\nu,1)} ={}&
(-1+d) \left(-1+d-j-\omega -2 \Delta _1\right) \alpha _1-\left(-1+d-j-2 \Delta _1\right) \left(d-j-2 \Delta _1\right) \alpha _2
\nonumber\\
&-2 (-1+d) \omega  \left(d-j-2 \Delta _1\right) \alpha _3-(-2+d) (-1+d) \omega  (1+\omega ) \alpha _4\,,
\label{eq:alpha_beta_vectors}\\
\frac{4\beta_3(\nu)}{\zeta(\nu,1)} ={}&
\left(-d+j+2 \Delta _1\right) \alpha _1+\left(d-j-2 \Delta _1\right) \alpha _2+\big((1+d-j) (-1+j)+(-1+d) \omega 
\nonumber\\
&+2 \left(2+d-2 j-2 \Delta _1\right) \Delta _1\big) \alpha _3+(-2+d) (1+\omega ) \left(-1+j+2 \Delta _1\right) \alpha _4\,,
\nonumber\\
\frac{4\beta_4(\nu)}{\zeta(\nu,1)} ={}&
\alpha _1-\alpha _2-2 \left(-1+j+2 \Delta _1\right) \alpha _3+\left(-(-1+j)^2-\omega -4 \Delta _1 \left(-1+j+\Delta _1\right)\right) \alpha _4\,.
\nonumber
\end{align}

%%%%%%%%%%%%%%%%%%%%%%%%%%%%%%%%%%%%%%%%%%%%%%%%%%%%%%%%%%%%%%%%%%%%%%%%%%%%%%%%%%%%%%%%%%%
\subsection{Correlator \texorpdfstring{$\bra T T \phi \phi \ket$}{<T T phi phi>}}
%%%%%%%%%%%%%%%%%%%%%%%%%%%%%%%%%%%%%%%%%%%%%%%%%%%%%%%%%%%%%%%%%%%%%%%%%%%%%%%%%%%%%%%%%%%

Now we compute the Fourier transform
\eqref{Bft}
that relates the $\beta_k(\nu)$ to the
$\alpha_k(\nu)$ in equations \eqref{eq:BtoCalB} and \eqref{ReggeA} for the case of two external stress-tensors.

The computation follows the same logic as in the previous section, so we will not repeat it here in full detail.
It can be simplified by first rewriting the first three operators of \eqref{DTensors_st} as
\begin{align}
\calD_1 ={}& \Big(z_1 \cdot z_2  + (z_1 \cdot \hat x) (z_2 \cdot \hat x)\Big)^2 - \frac{1}{d-1}\, \calD_5\,,
\nonumber\\
\calD_2 ={}& \Big(z_1 \cdot z_2  + (z_1 \cdot \hat x) (z_2 \cdot \hat x)\Big) (z_1 \cdot \nabla) (z_2 \cdot \nabla) - \frac{1}{d-1} \calD_9 + \frac{1}{(d-1)^2}
\calD_5 \nabla^2 - \frac{1}{d-1} \calD_1 \nabla^2 \,,
\nonumber\\
\calD_3 ={}&  \frac{1}{2} (z_1 \cdot \nabla) (z_2 \cdot \nabla) (z_1 \cdot \nabla) (z_2 \cdot \nabla) + \frac{1}{2} (z_2 \cdot \nabla) (z_1 \cdot \nabla) (z_2 \cdot \nabla) (z_1 \cdot \nabla) 
\label{eq:Diff123_simplified}
\\
&- \frac{1}{2(d-1)} \Big( \calD_9 (5-3d +2 \nabla^2) + 2 \calD_5 \nabla^2  \Big)
+ \frac{1}{(d-1)^2} \,\calD_5 (2-d+ \nabla^2) \nabla^2
\nonumber\\
&-\frac{1}{d+1} \left( \mathcal{D}_{2} +\frac{1}{d-1} \mathcal{D}_{1} \nabla^2 \right) (3-d+2 \nabla^2)\,.
\nonumber
\end{align}
Now the $\nabla^2$ can  be replaced by its eigenvalue $-\nu^2-(h-1)^2$ before doing the Fourier transformation.

The resulting relation between the $\beta_k(\nu)$ and the $\alpha_k(\nu)$ is
\bea
\frac{4\beta_{k=1,\ldots,10}(\nu)}{\zeta(\nu,5)} ={}& \sum_{i=1}^{10} \#_i \, \alpha_i (\nu_0)\,, \qquad
(\mathtt{\beta\  to\  \alpha} \text{ in Mathematica file})\,.
\eea{eq:beta_to_alpha_st}
Notice that 
the overall factor in this relation
is $\zeta(\nu,5)$ because the differential operators \eqref{DTensors_st} need to be multiplied by $x^{10}$,
in order to remove all powers of $x^2$ from the denominators.

%%%%%%%%%%%%%%%%%%%%%%%%%%%%%%%%%%%%%%%%%%%%%%%%%%%%%%%%%%%%%%%%%%%%%%%%%%%%%%%%%%%%%%%%%%%
\section{Functions \texorpdfstring{$f_2(\nu_0,L)$ and $f_4(\nu_0,L)$}{t2(L) and t4(L)}}
\label{sec:t2_and_t4}
%%%%%%%%%%%%%%%%%%%%%%%%%%%%%%%%%%%%%%%%%%%%%%%%%%%%%%%%%%%%%%%%%%%%%%%%%%%%%%%%%%%%%%%%%%%

The full expressions for the functions $f_2(\nu_0,L)$ and $f_4(\nu_0,L)$ which appear in \eqref{eq:tensors_n} are
\begin{align}
\,
&f_2(\nu_0,L) =
 \frac{4 e^L \left(\nu _0-i (h-1)\right)}{(2 h+1) \left(e^{2 L}-1\right)^2 \left(\nu _0-i\right) 
\, _2F_1\big(h-1,h+i \nu _0-1;i \nu _0+1;e^{-2 L}\big)} 
\nonumber\\
&\Bigg[
\left(h-i \nu _0-1\right) \, _2F_1\big(h,h+i \nu _0;i \nu _0+2;e^{-2 L}\big) 
\Big[i \nu _0 \Big(\left(h (7 h+6)-\nu _0^2+1\right) \sinh (3 L)
\nonumber\\
& 
+\left(h (11 h+14)+3 \nu_0^2+5\right) \sinh (L)\Big)
 +\left(h (h+1) (3 h+1)-(5 h+2) \nu _0^2\right) \cosh (3 L)
\nonumber\\
&+\left((5 h+2) \nu _0^2+h (h (29 h+28)+7)\right) \cosh (L)\Big]
+h \, _2F_1\big(h+1,h+i \nu _0;i \nu _0+2;e^{-2 L}\big)
\nonumber\\
& \times\Big(\left(h \left(-h (10 h+3)+6 \nu _0^2+4\right)+\nu _0^2+1\right) \cosh (3 L)
\nonumber\\
&-\left(h \left(h (54 h+29)+6 \left(\nu
   _0^2-2\right)\right)+\nu _0^2-7\right) \cosh (L)
   \nonumber\\
&+2 \left(h-i \nu _0-1\right) \sinh (L) \left(\left(h (7 h+6)-\nu _0^2+1\right) \cosh (2 L)+h (9 h+10)+\nu _0^2+3\right)\Big)
\Bigg]\,,
\nonumber
%\\
\end{align}
\begin{align}
&f_4(\nu_0,L) = 
\frac{2 e^L \left(\nu _0-i (h-1)\right)}{\left(e^{2 L}-1\right)^2 \left(\nu _0-i\right) \, _2F_1\big(h-1,h+i \nu _0-1;i \nu _0+1;e^{-2 L}\big)}
\\
&\Bigg[
2 \, _2F_1\big(h+1,h+i \nu _0;i \nu _0+2;e^{-2 L}\big) \times 
\nonumber\\
&\Big[2 h (2 h+1) \cosh (L) \Big(\left(h^2+h-\nu _0^2-1\right) \cosh (2 L)+3 h (h+1)+\nu _0^2-2\Big)
\nonumber\\
&-h \left(h-i \nu
   _0-1\right) \sinh (L) \left(\left(3 h (h+2)-\nu _0^2+2\right) \cosh (2 L)+5 h (h+2)+\nu _0^2+4\right)\Big]
   \nonumber\\
&+\left(-h+i \nu _0+1\right) \, _2F_1\big(h,h+i \nu _0;i \nu
   _0+2;e^{-2 L}\big)\times
   \nonumber\\
& \Big[i \nu _0 \Big(\left(3 h (h+2)-\nu _0^2+2\right) \sinh (3 L)+\left(7 h (h+2)+3 \left(\nu _0^2+2\right)\right) \sinh (L)\Big)
\nonumber\\
&+(h+1) \left(h (h+2)-3
   \nu _0^2\right) \cosh (3 L)+\Big(3 (h+1) \nu _0^2+h \big(h (15 h+29)+10\big)\Big) \cosh (L)\Big]
\Bigg]\,.
\nonumber
\end{align}

%%%%%%%%%%%%%%%%%%%%%%%%%%%
\section{Convexity of the leading Regge trajectory}
\label{sec:convexity}
%%%%%%%%%%%%%%%%%%%%%%%%%%%

The leading Regge trajectory is the set of operators of minimal dimension $\Delta(J)$ for each spin $J$. Here we focus on the trajectory with vacuum quantum numbers which includes the stress tensor   with $\Delta=d$ and $J=2$.
As shown in \cite{Caron-Huot:2017vep}, this trajectory can be continued to complex spin.
Below we shall prove, using the recent results of \cite{Caron-Huot:2017vep}, that the continuous trajectory is a monotonic convex function (for $J>1$)  as depicted in figure \ref{fig:leadingReggeTraj}. 
We emphasize that this proof applies to the leading Regge trajectory of a full non-perturbative CFT. In particular, in the case of large $N$ CFTs the exact trajectory can be different from the leading single-trace trajectory which plays a central role throughout this paper.

%%%%%%%%%%%%%%%%%%%%%%%%%%%
\subsection{Proof of convexity}
%%%%%%%%%%%%%%%%%%%%%%%%%%%

In a recent paper \cite{Caron-Huot:2017vep},  Caron-Huot showed how to "invert" the OPE. 
We will need the generating function
\beq
C(z,\beta) =\int_z^1 \frac{d\bar{z}}{\bar{z}^2} f_\beta(\bar{z})\, {\rm dDisc}\left[ \mathcal{A}(z,\bar{z})\right]\,,
\label{defCbeta}
\eeq
where the double discontinuity ${\rm dDisc}\left[ \mathcal{A}(z,\bar{z})\right] \ge 0$  for  $0\le z,\bar{z} \le 1$ and
\beq
f_\beta(\bar{z}) = \frac{\Gamma^2 \big(\frac{\beta}{2}\big)}{\Gamma(\beta)}
\bar{z}^\frac{\beta}{2}
\ _2F_1\!\left( \frac{\beta}{2},\frac{\beta}{2},\beta,\bar{z}\right)
=\int_0^1\frac{dt}{t(1-t)} \left(\frac{t(1-t)\bar{z}}{1-\bar{z}t}\right)^{\frac{\beta}{2}}\ .
\label{eq:fbetaintrep}
\eeq
We have normalized the function $f_\beta(\bar{z})$ in a way that will be convenient for our purposes.
Caron-Huot showed that the leading Regge trajectory controls  the small $z$ behavior of the function \beq
C(z,\beta)\approx c(\beta) z^{\frac{1}{2} \tau(\beta)} +\dots\,, \qquad z\to 0\,.
\eeq
The scaling dimensions $\Delta(J)$ of the operators of spin $J$ in the leading Regge trajectory are given by
\beq
\Delta(J)-J = \tau \big(\Delta(J)+J\big)\,.
\eeq

We will prove that 
\beq
0\le \tau'(\beta)\le 1\,,\qquad \tau''(\beta)\le 0\,,
\label{eq:betaconditions}
\eeq
for $J=\frac{1}{2}\big(\beta-\tau(\beta)\big)>1$ which was assumed in 
\cite{Caron-Huot:2017vep}. For our arguments below, we will only use $\beta>0$ so that the integral representation \eqref{eq:fbetaintrep} is valid.
Notice that \eqref{eq:betaconditions} implies that  
\beq
\frac{d\Delta}{dJ}=\frac{1+\tau'(\beta)}{1-\tau'(\beta)}\ge1\,,\qquad \frac{d^2\Delta}{dJ^2}=\frac{4\tau''(\beta)}{\big(1-\tau'(\beta)\big)^3}\le0\,.% ,\qquad {\rm for}\, J>1\,.
\eeq
This result extends the discrete convexity properties of the leading Regge trajectory derived in \cite{Nachtmann:1973mr, Komargodski:2012ek} to the continuation to any real value of spin $J>1$.

Our strategy to prove \eqref{eq:betaconditions} is to study
the small $z$ behavior of 
\begin{align}
\frac{\partial}{\partial \beta}\log C(z,\beta) &= \frac{1}{2}\tau'(\beta) \log z +O(z^0)\,, \label{dbetalogC}\\
\frac{\partial^2}{\partial \beta^2}\log C(z,\beta) &= \frac{1}{2}\tau''(\beta) \log z +O(z^0)\,. \label{d2betalogC}
\end{align}
The first derivative of $\log C(z,\beta)$ can be written as the average
\beq
\frac{\partial}{\partial \beta}\log C(z,\beta) = \left\langle 
\frac{\partial}{\partial \beta}\log f_\beta(\bar{z}) \right\rangle_{z,\beta}\equiv
\int_z^1 d\bar{z} \rho_{z,\beta}(\bar{z}) \frac{\partial}{\partial \beta}\log f_\beta(\bar{z})  \,,
\label{averagedbetalogf}
\eeq
where 
\beq
\rho_{z,\beta}(\bar{z})=\frac{1}{C(z,\beta)}
\frac{1}{\bar{z}^2}  f_\beta(\bar{z})  \, {\rm dDisc}\left[ \mathcal{A}(z,\bar{z})\right] 
\eeq
is a normalized non-negative distribution in the interval $\bar{z} \in [z,1]$.
From the series representation
\beq
f_\beta(\bar{z}) = \sum_{n=0}^\infty \frac{\Gamma^2\!\left(\frac{\beta}{2}+n\right)}{n! \Gamma(\beta+n)} 
\,\bar{z}^{\frac{\beta}{2}+n}\,,
\eeq
one obtains
\beq
\frac{\partial}{\partial \beta} f_\beta(\bar{z}) = \sum_{n=0}^\infty \frac{\Gamma^2\left(\frac{\beta}{2}+n\right)}
{n! \Gamma(\beta+n)} \,\bar{z}^{\frac{\beta}{2}+n}\left[
\frac{1}{2}\log \bar{z} -\psi(\beta+n)+\psi\!\left(\frac{\beta}{2}+n\right)
\right] ,
\eeq
where $\psi(x)\equiv \frac{d}{dx}\log\Gamma(x)$. Since $\psi(x)$ is a growing function of $x$  for $x>0$, we conclude that $\frac{\partial}{\partial \beta} f_\beta(\bar{z})<0$ for any $\bar{z}\in [0,1]$.
Together with  \eqref{dbetalogC}  this implies that $\tau'(\beta)>0$.

The function $\frac{\partial}{\partial \beta}\log f_\beta(\bar{z})$ is a smooth growing function of $\bar{z}\in [0,1]$. The only region where it diverges is for $\bar{z} \to 0$ where it behaves as
\beq
\frac{\partial}{\partial \beta}\log f_\beta(\bar{z}) =
\frac{1}{2}\log \bar{z} -\psi(\beta)+\psi\!\left(\frac{\beta}{2}\right)+ \frac{z}{4}+O(z^2)\,.
\eeq 
Indeed we have the bound 
\beq
\frac{\partial}{\partial \beta}\log f_\beta(\bar{z})>
\frac{1}{2}\log z -\psi(\beta)+\psi\!\left(\frac{\beta}{2}\right) ,
\eeq
 for all $\bar{z}\in [z,1]$, which means we can also bound the average
\beq
\left\langle 
\frac{\partial}{\partial \beta}\log f_\beta(\bar{z}) \right\rangle_{z,\beta}>
\frac{1}{2}\log z -\psi(\beta)+\psi\!\left(\frac{\beta}{2}\right) .
\eeq
Taking the limit $z\to 0$ and comparing with \eqref{dbetalogC} we conclude that $\tau'(\beta)<1$.

In fact, in order to obtain the $\log z$ divergence as $z \to 0$ of the average \eqref{averagedbetalogf} predicted by \eqref{dbetalogC}, the distribution $\rho_{z,\beta}(\bar{z})$ must have a finite weight localized in the small region $\bar{z} \sim z$ when $z\to 0$. This means we can write
\beq
\tau'(\beta) = \lim_{a\to \infty} \lim_{z\to 0} \frac{
\int_z^{a z} \frac{d\bar{z}}{\bar{z}^2} f_\beta(\bar{z})\, {\rm dDisc}\left[ \mathcal{A}(z,\bar{z})\right]
}{
\int_z^1 \frac{d\bar{z}}{\bar{z}^2} f_\beta(\bar{z})\, {\rm dDisc}\left[ \mathcal{A}(z,\bar{z})\right]
}\,,
\eeq
with this order of limits. This also proves the first inequalities in \eqref{eq:betaconditions}.

We shall now prove that 
\beq
\frac{\partial^2}{\partial \beta^2}\log C(z,\beta)=
\frac{C(z,\beta)\frac{\partial^2}{\partial \beta^2}C(z,\beta)-\left[\frac{\partial}{\partial \beta}C(z,\beta)\right]^2}{\left[C(z,\beta)\right]^2} \ge 0 \,. \label{posd2logC}
\eeq
In the limit $z\to 0$, this implies that $\tau''(\beta)\le0$ using \eqref{d2betalogC}.
Using the definition \eqref{defCbeta}, we can easily write
\begin{align}
&C(z,\beta)\frac{\partial^2}{\partial \beta^2}C(z,\beta)-\left[\frac{\partial}{\partial \beta}C(z,\beta)\right]^2=\\
=&\int_z^1 \frac{d\bar{z}}{\bar{z}^2}
\int_z^1 \frac{d\bar{w}}{\bar{w}^2}\, {\rm dDisc}\left[ \mathcal{A}(z,\bar{z})\right]\, {\rm dDisc}\left[ \mathcal{A}(z,\bar{w})\right] \frac{1}{2} K_\beta( \bar{z},\bar{w}) \,,
\nonumber
\end{align}
where
\begin{align}
K_\beta( \bar{z},\bar{w})&=
  f_\beta(\bar{z})\frac{\partial^2}{\partial \beta^2} f_\beta(\bar{w})
+f_\beta(\bar{w})\frac{\partial^2}{\partial \beta^2} f_\beta(\bar{z})-
2\frac{\partial}{\partial \beta} f_\beta(\bar{z})\frac{\partial}{\partial \beta} f_\beta(\bar{w})\\
&=\frac{1}{4}\int_0^1\frac{dt}{t(1-t)} \frac{ds}{s(1-s)}\left(\frac{t(1-t)\bar{z}}{1-\bar{z}t}\right)^{\frac{\beta}{2}}
 \left(\frac{s(1-s)\bar{w}}{1-\bar{w}s}\right)^{\frac{\beta}{2}}\times\\
 &\qquad\times
 \left[\ln\left(\frac{t(1-t)\bar{z}}{1-\bar{z}t}\right) -\ln \left(\frac{s(1-s)\bar{w}}{1-\bar{w}s}\right)
 \right]^2 \ge 0\,.
\end{align}
Therefore, condition \eqref{posd2logC} follows from positivity of the double discontinuity together with positivity of $K_\beta( \bar{z},\bar{w})$.

\bibliographystyle{JHEP}

\bibliography{regge_causality}

\end{document}